\documentclass[fleqn]{2020SCGE}

\usepackage[T1]{fontenc}
\usepackage{footmisc}
\usepackage{subfigure}
\usepackage{diagbox}
\usepackage{amsfonts,amsmath,amstext,amssymb,mathrsfs,gensymb}
\usepackage{makecell,multirow}
\usepackage{ulem}

\newcommand{\uvec}[1]{\hat{#1}}
\newcommand{\mat}[1]{\textbf{#1}}

\newcommand{\aap}{Astron. \& Astrophys.}
\newcommand{\aj}{Astron. J.}
\newcommand{\apj}{Astrophys. J.}
\newcommand{\apjl}{Astrophys. J. Lett.}
\newcommand{\apjs}{Astrophys. J. Suppl.}
\newcommand{\araa}{Annu. Rev. Astron. Astrophys.}
\newcommand{\baas}{Bull. Am. Astron. Soc.}
\newcommand{\mnras}{Mon. Not. R. Astron. Soc.}
\newcommand{\pasa}{Publ. Astron. Soc. Aust.}
\newcommand{\prd}{Phys. Rev. D}
\newcommand{\raa}{Res. Astron. Astrophys.}

\newcommand{\m}{\ensuremath{\,{\rm m}}}
\newcommand{\K}{\ensuremath{\, {\rm K}}}
\newcommand{\kHz}{\ensuremath{\, {\rm kHz}}}
\newcommand{\MHz}{\ensuremath{\, {\rm MHz}}}
\newcommand{\GHz}{\ensuremath{\, {\rm GHz}}}
\newcommand{\Jy}{\ensuremath{\,{\rm Jy}}}
\newcommand{\dB}{\ensuremath{\,{\rm dB}}}

\begin{document}
\ensubject{subject}
\ArticleType{Article}
\Year{2020}
\Month{}
\Vol{}
\No{}
\DOI{}
\ArtNo{000000}
\ReceiveDate{}
\AcceptDate{}

\title{\vspace{-1cm}
The Tianlai Cylinder Pathfinder Array: \\
System Functions and Basic Performance Analysis}{}

\author[1,2]{Jixia LI}{}
\author[3,1]{Shifan ZUO}{}
\author[1]{Fengquan WU}{}
\author[1]{Yougang WANG}{}
\author[4]{Juyong ZHANG}{}
\author[1,2]{Shijie SUN}{}
\author[1]{\\Yidong XU}{}
\author[1,2]{Zijie YU}{}
\author[5]{Reza ANSARI}{}
\author[6]{Yichao LI}{}
\author[7]{Albert STEBBINS}{}
\author[8]{Peter TIMBIE}{}
\author[1,2]{\\Yanping CONG}{}
\author[9]{Jingchao GENG}{}
\author[10]{Jie HAO}{}
\author[1]{Qizhi HUANG}{}
\author[1]{Jianbin LI}{}
\author[11]{Rui LI}{}
\author[1]{\\Donghao LIU}{}
\author[1,2]{Yingfeng LIU}{}
\author[4]{Tao LIU}{}
\author[7]{John P. MARRINER}{}
\author[1]{Chenhui NIU}{}
\author[12]{Ue-Li PEN}{}
\author[13]{\\Jeffery B. PETERSON}{}
\author[1]{Huli SHI}{}
\author[10]{Lin SHU}{}
\author[10]{Yafang SONG}{}
\author[14]{Haijun TIAN}{}
\author[9]{Guisong WANG}{}
\author[14]{\\Qunxiong WANG}{}
\author[4]{Rongli WANG}{}
\author[11]{Weixia WANG}{}
\author[15]{Xin WANG}{}
\author[1,2]{Kaifeng YU}{}
\author[16]{\\Jiao ZHANG}{}
\author[1]{Boqin Zhu}{}
\author[4]{Jialu ZHU}{}
\author[1,2,17]{Xuelei CHEN \thanks{Corresponding author: xuelei@cosmology.bao.ac.cn}}{}

\AuthorMark{Li J X}
\AuthorCitation{Li J X et al}

\address[1]{~National Astronomical Observatories, Chinese Academy of Sciences, Beijing 100101, China}
\address[2]{~School of Astronomy and Space Science, University of Chinese Academy of Sciences, Beijing 100049, China}
\address[3]{~Center for Astrophysics and Department of Astronomy, Tsinghua University, Beijing 100084, China}
\address[4]{~School of Mechanical Engineering, Hangzhou Dianzi University, Hangzhou 310018, China}
\address[5]{~IJC Lab, CNRS/IN2P3 \& Universit\'e Paris-Saclay,  91405 Orsay, France}
\address[6]{~Department of Physics and Astronomy, University of the Western Cape, Belville 7535, Republic of South Africa}
\address[7]{~Fermi National Accelerator Laboratory, P.O. Box 500, Batavia IL 60510-5011, USA}
\address[8]{~Department of Physics, University of Wisconsin Madison, Madison WI 53703, USA}
\address[9]{~The 54th Research Institute, China Electronics Technology Group Corporation, Shijiazhuang, Hebei 050051, China}
\address[10]{~Institute of Automation, Chinese Academy of Sciences, Beijing 100190, China}
\address[11]{~Xinjiang Astronomical Observatory, Chinese Academy of Sciences, Urumqi, Xinjiang 830001, China}
\address[12]{~Canadian Institute for Theoretical Astrophysics, University of Toronto, Toronto, M5S 3H8, Canada }
\address[13]{~Department of Physics, Carnegie Mellon University, Pittsburgh, PA 15213, USA}
\address[14]{~Center for Astronomy and Space Sciences, China Three Gorges University, Yichang 443002, China}
\address[15]{~School of Physics and Astronomy, Sun Yat-Sen University, Guangzhou 510297, China}
\address[16]{~College of Physics and Electronic Engineering, Shanxi University, Taiyuan 030006, China}
\address[17]{~Center of High Energy Physics, Peking University, Beijing 100871, China}

\abstract{The Tianlai Cylinder Pathfinder is a radio interferometer array designed to test techniques for 21~cm intensity mapping in the post-reionization Universe, with the ultimate aim of mapping the large scale structure and measuring cosmological parameters such as the dark energy equation of state.  Each of its three parallel cylinder reflectors is oriented in the north-south  direction, and the array has a large field of view. As the Earth rotates, the northern sky is observed by drift scanning. The array is located in Hongliuxia, a radio-quiet site in Xinjiang, and saw its first light in September 2016. In this first data analysis paper for the Tianlai cylinder array, 
we discuss the sub-system qualification tests, and present basic system performance obtained from preliminary analysis  of the commissioning observations during 2016-2018. We show typical interferometric visibility data, from which we derive the actual beam profile in the east-west direction and the frequency band-pass response. 
We describe also the calibration process to determine the complex gains for the array elements, either using bright astronomical point sources, or an artificial on site calibrator source,  and discuss the instrument response stability, crucial for transit interferometry.  Based on this analysis,  we find a system temperature of about 90 K, and we also estimate the sensitivity of the array.
}


\maketitle

\begin{multicols}{2}

\section{Introduction}
\label{sec:introduction}

\begin{figure*}
  \centering
  \includegraphics[width=0.9\textwidth,trim = 0 150 0 50, clip]{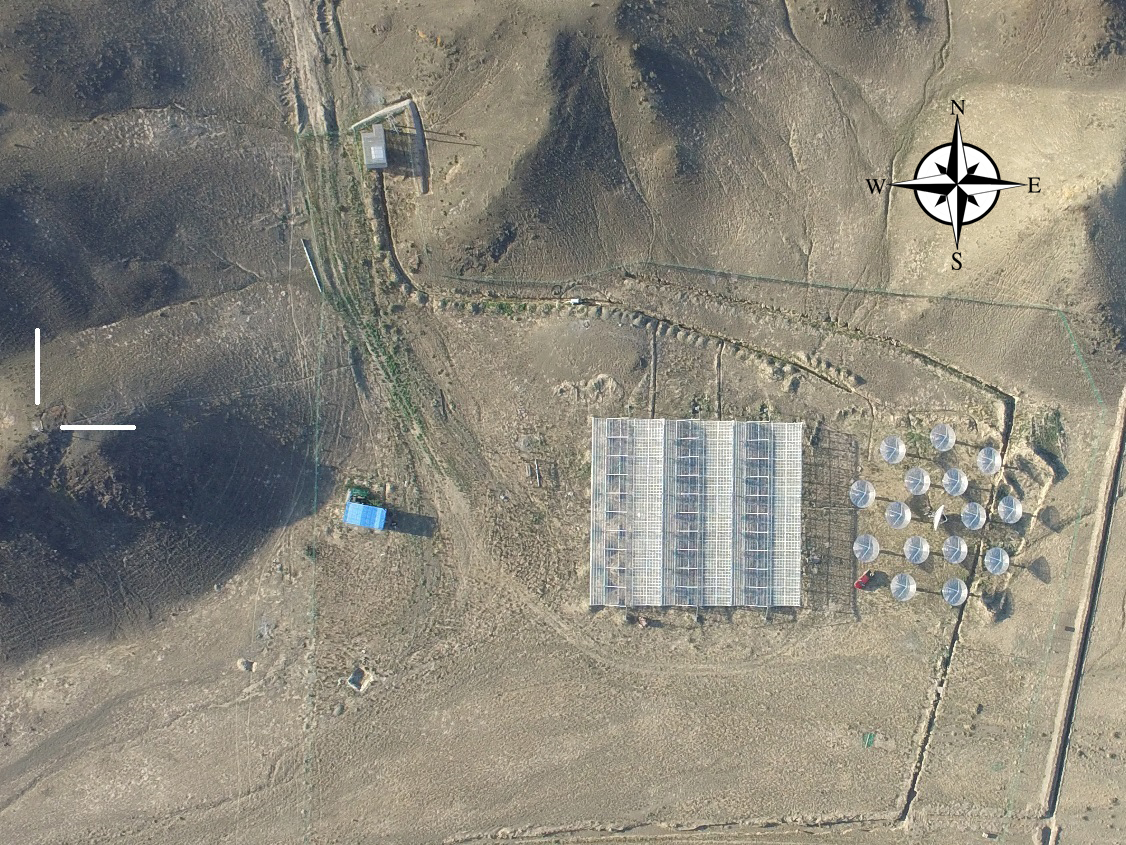}
  \caption{A top view photograph of the Tianlai cylinder and dish arrays. The dish array is located on the east side of the cylinder array, and the calibrator noise source (CNS) is located on a hill to the west of the arrays, as marked on the left side of the photograph. There is a blue cabin for storage between the array and the CNS. }
  \label{fig:topview}
\end{figure*}

This paper describes our analysis for the performance characteristics of the Tianlai Cylinder Pathfinder.  This array consists of three parallel cylindrical reflectors, each with a width of 15 meters and a length of 40 meters, with the long axis aligned in the north-south (N-S) direction. It is co-located with the Tianlai Dish Pathfinder, an interferometer array consisting 16 dishes of 6 meter aperture. These complementary designs were chosen specifically for testing different approaches to 21~cm intensity mapping.  Both saw their first light in 2016. An aerial photograph of the Tianlai arrays site is shown in Fig.~\ref{fig:topview}.

Intensity mapping is a technique for making rapid, low angular resolution surveys of the large scale structure using a redshifted spectral line \cite{Kovetz2019}, without attempting to resolve individual galaxies. The 21~cm line emitted by neutral hydrogen is of particular interest \cite{Liu&Shaw2019,Morales&Wyithe2010}.  So far, the 21~cm signal has only been marginally detected by this method in cross-correlation with optical galaxy redshift surveys by two instruments: the Green Bank Telescope (GBT) \cite{Masui2013,Switzer2013} and the Parkes Observatory \cite{Anderson2018}. Nevertheless, 21~cm intensity mapping is considered to be a potentially powerful probe of cosmology:  measuring the equation of state of dark energy from the baryon acoustic oscillation (BAO) features (e.g. \cite{2012A&A...540A.129A,Xu2015}), constraining models of inflation  (e.g. \cite{Xu2016}), as well as studying the Epoch of Reionization (EoR) ($6\lesssim z \lesssim 20$), Cosmic Dawn ($z\sim 20$), and the dark ages ($z\gtrsim 24$).  A primary observable for cosmological studies is the three-dimensional power spectrum of the underlying dark matter, which is traced by neutral hydrogen. Intensity mapping provides a natural means to compute this spectrum over a broad range of wavenumbers, $k$, in which the perturbations are in the linear regime.

While HI intensity mapping is being used to study the EoR and Cosmic Dawn by a variety of meter wavelength instruments, including LOFAR \cite{LOFAR2013}, MWA \cite{MWA2013}, HERA \cite{HERA2017}, PAPER \cite{PAPER2010}, and LWA \cite{LWA2018}, this paper focuses on measurements of the post-reionization epoch. Several dedicated instruments have been constructed, or are under development, to detect the 21~cm signal: CHIME \cite{CHIME}, Tianlai \cite{2012IJMPS..12..256C,chen2015AAPPS,Xu2015,2016RAA....16..158Z,Das2018}, HIRAX \cite{HIRAX}, OWFA \cite{OWFA}, and BINGO \cite{BINGO}. Other instruments are being designed and built to test the technique, including BMX and PAON-4 \cite{PAON4_Zhang_2016, Ansari2019}. With the exception of BINGO, they are all interferometer arrays with large numbers of receivers in order to provide enough mapping speed to detect the weak 21~cm signal; and all are laid out in a compact configuration in order to provide sensitivity at the relatively large scales ($0.5\gtrsim k \gtrsim 0.05$) where the BAO features appear in the power spectrum.

The 21~cm signal is $4\sim 5$ orders of magnitude lower than the foreground emission (primarily synchrotron radiation) from Galactic and extra-galactic radio sources (see e.g., Ref.\cite{Huang:2018ral}), making the detection a great challenge. Extracting the 21~cm signal generally relies on the fact that foreground emissions are smooth functions of frequency, while the 21~cm spectrum has a structure arising from the large-scale distribution of matter along the line of sight \cite{Liu&Shaw2019}. However, instrumental effects can introduce structure into the spectrum of otherwise smooth foregrounds. In particular, the antenna and array beam patterns are frequency-dependent, which aliases the angular dependence of the bright foregrounds into frequency dependence of the visibilities. Removing these so-called ``mode-mixing'' effects requires detailed understanding and measurement of the frequency-dependent gain patterns of the antennas
and the calibration of the gain and phase of the instrument's electronic responses.

While many radio interferometers track targeted regions of the sky, most of the post-EoR instruments observe the sky by drift scanning.  This observing strategy allows for large sky coverage using simple and inexpensive instrument designs but complicates the calibration strategy. Tracking instruments can calibrate continuously on bright sources in or near the field they are mapping. Drift scanning instruments such as Tianlai must wait for bright sources to pass through the field, or attempt to calibrate on dimmer sources. It is important to check the stability of the system, which greatly affects the
sensitivity of the 21~cm signal.

Before using the array to make astronomical observations, it is necessary to first check and understand its basic performance characteristics. In particular, it is important to assess the stability of the system, the precision of calibration, the non-smooth instrumental effects, etc., which all affect the detection of 21~cm signal. In this work, we make an analysis of basic performance characteristics of the Tianlai Cylinder Pathfinder array based on test observation data. We first review the instrument design of the system, its observations during the 2016-2018 test run, and the data processing procedure in Section  \ref{sec:overview}, then present some system tests in Section \ref{sec:system_checks}. We examine the raw output data from the array in Section \ref{sec:visibility_preview}, and derive the beam pointing and width from the transit of a strong sources in Section \ref{sec:beam}. In Section \ref{sec:calibration} we describe the calibration of the system and study its stability. In Section \ref{sec:sensitivity} we estimate the system temperature of the radio telescope. Finally, we summarize our results in Section \ref{sec:conclusion}.

\section{An Overview of Instruments, Observations and Data Processing Procedures}
\label{sec:overview}

An interferometer array combines signals from a number of receiving units (antenna/feed combination) distributed in space, each of them generating an electric voltage signal,
\begin{equation}
  \mathcal{E}_a (\nu) = g_a \int \mathrm{d}^2 \vec{n} A_a (\vec{n},\nu) E(\vec{n},\nu) e^{-i 2\pi \frac{\nu}{c} \vec{n} \cdot \vec{x}_a} + n_a(\nu)
\end{equation}
where $A_a(\vec{n},\nu) $ is the antenna response to the electric field $E(\vec{n},\nu)$ from the direction $\vec{n}$ at frequency $\nu$,  $g_a$ is the receiver gain, $\vec{x}_a$ is the position of the antenna, and $n_a$ is the noise.  The short time-averaged correlation between the voltage outputs is called an (interferometric) visibility,
$$ V_{ab} (\nu) \equiv \langle \mathcal{E}_a^* \mathcal{E}_b \rangle_t .$$
The visibility is related to the sky intensity by
\begin{equation}
  V_{ab}\propto g_a^* g_b \int \mathrm{d}^2\vec{n} B_{ab}(\vec{n},\nu) I_{\nu}(\vec{n}) \exp[-i2\pi\vec{u}_{ab} \cdot \vec{n}] +\langle n_a^* n_b\rangle_t,
  \label{eq:radiometer}
\end{equation}
where $\langle E^*(\vec{n}) E(\vec{n}^\prime)\rangle \propto I_{\nu}(\vec{n}) \delta^2(\vec{n}-\vec{n}^\prime)$ is the sky intensity in the direction $\vec{n}$ at frequency $\nu$ and
$$\vec{u}_{ab}= \frac{\nu}{c} (\vec{x}_b - \vec{x}_a) $$
is the baseline vector between antenna or feed $a$ and $b$ in units of wavelength. $B_{ab}(\vec{n},\nu) = A_a^*(\vec{n},\nu) A_b (\vec{n},\nu)$ is the primary beam response for the pair, and can often be simplified as $B(\vec{n},\nu)$ if the array is uniformly constructed and each antenna/feed is pointing in the same direction.  The uncorrelated noise satisfies $\langle n_a^* n_b\rangle = \delta_{ab} T_{\rm sys} $.
This equation can also be generalized to the polarized case. The interferometer instrument is designed to produce this visibility data. From the measured visibilities, the sky intensity $I_{\nu}(\vec{n})$ can be recovered by synthesis \cite{2017isra.book.....T}.

\subsection{Instrument}
\label{subsec:instrument}

\begin{figure*}
  \centering
  \includegraphics[width=0.8\textwidth]{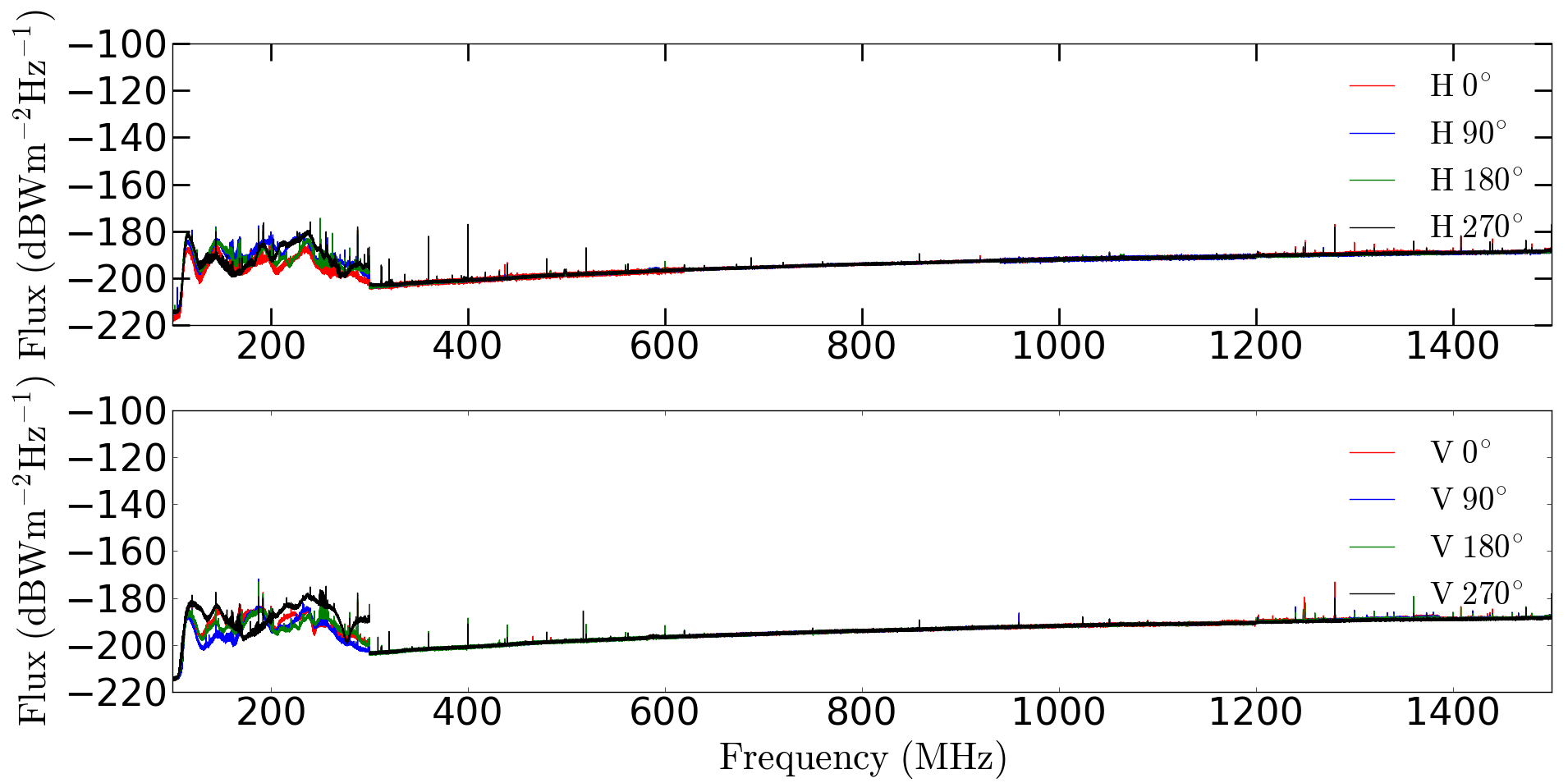}
  \caption{Radio environment measured at the Hongliuxia site. H and V correspond to horizontal and vertical polarization, respectively. Angles correspond to the orientation of the test antenna. The change at 300 MHz is due to different amplifier used in the measurement.}
  \label{fig:radio_env}
\end{figure*}

The Tianlai arrays are located at $91^\circ 48' \rm{E}, 44^\circ 09' \rm{N}$, near Hongliuxia village, Balikun county in Xinjing. The site is chosen for its very quiet radio environment \cite{wu2014site}. A calibrated site measurement was carried out between May $2^{nd}$ and May $4^{th}$, 2015. The result is shown in Fig. \ref{fig:radio_env}. In the L-band ($>1\GHz$), the primary radio frequency interferences (RFIs) are navigation signals ($\sim 1227 \MHz$ and $\sim 1575 \MHz$), digital audio broadcasting satellite signals (1452--1492~MHz), and satellite communication signals (1525--1559~MHz). In the UHF band (300--1000~MHz), the RFI is very weak, especially in the 700--800~MHz band where we are currently observing. At the time of this measurement, the power supply and a few other hardware devices at the site generated some RFIs, but these were subsequently replaced with quieter models.

\begin{figure}[H]
  \centering
  \includegraphics[width=0.4\textwidth]{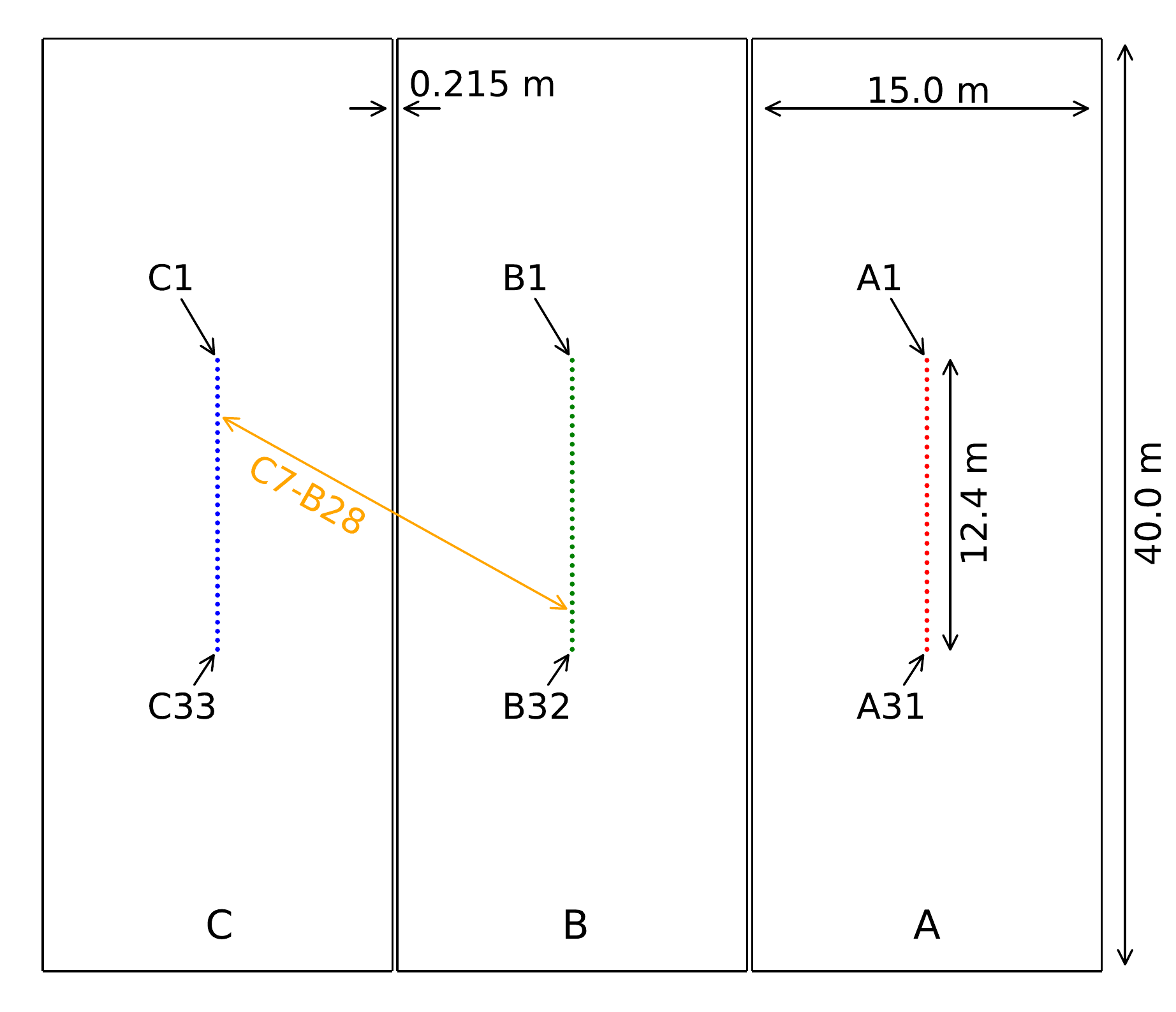}
  \caption{The Tianlai Cylinder Array. The cylinders are aligned in the N-S direction, with a gap of 0.215~m between adjacent ones. The three cylinders are designated as A, B, C from east to west, and have 31, 32, and 33 feeds respectively. The feeds in each cylinder are evenly distributed, with the ones at both ends
 (A1, B1, C1 in the north and A31, B32, C33 in the south) aligned with each other. The baseline C7-B28 is depicted by the orange double end arrow as an example.}
  \label{fig:cyl_array_schematic}
\end{figure}

The Tianlai project is now running in a pathfinder stage. The cylinder array consists of three adjacent parabolic cylinder reflectors, each $40 {\rm m} \times 15 {\rm m}$, with their long axes oriented in the N-S direction. Dual-polarization dipole feeds are placed along the focus line of each cylinder \cite{chen2016design}. The cylinders are closely spaced in the East-West (E-W) direction, as shown in  Fig. \ref{fig:cyl_array_schematic}.   The reflectors focus the incoming radio signal in the E-W direction, while allowing a wide field of view (FoV) in the N-S direction. The Tianlai cylinder reflector is fixed on the ground, and at any moment its FoV is a narrow strip running from north to south through the zenith.  While the reflectors allow the FoV to run from horizon-to-horizon in the N-S direction, the beam of the feeds limits the strip to $\pm 60^\circ$ from zenith \cite{Cianciara2017}. As the Earth rotates, the beams scan the northern celestial hemisphere. (The latitude of the telescope site is $44^\circ$, so the FoV extends from $-16^\circ$ declination up to $+90^\circ$ and back down to $+76^\circ$ on the other side).

From east to west, the 3 cylinders are denoted cylinder A, B, C respectively. Each has been installed with a slightly different number of feeds, 31, 32 and 33, respectively. From north to south, the feeds in each cylinder are labeled in numbers $1, 2, 3,  ...$. The northernmost (or southernmost) feeds A1, B1, C1 (or A31, B32, C33) are aligned, and the distance between the northernmost and southernmost feeds is 12.4 m.  The currently installed feeds occupy less than half of the cylinder; the remaining space is reserved for additional feeds for future upgrades. Since the feeds are evenly distributed, this results in different feed spacings for each different cylinder: 41.33 cm, 40.00 cm and 38.75 cm, respectively. This arrangement is made to reduce the grating lobe, which is generated due to the degeneracy in arrival time for signals from different directions when the spacings between adjacent feeds are larger than half a wavelength. As the spacings of the three cylinders are slightly different, their grating lobes are also slightly different, so the grating lobe is reduced \cite{2016RAA....16..158Z}.
Each dual linear polarization feed generates two signal outputs. We will use X to denote the output for the polarization along the N-S direction and Y along E-W direction. Each signal channel is designated by its cylinder, feed number, and polarization basis. For example, the E-W polarized output of the 2nd feed in the middle cylinder will be noted as B2Y. The baseline between two feeds is denoted by its two components linked with a hyphen; for example the baseline C7-B28 is shown in Fig. \ref{fig:cyl_array_schematic}, and the cross-correlation between their X-polarization channels is denoted as C7X-B28X.

\begin{figure*}
  \centering
  \includegraphics[width=0.9\textwidth]{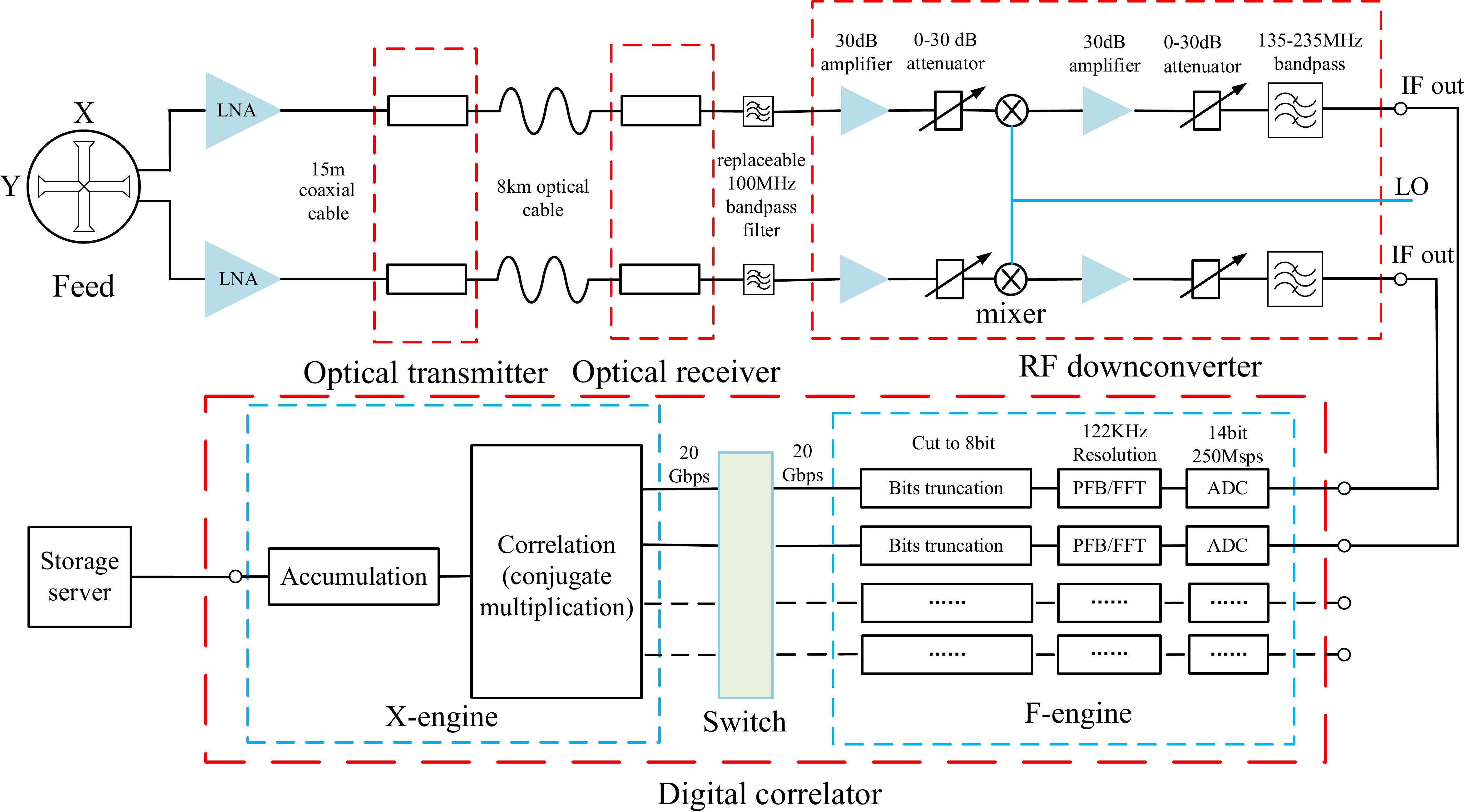}
  \caption{A schematic of the Tianlai Cylider Array analog and digital systems. }
  \label{fig:schematic}
\end{figure*}

The system consists of the antennas (cylinder reflectors and feeds) and the optical communication system (optical transmitter/receiver and cable), which converts the radio frequency (RF) electric signal to optical signals sent via optical fiber to the station house, which is located about 6~km away in the nearby Hongliuxia village. In the station, the system is housed in separate analog and digital electronics rooms. The optical signal is converted back to the RF electric signal, then down-converted to the intermediate frequency (IF). The IF signal is then digitized and processed in the digital correlator. The whole system is designed to operate over a wide range of frequencies (400--1500~MHz), while the working frequency band is set by replaceable bandpass filters. At present, the bandpass is set to 700--800~MHz, corresponding to redshift $1.03>z>0.78$ for the 21~cm line. A summary of the design parameters of the cylinder array is given in Table \ref{tab:cylinder_parameter}, and a schematic of the system is shown in Fig.~\ref{fig:schematic}.

\begin{table}[H]
  \caption{Main design parameters of Tianlai Cylinder Array.}
  \centering
  \begin{tabular}{l l}
    \hline
    Parameter & Value \\
    \hline
    Number of cylinders & 3 \\
    Reflector N-S length & 40.0 m \\
    Reflector E-W diameter & 15.0 m \\
    f/D & 0.32 \\
    Surface error (design) & $ <\lambda/20$ at 21 cm\\
    Number of feeds per cylinder & 31(A), 32(B), 33(C) \\
    Feed spacing (cm)  & 41.33, 40.00, 38.75 \\
    Feed illumination angle  &  $152^\circ$ \\
    Number of data channels & 192 \\
    Antenna gain (design) & $>20 $ dBi \\
    Adjustable frequency range & 500--1500 MHz\\
    Current frequency range & 700--800 MHz \\
    Frequency resolution & 122 kHz \\
    X-pol(N-S) FWHM @750 MHz & \begin{tabular}[c]{@{}l@{}}$~1.6^\circ$ (H-plane),\\$62.2^\circ$(E-plane) \end{tabular} \\
    Y-pol(E-W) FWHM @750 MHz & \begin{tabular}[c]{@{}l@{}}$~1.8^\circ$ (E-plane),\\$71.4^\circ$(H-plane) \end{tabular} \\
    Location & $91^\circ48'$ E, $44^\circ 09'$ N\\
    \hline
  \end{tabular}
  \label{tab:cylinder_parameter}
\end{table}

In detail, the wide band low noise amplifiers (LNAs) are mounted on the back plate of the feed. The amplified RF signals pass through 15-meter long coaxial cables to the optical transmitters mounted under the cylinder antennas. In the transmitter box, the RF signal amplitude modulates a distributed feedback laser diode (DFB-LD) to generate the RF optical signals. The typical intensity of the laser output power is 3 dBmW, and a built-in thermal electric cooler module controls the temperature of the laser diode, to keep the output intensity at a stable value. The output laser is coupled into a single-mode fiber pigtail and ferrule connector/angled physical contact (FC/APC) connector, which connects the optical fiber link. The optical cable carries a bundle of 288 fibers a distance of 8 km. The typical loss in the  optical fiber link is 4.8 dB at the operating wavelength of $1.55\mu \rm{m}$. And to compensate for this loss, the analog optical transceiver system is designed with an end-to-end gain of 13 dB, which varies less than 2 dB within the band. At the receiving end, a PIN photo-detector demodulates the RF signal. The receiver modules are housed in box cases; each handles 32 signal channels. The typical noise figure for the optical link system is 17 dB. However, this does not significantly increase the noise temperature for the signal which has already been amplified by the LNA.

An analog mixer down-converts the RF signal to the IF band of 125--235~MHz. The observing frequency band can be adjusted by shifting the frequency of the local oscillator (LO). Currently, the observing frequency band is 700--800~MHz. The LO is fixed at 935 MHz with a Rubidium standard frequency reference, with maximum phase noise of  $-90$ dBc/Hz at 10 kHz and $-90$ dBc/Hz at 100 kHz. The image frequency rejection ratio for the down-converter is larger than 60 dB. Replaceable cavity bandpass filter modules are placed between the optical receiver RF output and the mixer input to reject out-of-band noise; its insertion loss is less than 0.7~dB. The IF channel output power level can be adjusted within the range of $0\sim 60 \dB$ by attenuators, with gain adjustment steps of 1 dB controlled by a serial communication port. The gain flatness is less than 2~dB for the IF bandwidth. For the IF channel output, the minimum 1 dB compression point is 15~dBm.

Finally, the IF signal is sent to the digital backend system through bulkhead connectors between the analog and digital rooms.

The digital backend system has a commonly used FX (Fourier transformation and correlation) architecture. It includes three parts: the F-engine, which performs the Fourier transform, the X-engine, which performs cross-correlations of the Fourier-transformed data, and the network switch, which performs the so-called ``corner turning'', i.e., it distributes the data from different input channels of the same frequency to a unit on the X-engine for computing the cross-correlations.

\begin{figure}[H]
  \centering
  \includegraphics[width=0.45\textwidth]{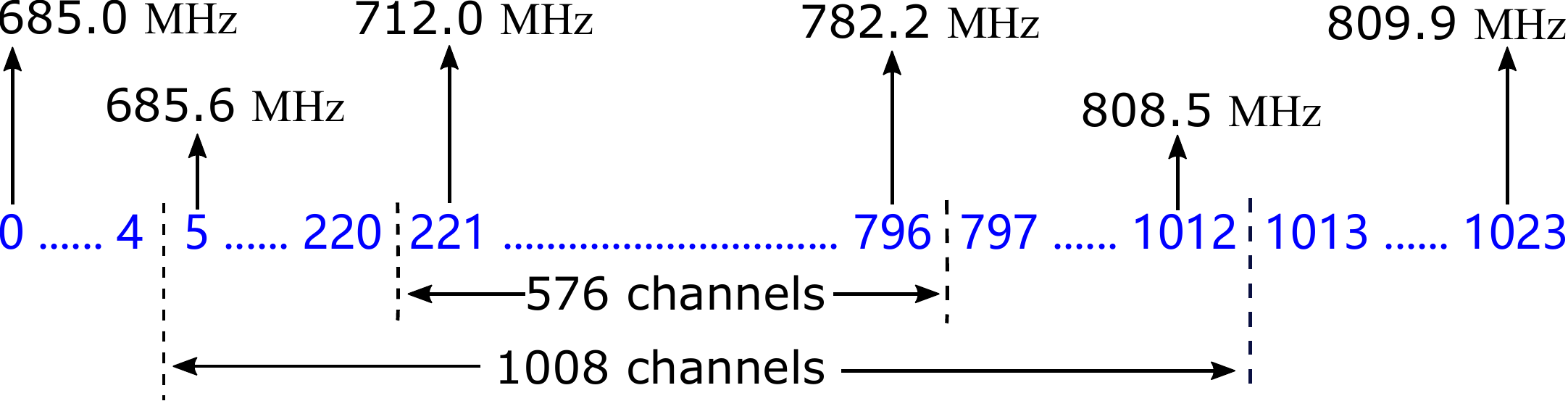}
  \caption{The relationship between the FFT channels and the radio frequency. Though the correlator does a 2048-point FFT, with 1024 positive frequencies, only 1008 of these are cross-correlated. Due to limited X-engine computation ability, currently only 576 channels have been computed. }
  \label{fig:freq_axis}
\end{figure}

The F-engine consists of twelve FPGA boards, each equipped with two 8-input analog-to-digit converters (ADCs) of the type ADS62P49. These ADCs digitize simultaneously the 192 input channels, and work at a sampling rate of 250~Mega samples per second (Msps) and with a length of 14~bits. The sampler is synchronized by a 10~MHz reference signal output from a GPS receiver. The time-sequential data is Fast Fourier Transformed (FFT) with a block length of 2048, so that there should be 1024 positive frequencies, though due to hardware limitations, 16 of the frequency channels near the edge are discarded and only 1008 frequency channels are recorded. Given the 250~Msps sampling rate, each frequency channel has a width of $\delta \nu = 125/1024 \MHz=122.07 \kHz$. However, due to some practical problems with the correlator hardware, in most runs we only take data for the central 576 frequency channels, which span a bandwidth of about 70~MHz from 712.0~MHz to 782.2~MHz, as shown in Fig.\ref{fig:freq_axis}.

The FFT results are packed in several frequency blocks and sent to a 16~Gb$\times 128$ RapidIO high-speed switching network. The data of the same frequency bands from all input channels are switched to the same DSP boards in the X-engine.  Each DSP board undertakes the cross-correlation of one block of frequency bands for all input channels.  A total of 27 DSP boards are used. The correlation results from all of the DSP boards are integrated on an FPGA board, which outputs the visibility to a storage server and the data is dumped to hard drives in \texttt{HDF5} format. The integration time is adjustable; typically we adopt 4 seconds (more precisely 3.995 s for integer multiples of integration cycle), which the system can handle well and during which the drift of the sky due to Earth rotation is negligible for the angular resolution of our system.  The raw visibilities are arranged in the form of [time, frequency, correlation], with correlation being the fastest-changing index. The cross correlations between all of the 192 inputs are computed, thus forming a total of 18528 correlation pairs (192 auto-correlations plus 18336 cross-correlations).

To monitor the variation of instrumental phase, an artificial calibrator is installed on a small hill to the west of the array to broadcast wide-band RF noise periodically. Below we shall call it the Calibrator Noise Source (CNS). The device consists of a small broadband disc-cone antenna erected on a wooden pole, and a diode noise generator enclosed in a temperature-controlled box. Its temperature is kept at $21^\circ$C by a thermostat, with a maximum variation of ~$\pm0.1^\circ$C per day or ~$\pm2^\circ$C in a year. The CNS is placed at the top of a hill about 131.5  meters from the center of the array in the North-West direction, and its elevation relative to the center of the array (including the hill and wooden pole height) is about 13.1 meters.

The CNS is turned on and off with a period of 240 seconds. Initially the noise source broadcasted for 20 seconds, but after December 2017, this was reduced to 4 seconds; giving a duty cycle of 8.3\% and 2.1\%, respectively. The amplitude of the CNS is adjusted with an attenuator, so that as received by the feed near the center of the cylinder array, the CNS induced auto-correlation is about 5~dB stronger than the Sun for the X-polarization and 10~dB for Y-polarization.

\subsection{Observations}
\label{subsec:observations}

The Tianlai Cylinder Array does not have moving parts and is designed to run continuously once powered on, though in practice it runs a few days at a time. Once the digital correlator has started up properly, one can run the data acquisition program. The data are automatically collected and saved to hard drives. The drives are shipped to NAOC for off-line processing and analysis.

The Tianlai Cylinder Array saw its first light in September, 2016. Test observation data were collected in 2016 -- 2018. The observations were usually halted after a few days when some problems were found, or when some hardware malfunction prevented normal operation. Currently, we have collected a total of 114 days' of data, as listed in Table \ref{tab:obs_data_log}.

\begin{table}[H]
  \centering
  \caption{List of Tianlai Cylinder Array Data Sets. }
  \centering
  \small
  \begin{tabular}{l r p{3cm}}
    \hline
    Data set & Lengths & Malfunction Channels\\
    \hline
    2016/09/27 20:15:37 &  5 days & A18Y,A28,B19,C7Y, C9X,C14X,C16,C29Y\\
    2016/10/11 00:49:48 &  5 days & A18Y,A28,B19,C7Y, C9X,C14X,C16,C29Y\\
    2016/12/31 20:51:54 & 17 days & Many\\
    2017/02/13 19:23:07 &  5 days & A18Y,A24Y,B6Y,B26Y, C15X,C7Y\\
    2017/02/24 19:31:07 &  9 days & A18Y,A24Y,B6Y,B26Y, C15X,C7Y,C18\textasciitilde C33\\
    2017/08/21 21:26:39 &  3 days & A26X,B21X,C3X,C16 A17\textasciitilde A24,C26\textasciitilde C33\\
    2017/09/03 14:32:17 &  9 days & A26X,B21X,C3X,C16 A17\textasciitilde A24,C26\textasciitilde C33\\
    2017/09/22 01:33:18 &  7 days & A26X,B21X,C3X,C16 A17\textasciitilde A24,C26\textasciitilde C33\\
    2017/09/29 21:42:59 & 13 days & A26X,B21X,C3X,C16 A17\textasciitilde A24,C26\textasciitilde C33\\
    2017/12/09 19:21:54 & 10 days & A18Y,A24Y,A26X,B26Y, B31X\\
    2017/12/20 19:22:02 &  4 days & A18Y,A24Y,A26X,B26Y, B31X\\
    2018/01/21 00:05:35 & 14 days & A18Y,A24Y,A26X,B26Y, B31X,C28Y\\
    2018/03/22 18:07:58 &  9 days & A18Y,A24Y,A26X,B26Y, B31X,C28Y\\
    2018/03/31 17:08:12 &  4 days & A18Y,A24Y,A26X,B26Y, B31X,C28Y\\
    \hline
    \makecell[r]{Total} & 114 days \\
    \hline
  \end{tabular}
  \label{tab:obs_data_log}
\end{table}

The data sets in Table \ref{tab:obs_data_log} are designated by the starting time of observation in the Beijing standard time (UTC+08h), and the duration of each observation run is listed in the second column of the table. In the third column, we also list the malfunctioning channels during that run. These malfunctions are limited to those particular channels and do not affect the running of the whole system. Some malfunctions were identified during the run (e.g. if their output is too weak) while some were only found after offline analysis (e.g. the output is too noisy). The visibilities involving these malfunctioning units are ignored in the analysis. There were some problems in the correlator for the 2016/12/31 data set which render much of the data set unusable.

The present work is based on the analysis of the first light data set, which started on September 27th, 2016, (below we denote it as 2016/09/27) and was used in our early analysis \cite{Zuo2019}, and a second data set, which started on March 22nd, 2018 (denoted as 2018/03/22). These two data sets consist of a total of 10 days' of observation, and cover both the autumn and spring sky. For the 2016/09/27 data set, 20 frequency channels are processed while for the 2018/03/22 data set all 576 frequency channels are processed.

\subsection{Data Processing Procedures}
\label{subsec:data_process}

\begin{figure}[H]
  \centering
  \includegraphics[width=0.4\textwidth]{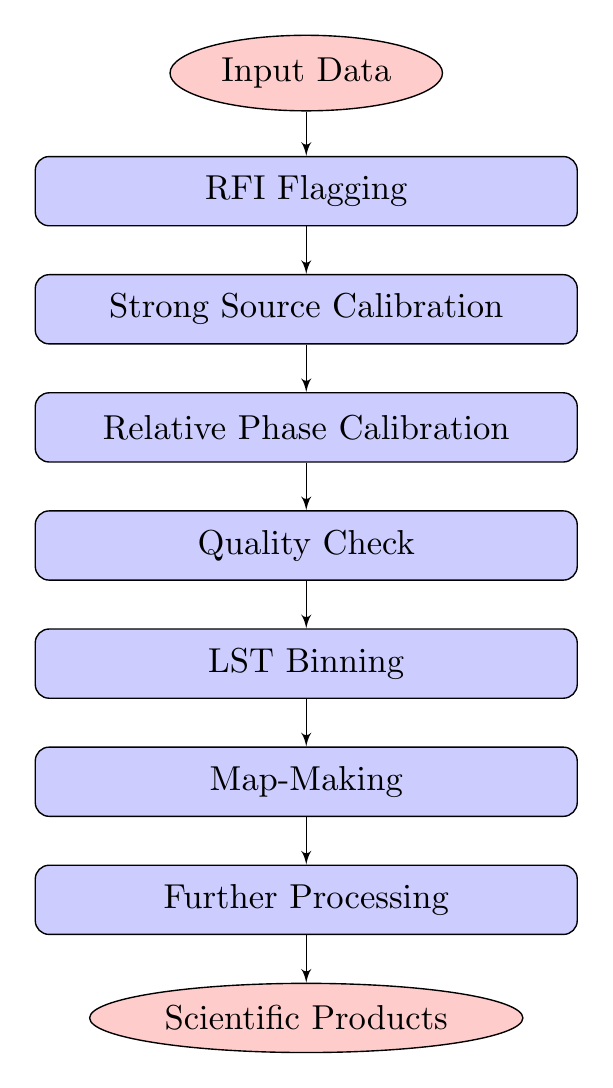}
  \caption{An outline of the offline data processing procedures.}
  \label{fig:offline_proc_flow}
\end{figure}

The offline data processing procedure is outlined in Fig.~\ref{fig:offline_proc_flow}.

The first step of the data processing is radio frequency interference (RFI) flagging. The RFI is often stronger than the astronomical sources or noise, and have characteristic distributions in the time and frequency domains. These are recognized and recorded in a binary mask file and ignored in subsequent analysis. The Tianlai site is a very radio quiet site, but in some of the data sets, we found a significant amount of  self-generated RFI, primarily from the power supplies for the array and the motors of the nearby dish antennas. Later, we installed quieter power supplies and turned off the motors when making drift-scan observations, greatly reducing the amount of RFI.

Two types of calibration are performed:
(1) Absolute calibration. This is performed when a strong astronomical point source such as Cygnus A (Cyg~A) or Cassiopeia A (Cas~A) is transiting through the primary beam. The magnitude and phase of the complex gain of each element can be determined in such a calibration, up to a common offset. We use an eigenvector decomposition method to solve for these gains.
(2) Relative calibration. This is performed by using the CNS signal to track the variation of the phase of complex gains when there are no strong astronomical sources available.

We can check the quality of the data before and after the calibrations. The calibrated data which pass the quality checks can be combined with other good data taken at the same local sidereal time (LST) but on different days.  This co-added data will have an improved signal-to-noise ratio (SNR).

One can then use the LST-binned data to produce sky maps and then make further analyses such as foreground removal and power spectrum estimations, which are the ultimate science products for the 21~cm intensity mapping experiment. A python-based data processing pipeline has been developed \cite{Zuo2020}, which provides a general framework for the data processing outlined above with at least one and sometimes several optional algorithms implemented for each step.

In the present paper, we shall focus on examining the general  quality of the raw data and investigating the precision and stability of the data achieved in the calibrations. The details of the RFI mitigation, and the methods and results of map-making will be described elsewhere.

\section{Sub Systems Performance }
\label{sec:system_checks}

\subsection{LNA}
\label{subsec:LNA}

\begin{figure}[H]
  \centering
  \includegraphics[width=0.45\textwidth]{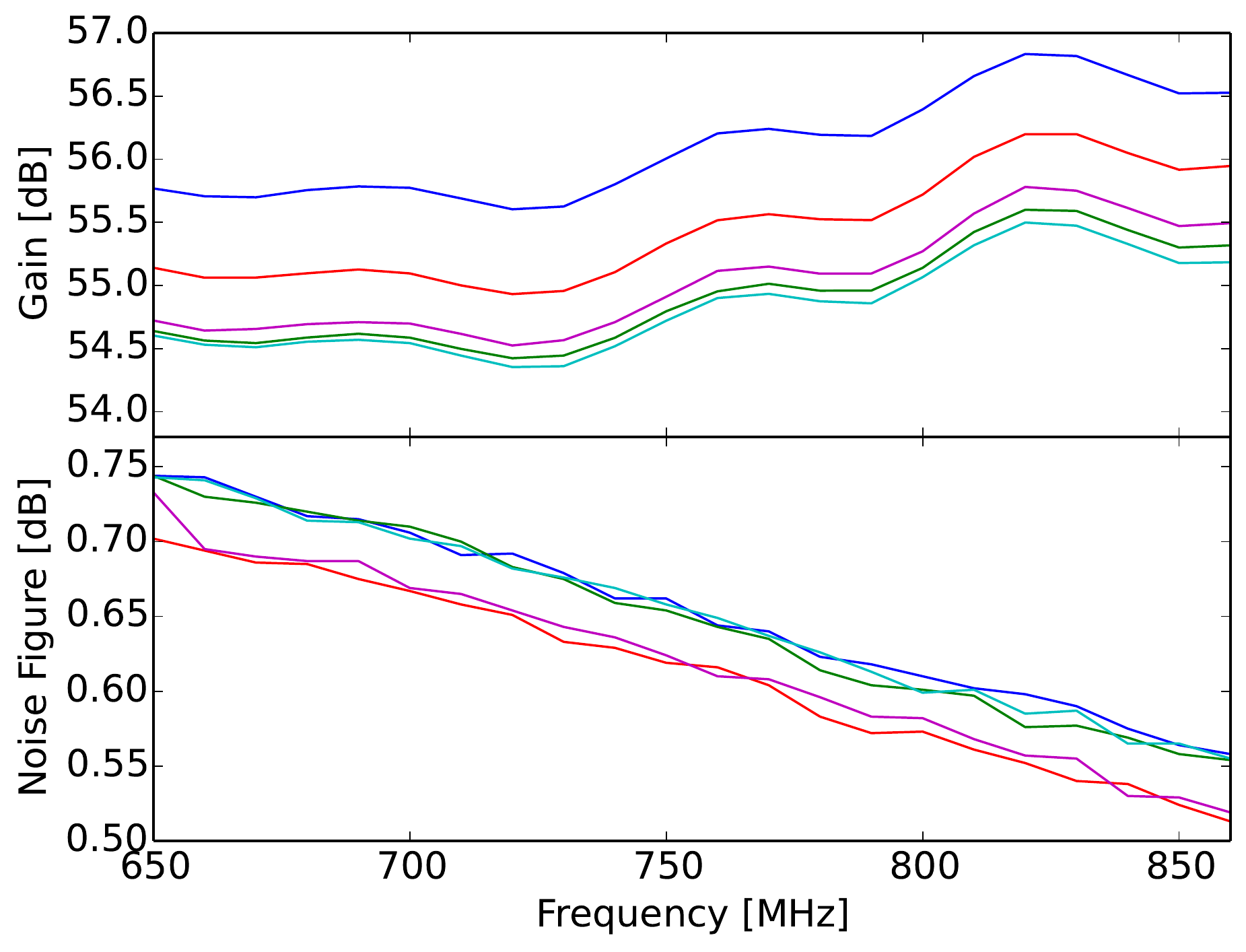}\\
  \caption{The measured gain (top) and Noise Figure (bottom) of 5 randomly chosen LNAs.}
  \label{fig:LNA_gain_NF}
\end{figure}

The receiver noise is determined to a large extent by the LNA. Because the operating frequency is relatively low, where the sky is relatively bright, the Tianlai receivers work at ambient temperature. Cooling would be expensive and not very effective.  The gain (G) and noise figure (NF) of the LNAs are measured in a frequency range from 650~MHz to 860~MHz,  and a few examples are shown in Fig. \ref{fig:LNA_gain_NF}. There are some variations in the value of the gain, but their frequency responses are similar. The NF variation in the 700--800~MHz band is about 0.1~dB, with a mean NF of 0.65~dB. The noise figure is related to the receiver noise temperature $T$ by
\begin{equation}
  T = \left(10^{\frac{\rm NF}{10}-1}\right) \times T_{0},
  \label{eq:T_from_NF}
\end{equation}
where $T_{0}$ is the environment temperature of the receiver. For example, if ${\rm NF} = 0.65 \dB$, and we take the IEEE standard reference temperature $T_0 = 290 \K$, the corresponding noise temperature of the LNA is about $47 \K$.

\subsection{Linearity}
\label{subsec:linearity}

Interferometry requires that the electronics respond linearly.  However, real analog and digital systems are only linear in their response for a finite range of input power. The design of the RF front-end systems (LNA, optical transmission system) should ensure that it operates in the linear regime during regular observations, and also produces amplified signals adapted to the digital system input requirements. We have estimated the required amount of amplification based on the typical radio frequency power level, the antenna gain, and the required input voltage for the ADC.  To allow adjustment of power levels, two sets of 30~dB variable attenuators are placed before and after the mixer. We adjust them so that the IF output power level matches the requirement of the ADC.

The digital system also requires an appropriate level of input power to produce outputs which are linearly related to the input. The Tianlai Cylinder Array ADCs have a sampling length of 14 bits, i.e. the output is an integer in the range of -8192 to +8191 decimal. Better precision can be obtained if the digital outputs have more non-zero bits, but on the other hand it is desirable not to saturate the output when the signal is very strong.

\begin{figure}[H]
  \centering
  \includegraphics[width=0.4\textwidth]{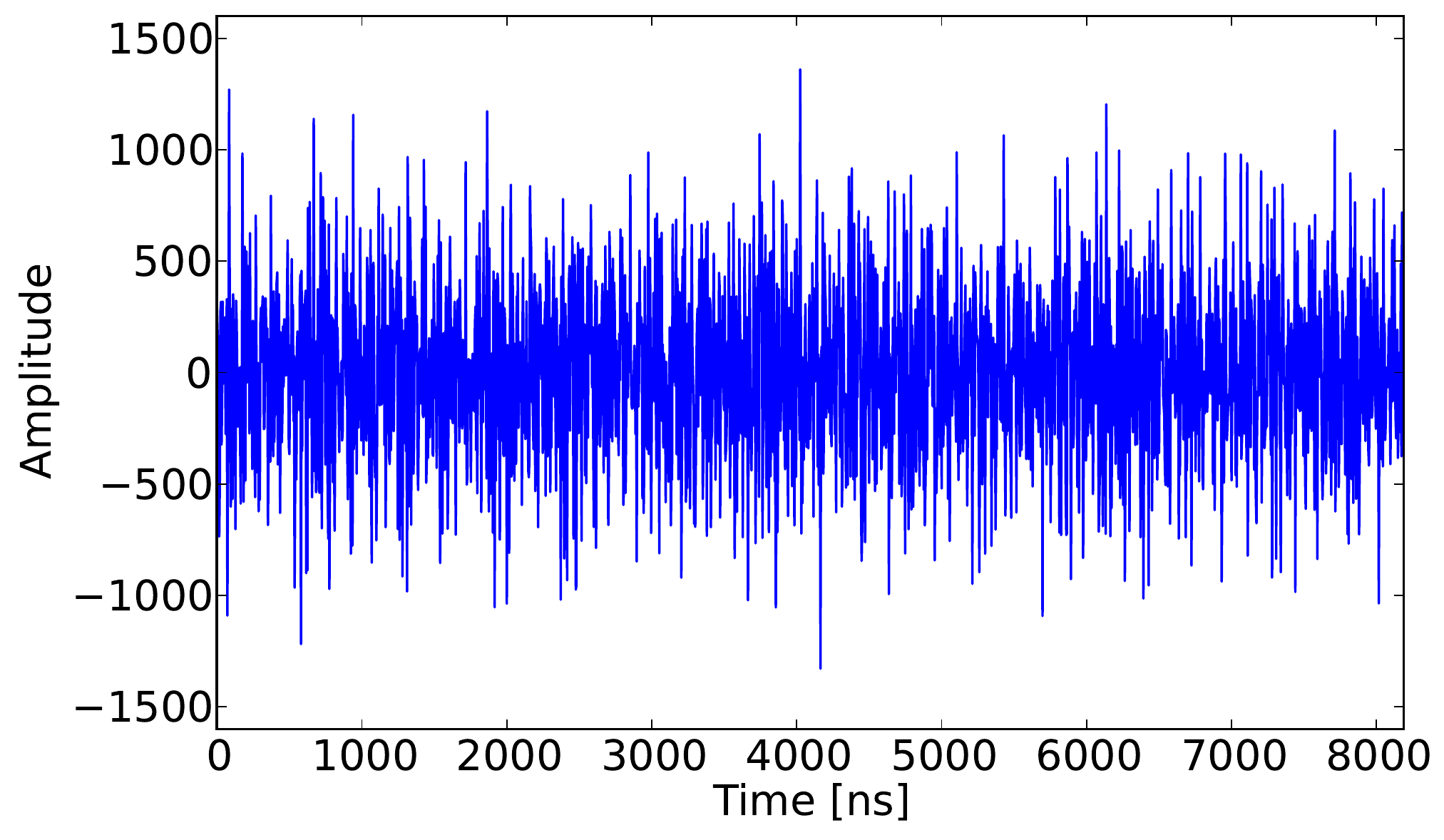} \\
  \includegraphics[width=0.4\textwidth]{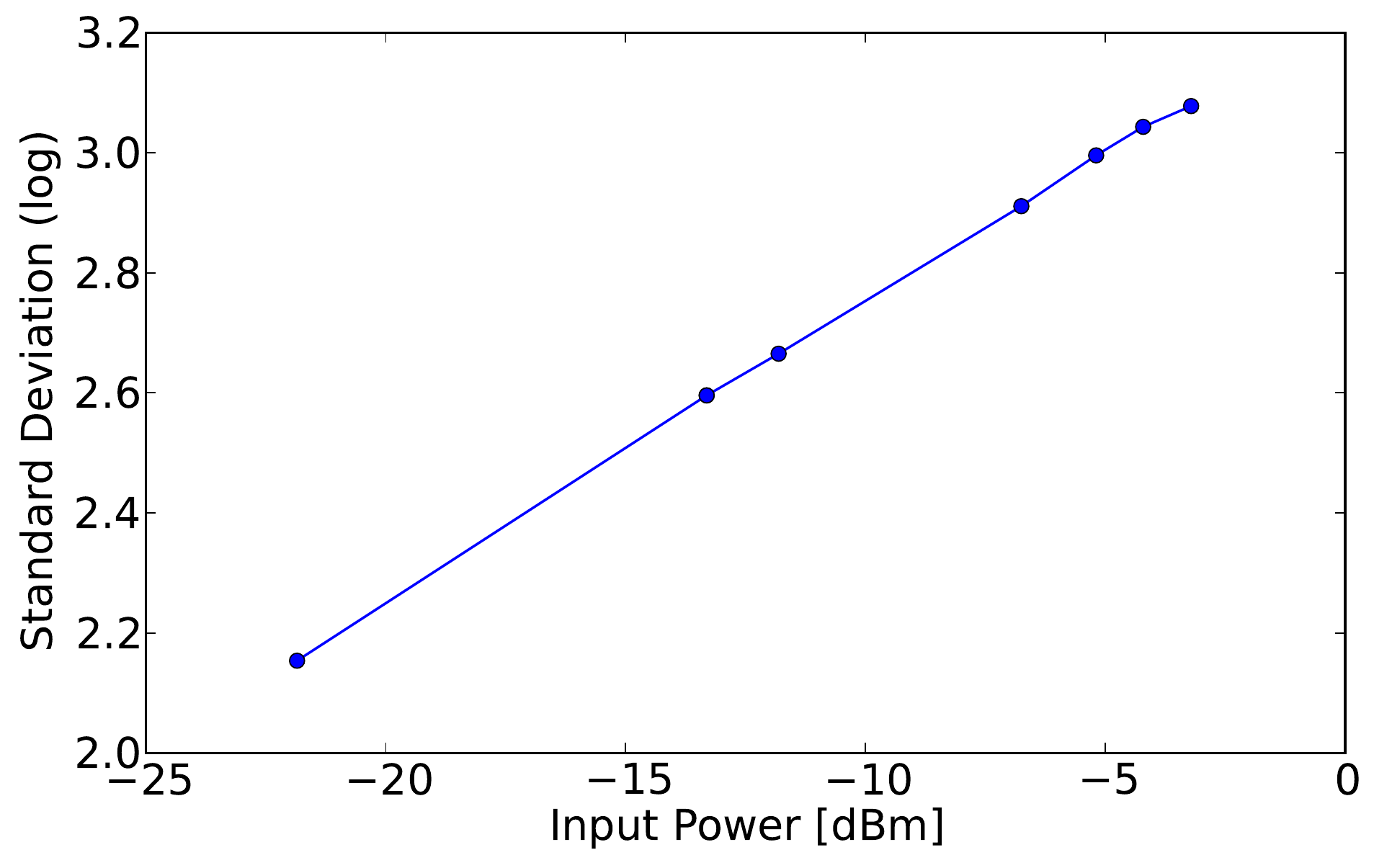}
  \caption{Top: ADC raw data samples taken over 8192~ns when the input power level is -13.31~dBmW. Bottom: the logarithm of the standard deviation of the ADC output at different input power levels. Currently, the IF level we set is about -13~dBmW within our observation band.}
  \label{fig:ADC_sample}
\end{figure}

A typical sample of the ADC output is shown in the top panel of Fig. \ref{fig:ADC_sample}, and the standard deviation (STD) of the output is shown as a function of the input power level in the bottom panel. The IF output power level is adjusted to about -13~dBmW when the array is observing a part of the sky with no bright sources during nighttime, so that the ADC output has a standard deviation of 400, about 5\% of the total range allowed by the sampling length,  which is a good level for quantization and avoids saturation. The brightest radio astronomical source is the Sun. When it is transiting through the telescope FoV, the maximum power level is about 5~dB stronger, still well within the range of the ADC. Occasionally RFI may generate strong signals that saturate the correlator, but as we are generally not so interested in RFI and it are generally filtered out, non-linear response due to saturation by RFI is not a big problem for data processing.

\begin{figure}[H]
  \centering
  \includegraphics[width=0.4\textwidth]{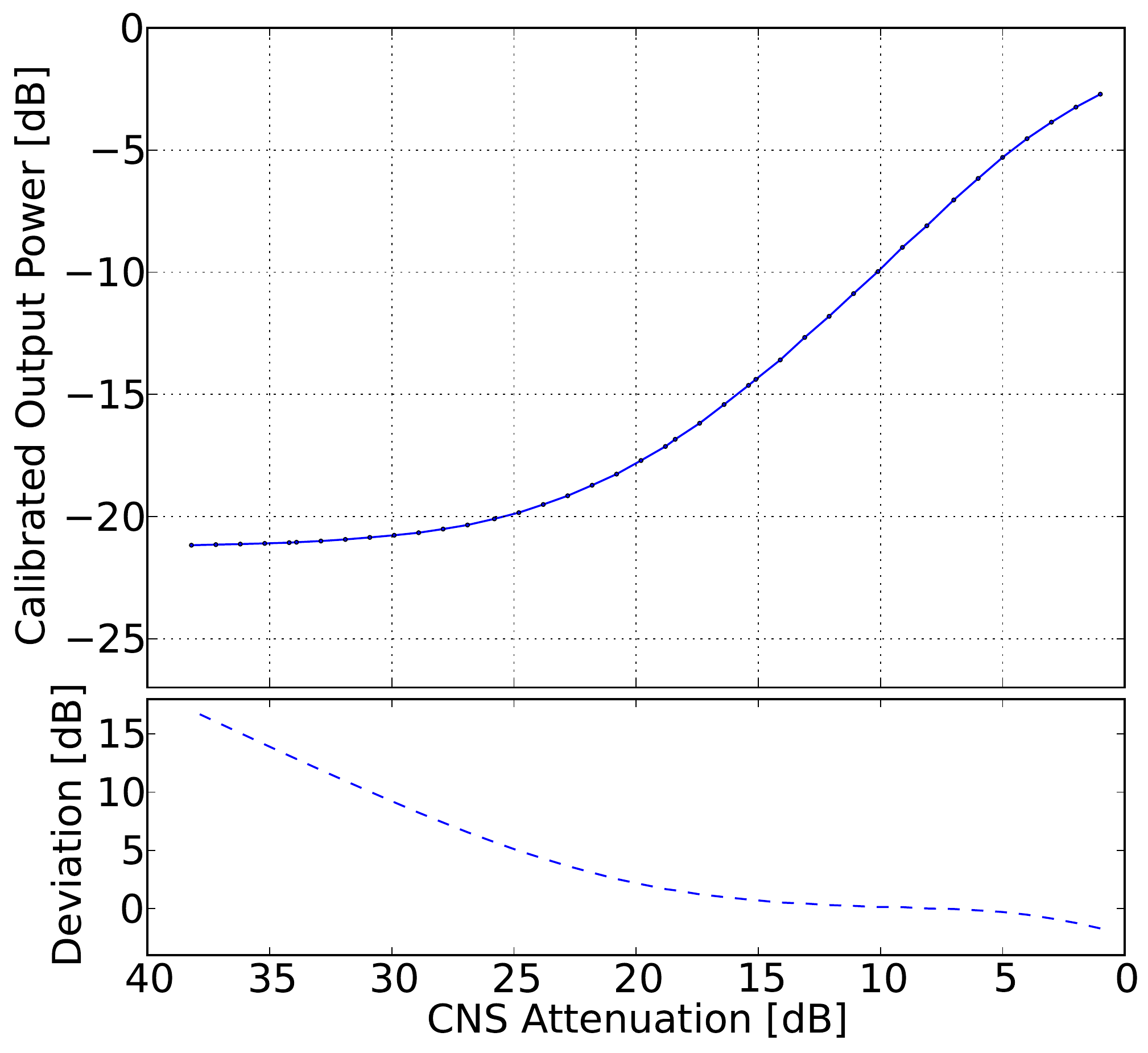}\\
  \caption{The auto-correlation output as a function of the power level of the CNS (top panel) and nonlinear residual (bottom panel).}
  \label{fig:system_linearity_auto}
\end{figure}

We use the CNS to check the signal chain response. The CNS generates a wide band noise signal in the 700--800~MHz frequency band and its power level is adjusted by an attenuator. The auto-correlation signal is shown in Fig.\ref{fig:system_linearity_auto} as a function of the attenuation in CNS power.  The auto-correlations have a non-zero noise floor at about -20~dB, due to typical sky signal and receiver noise; this is, however, not an indication of a breakdown of linearity. The cross-correlation power, which has a zero noise floor, is linear all the way down to  -40~dB. This test is limited by the sky signal, which is always present and contributes non-zero cross-correlations.

\section{Quick Look Analysis of the Visibilities}
\label{sec:visibility_preview}

\begin{figure}[H]
  \centering
  \includegraphics[width=0.4\textwidth]{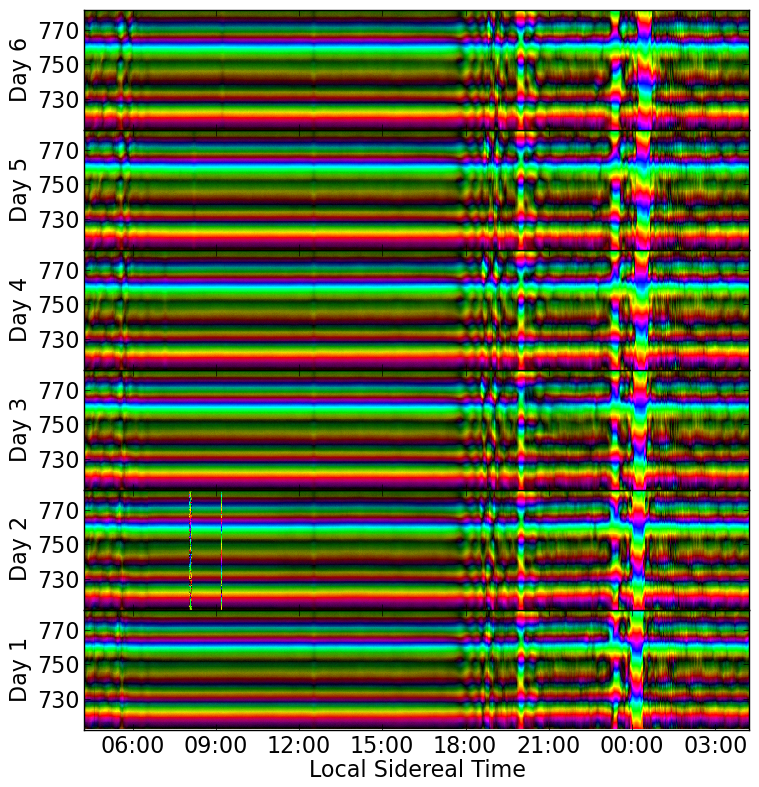}
  \includegraphics[width=0.4\textwidth]{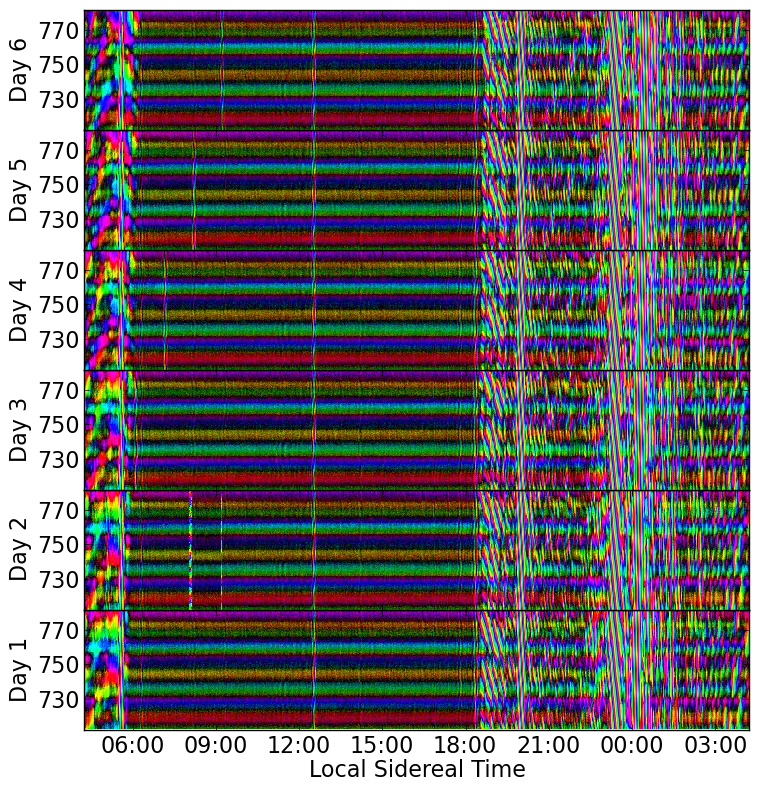}
  \includegraphics[width=0.4\textwidth]{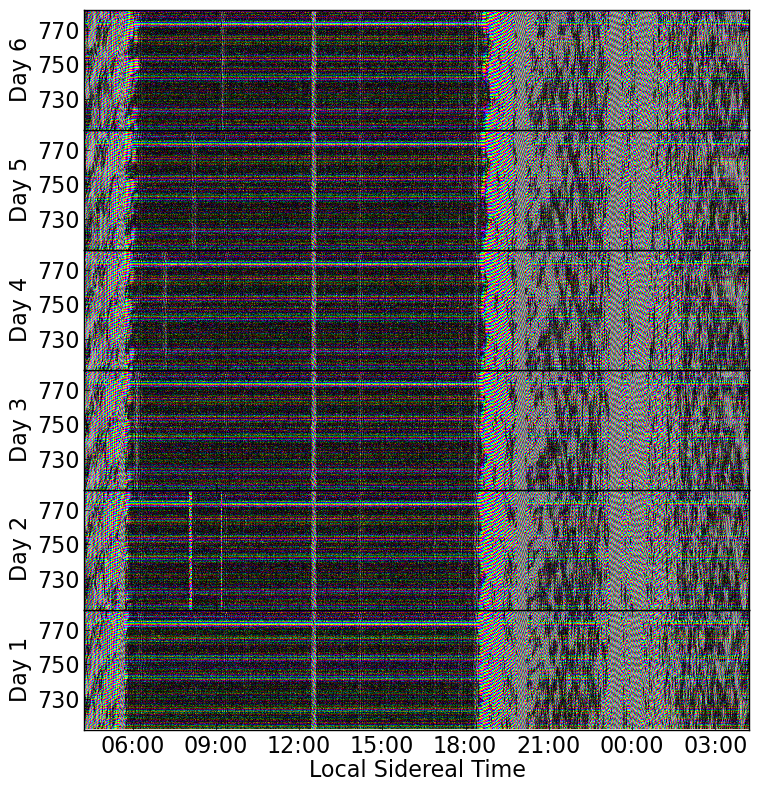}
  \caption{Typical raw visibilities as a function of LST and frequency for 6 days starting from 2018/03/22. The data are rebinned to 488~kHz and 20~s resolution. Top panel: baseline A3Y-A15Y; Middle panel: baseline A3Y-B18Y; Bottom panel: baseline A2Y-C2Y.}
  \label{fig:raw_vis_hsv_180322}
\end{figure}

We first make a quick look analysis of the raw visibilities. As an example, in Fig.~\ref{fig:raw_vis_hsv_180322} the raw visibilities of three baselines are plotted as a function of LST and frequency. From top to bottom, the three subplots show, respectively, a baseline for two feeds on the same-cylinder, on two adjacent cylinders, and on two non-adjacent cylinders. Each subplot shows the result of six consecutive days starting from 2018/03/22; each day is a sub-panel from bottom to top. The data are rebinned to 488~kHz (4 original frequency bins) and 20 seconds (5 original time bins) to suppress noise. We have removed the data affected by the periodic broadcasting of the CNS. In this two dimensional plot, the brightness is proportional to the amplitude of the visibility, while the hue is used to represent the phase, from red, through blue, to violet for phase angle varying from 0 to $2\pi$. The saturation is set as 1. We set the brightness to saturate at two times the median amplitude, to allow the display of fainter features.

All panels show horizontal stripes that are fairly stable over time. This ``background''  has some structure in frequency and is stable with time. This is correlated noise arising from the coupling between feeds. Indeed, it is particularly strong for the top panel, which shows the visibility of a medium length baseline on the same cylinder. The feeds on the same cylinder are relatively close to each other, and radiation from one can be reflected by the cylinder to the other feed. It is much weaker for feeds on adjacent cylinders (middle panel), and even weaker for the feeds on non-adjacent cylinders (bottom panel).

\begin{table}[H]
  \centering
  \caption{List of sources that can be recognized in Fig. \ref{fig:raw_vis_hsv_180322_rm_night_5097} and Fig. \ref{fig:raw_vis_hsv_160927_rm_night_1065}, including their right ascension (R.A.), Zenith Angle (ZA), and flux density at 750~MHz as given in \url{https://ned.ipac.caltech.edu}. The list is in the order of R.A..  }
  \begin{tabular}{r|r|r|r}
  \hline
    Source      &  R.A. &  ZA (deg)  & Flux (Jy) \\
  \hline
    3C 010      & 00:25 & 20.0 &   62 \\
    3C 058      & 02:05 & 20.7 &   34 \\
    IC 1805     & 02:32 & 17.4 &   -- \\
    3C 084      & 03:20 &  2.7 &   22 \\
    3C 123      & 04:37 & 14.5 &   76 \\
    M 42        & 05:35 & 49.5 &   -- \\
    IC 443      &  6:16 & 21.6 &  190 \\
    3C 196      &  8:13 &  4.1 &   23 \\
    Hydra A     &  9:18 & 56.2 &   81 \\
    M 82        &  9:55 & 25.5 &   11 \\
    M 87        & 12:30 & 31.8 &  353 \\
    3C 286      & 13:31 & 13.6 &   19 \\
    3C 295      & 14:11 &  8.1 &   37 \\
    Hercules A  & 16:51 & 39.2 &   88 \\
    3C 353      & 17:20 & 45.1 &   88 \\
Galactic Center & 17:45 & 73.0 &   -- \\
    3C 380      & 18:29 &  4.6 &   23 \\
    3C 392      & 18:56 & 42.8 &  242 \\
    3C 400      & 19:23 & 30.0 &  673 \\
    Cyg A       & 19:59 &  3.4 & 2980 \\
    Cyg X       & 20:28 & 41.2 &   -- \\
   NRAO 650     & 21:12 &  8.3 &   48 \\
    3C 433      & 21:24 & 19.1 &   21 \\
    Cas A       & 23:23 & 14.7 & 2861 \\
  \hline
  \end{tabular}
  \label{tab:source_list}
\end{table}

The vertical features in the figure show strong sources transiting through the telescope field of view. This observation was taken near the Vernal Equinox, when the Sun transit (at noon) takes place near 0:00 LST.  With a few exceptions, the visibility is very similar on each day. The most prominent feature in these plot is the day and night contrast. Sunrise (18h30m LST) and sunset (6h00m LST) can be clearly seen.  The strong fringes of the Sun shift slightly each day, due to the daily motion of the Sun along the ecliptic.

Besides the Sun, a few other transits can also be recognized. There is a strong fringe at 12h30m produced by the transit of Virgo A (M87). Another obvious source which has its transit on 6h8m (Day 3), then 7h10m (Day 4), 8h12m (Day 5) and 9h14m (Day 6) is the Moon, which moves 1h02m in R.A. each day during this observation. The two bright vertical fringes on day 2 at about 8h10m and 9h15m are from RFI.

\begin{figure}[H]
  \centering
  \includegraphics[width=0.42\textwidth]{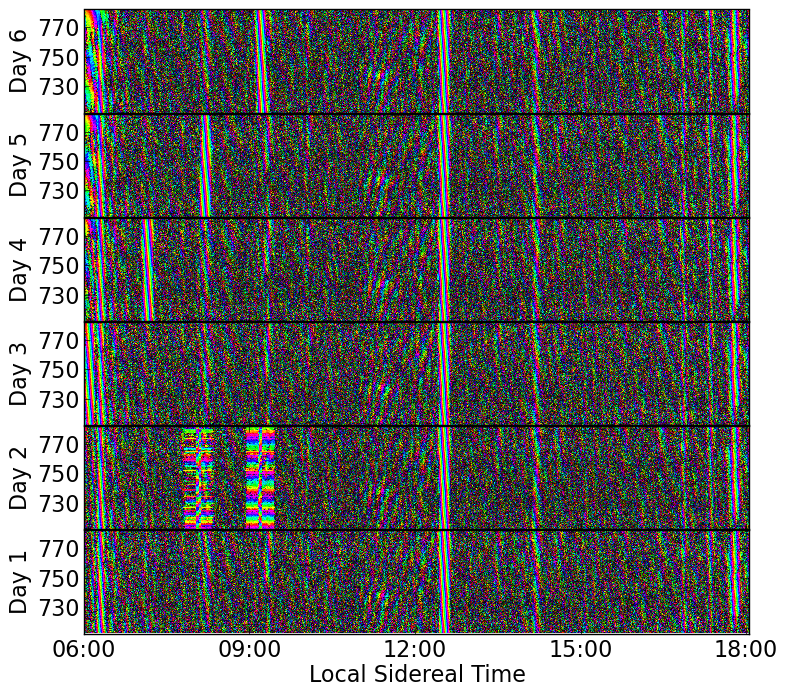}
  \caption{A typical nighttime visibility (A3Y-B18Y) for data from 2018/03/22, with the time-averaged band structure subtracted. }
  \label{fig:raw_vis_hsv_180322_rm_night_5097}
\end{figure}
\begin{figure}[H]
  \centering
  \includegraphics[width=0.42\textwidth]{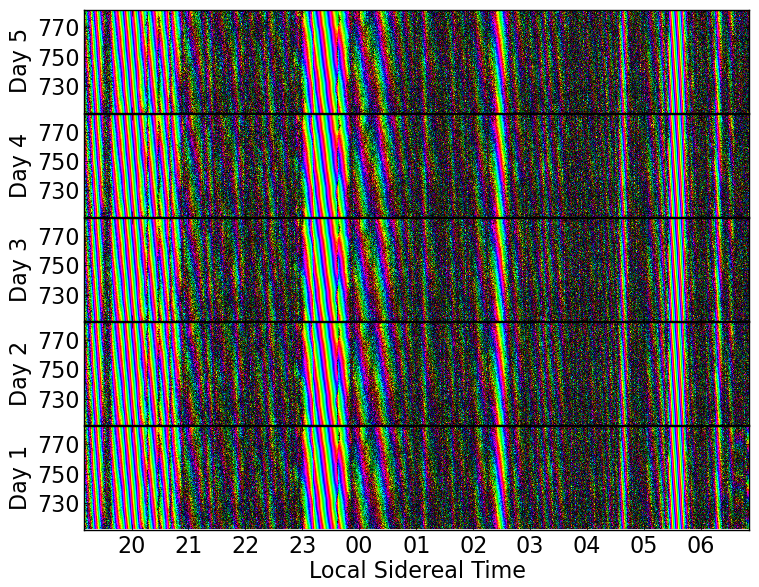}
  \caption{A typical nighttime visibility (A1Y-B2Y)  for data from 2016/09/27, with the time-averaged band structure subtracted.}
  \label{fig:raw_vis_hsv_160927_rm_night_1065}
\end{figure}

\begin{figure*}
  \centering
  \subfigure[E-plane of Y-polarization.]{  \includegraphics[width=0.45\textwidth]{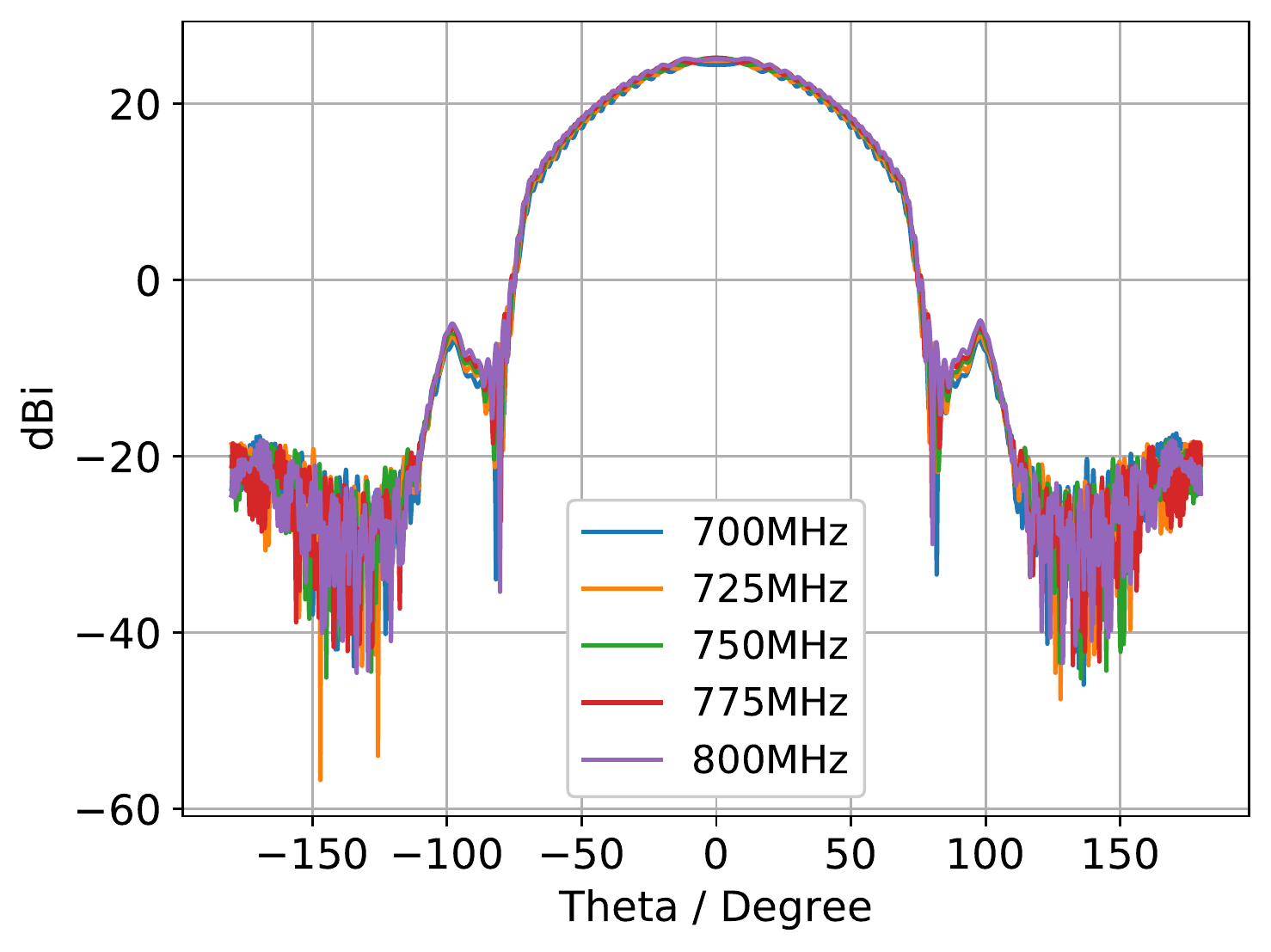}}
   \subfigure[H-plane of Y-polarization.]{ \includegraphics[width=0.45\textwidth]{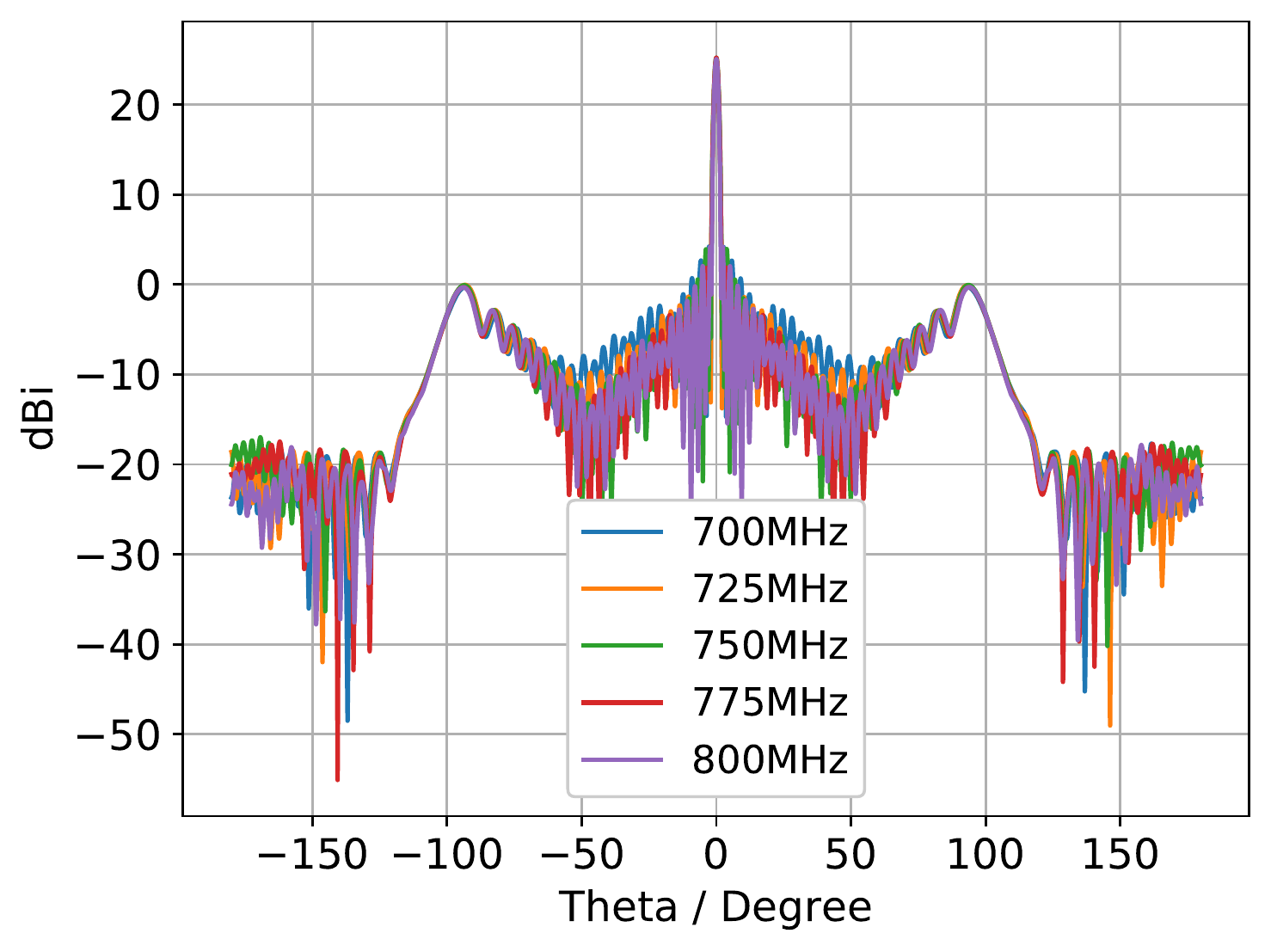}} \\
  \subfigure[E-plane of X-polarization.]{  \includegraphics[width=0.45\textwidth]{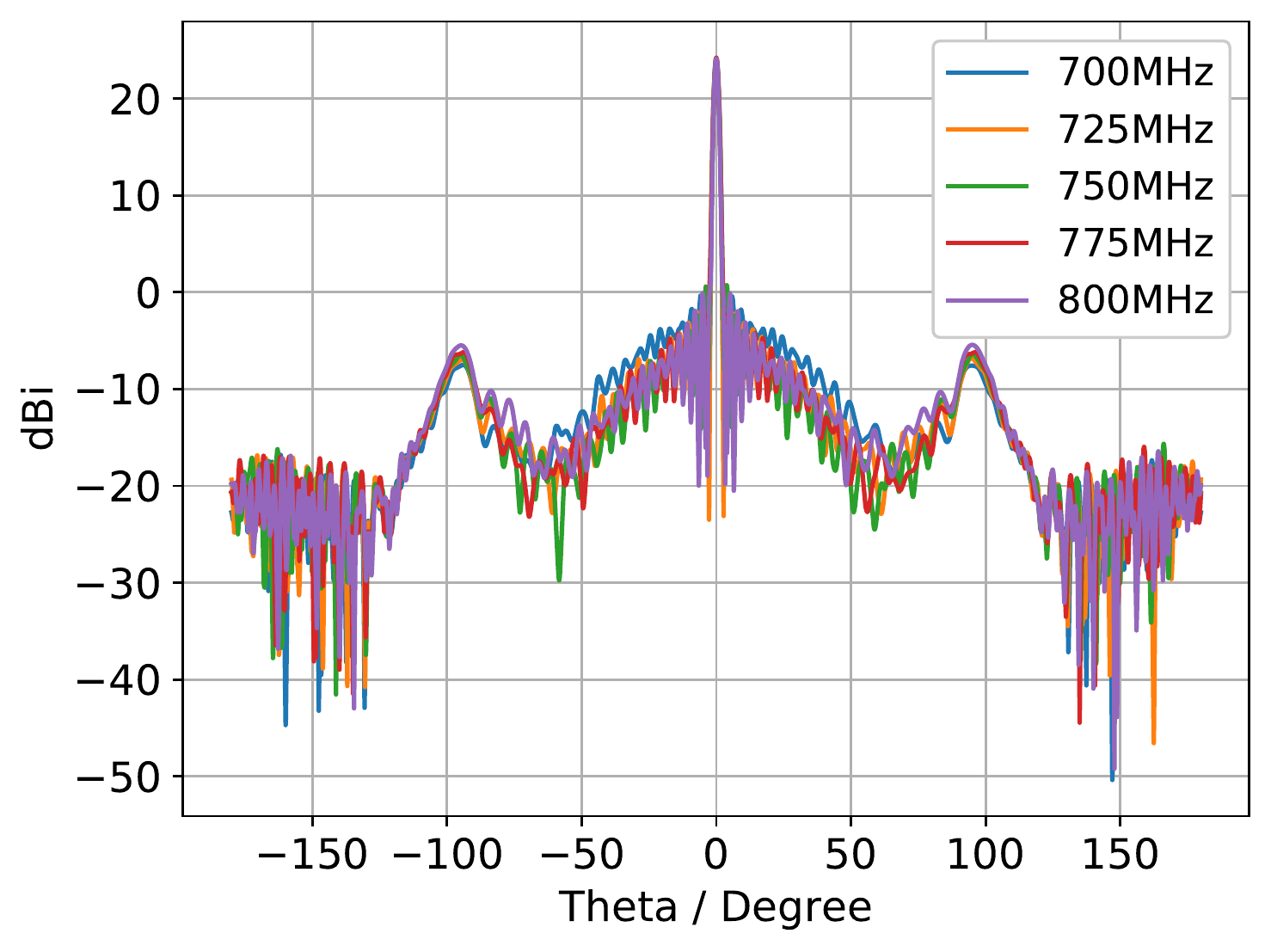}}
   \subfigure[H-plane of X-polarization.]{  \includegraphics[width=0.45\textwidth]{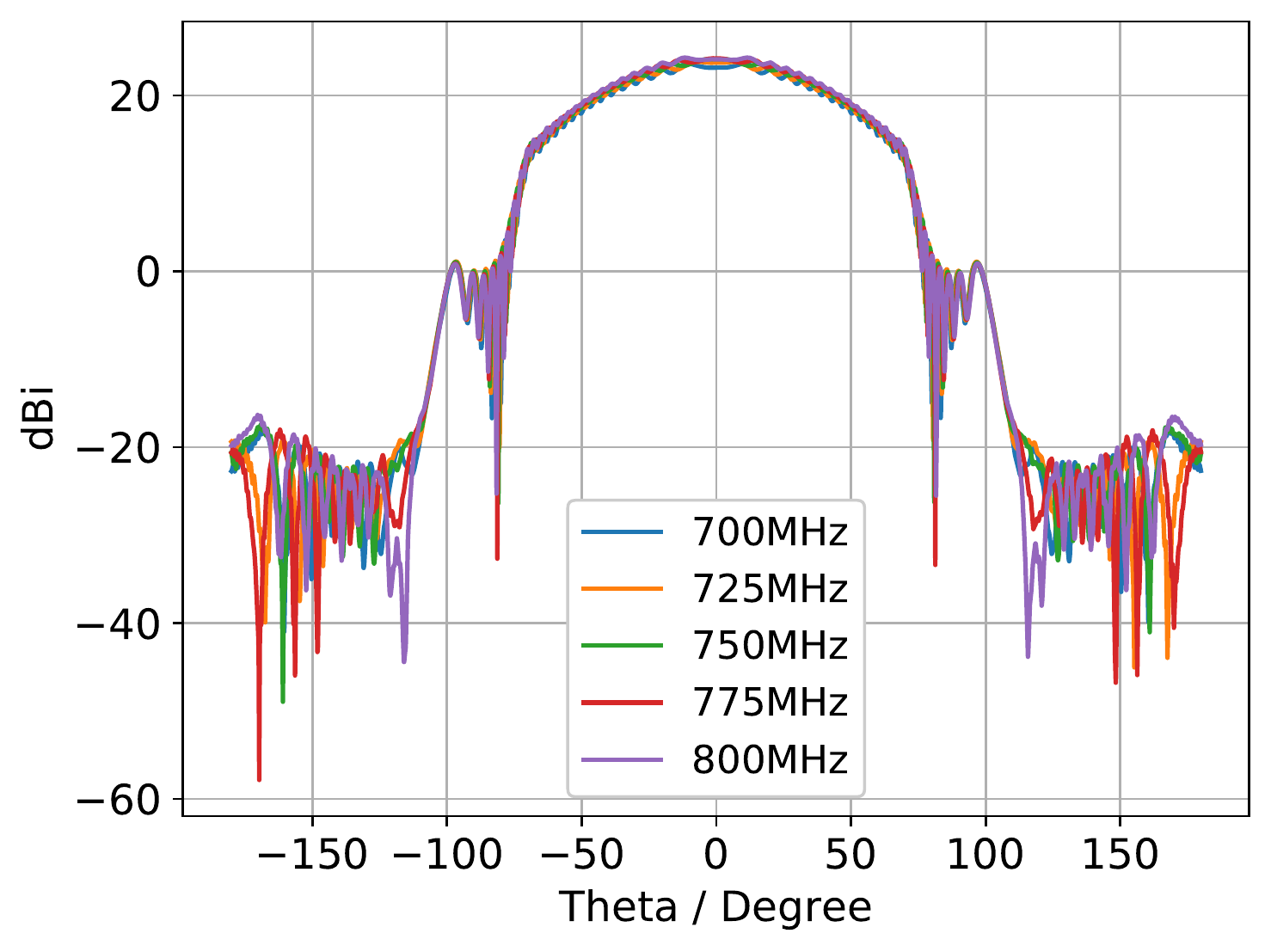}} 
  \caption{The simulated beam profiles at 700--800 MHz. The directivity is 25.2 dBi for the X-polarization and 24.2 dBi for the Y-polarization.}
  \label{fig:beamsim}
\end{figure*}

For the longer E-W baseline A2Y-C2Y, the angular resolution is higher and adjacent sources unresolved by the shorter baselines may be resolved by the longer ones.  The fringes are clearer and the phase varies more rapidly. The horizontal band structure is less significant in the longer baselines, perhaps because the larger physical separation of the feeds of longer baselines leads to smaller cross-coupling and weaker correlated noise.

The horizontal stripes are very stable, and are removed by subtracting off a moving average over 1 hour. As an example, we show the processed data during nighttime for the baseline A3Y-B18Y in Fig. \ref{fig:raw_vis_hsv_180322_rm_night_5097}.
Many less prominent sources can now be seen and they repeat each sidereal day. We list the sources that can be seen directly in this visibility and identified with well known strong radio sources in Table \ref{tab:source_list}, where both day and nighttime sources are included. Some sources are complex ones, which would be resolved into multiple components with higher angular resolution. The flux density for the entire complex set is not available. Fainter fringes could be spotted in the visibility, however are too numerous for all of them to be listed.

In Fig.~\ref{fig:raw_vis_hsv_160927_rm_night_1065}, we plot the visibility for A1Y-B2Y for a different observation starting from 2016/09/27. Here the Sun is positioned on the opposite part of the ecliptic. There happen to be many more bright radio sources during the night sky. We can see numerous fringes in this case, even without removing the Sun.

\section{The Antenna Beam Response Pattern}
\label{sec:beam}

The beam profile for a feed on the Tianlai array is simulated using the {\tt CST Microwave Studio} software.  The simulated beam profiles at 650~MHz, 1050~MHz and 1420~MHz were reported earlier in \cite{Cianciara2017}, which demonstrated that the Tianlai cylinder antennas and feeds have a quite good broadband response. Here, we show the simulated beam profiles in the 700--800~MHz range in Fig.~\ref{fig:beamsim}. As expected, the beam has a narrow peak  with a full width half maximum (FWHM) of $1.6^\circ (1.8^\circ)$ for X (Y) polarization at 750~MHz along the E-W direction (H(E)-plane for the X(Y) polarization), and a very broad peak with FWHM about $62.2^\circ(71.4^\circ)$ along the N-S direction (E(H)-plane for the X(Y) polarization). The broad beam along the N-S direction is largely determined by the beam profile of the feed.

We can verify the simulated beam profile in the E-W direction by observing the transit of strong sources. We use Cyg~A as our calibrator. The declination of Cyg~A is $+40.7^\circ$ and the zenith angle is only $3.4^\circ$ at the peak of this transit.  Because its track is quite close to the E-W great circle passing through the zenith, it provides a good determination of the E-W beam profile near the center of the beam.

\begin{figure}[H]
  \centering
  \includegraphics[scale=0.4]{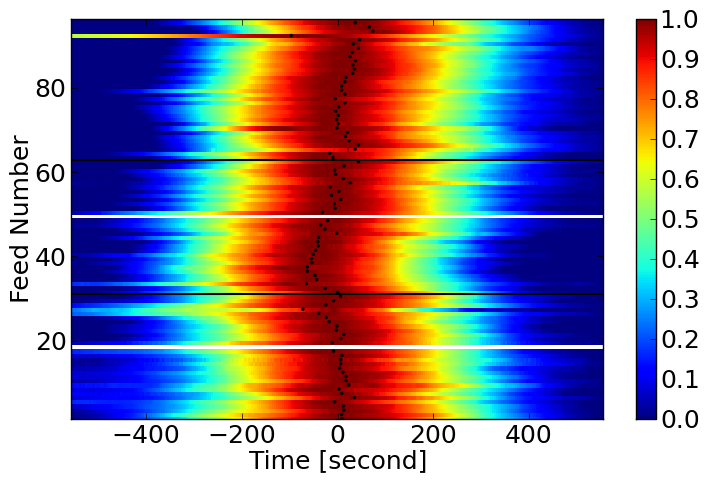}
  \includegraphics[scale=0.4]{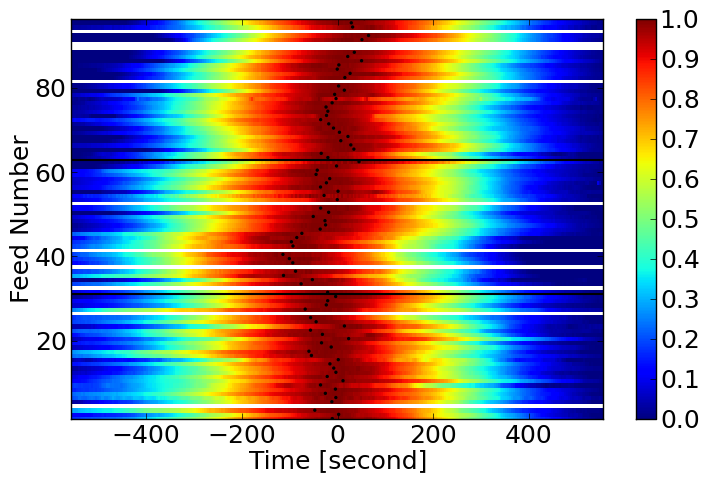}
  \caption{The auto-correlations of all X (top) and Y (bottom) polarizations vs time during Cyg A transit. The black dots mark the peak time of the transit curve fitted by a Gaussian function. Malfunctioning feeds are masked and appear as horizontal white lines.}
  \label{fig:transit_2D_auto}
\end{figure}

The auto-correlations of the cylinder array units during the peak of the Cyg~A transit event is plotted in Fig.~\ref{fig:transit_2D_auto} as a function of time. Here the background has been removed, so that the auto-correlation reflects what is induced by Cyg~A.  The amplitude is averaged over the full frequency band to improve the SNR. These curves are normalized at their peak values and arranged according to their relative positions on the cylinders. The feed numbers given here are $1\sim 31$ for cylinder A, $32\sim 63$ for cylinder B, and $64 \sim 96$ for cylinder C.

The direction angle alignment of different feeds inevitably has some errors, so the time of the signal peak during the Cyg~A transit is different for each feed. The standard deviation of the transit times are $\sigma_Y = 38.8 {\rm s}$ and $\sigma_X = 31.1 {\rm s}$, corresponding to $0.123^\circ$ and $0.099^\circ$ in angle respectively. These errors in angle are reasonable values for an installation with simple mechanical tools and adjustment. The difference of the average time of the two polarizations is 18.6 s (or $0.059^\circ$ in angle), which is very small.

While the auto-correlation directly gives the response for each feed, it has a large receiver noise and is sensitive to the contribution of the mean sky brightness. Therefore, we also study the pointing of the beams using cross-correlations. We derive an average beam profile by averaging the cross-correlation amplitudes for all of the E-W baselines which span different cylinders during the transit of Cyg~A. This is averaged over all frequencies.

\begin{figure}[H]
  \centering
  \includegraphics[width=0.4\textwidth]{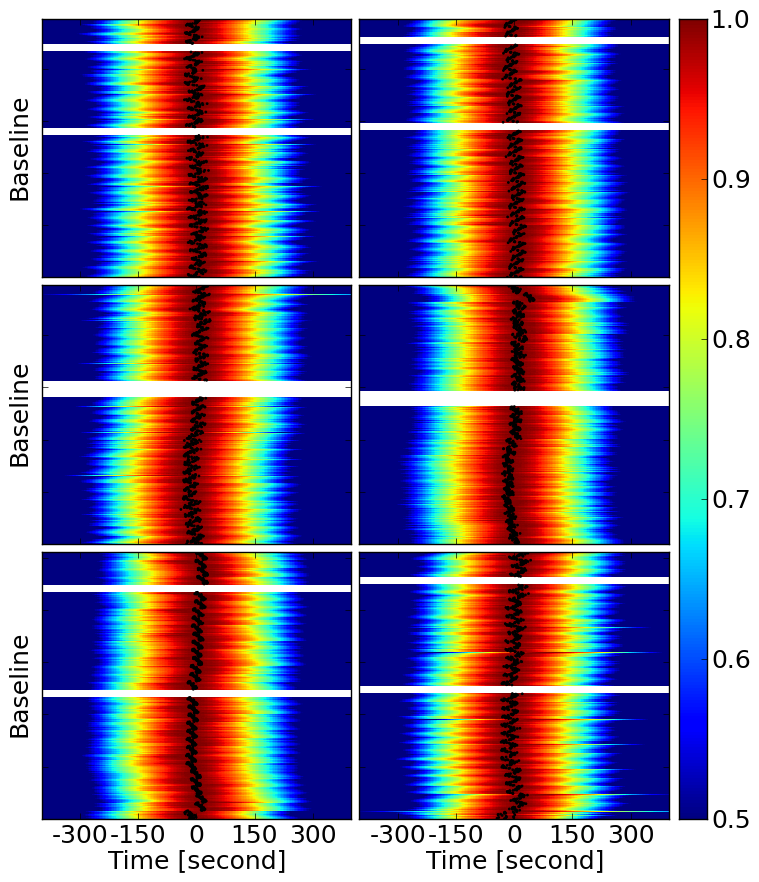}
  \caption{Distribution of the transit amplitude curves for the X-polarization derived from averaging the cross-correlations of all of the E-W baselines. Top panels: $A_B$ (left) and $A_C$ (right); Center panels: $B_C$, $B_A$; Bottom panels: $C_A$, $C_B$. The black dots show the transit peaks fitted by a Gaussian function. }
  \label{fig:transit_2D_cross_V}
\end{figure}

\begin{figure}[H]
  \centering
  \includegraphics[width=0.4\textwidth]{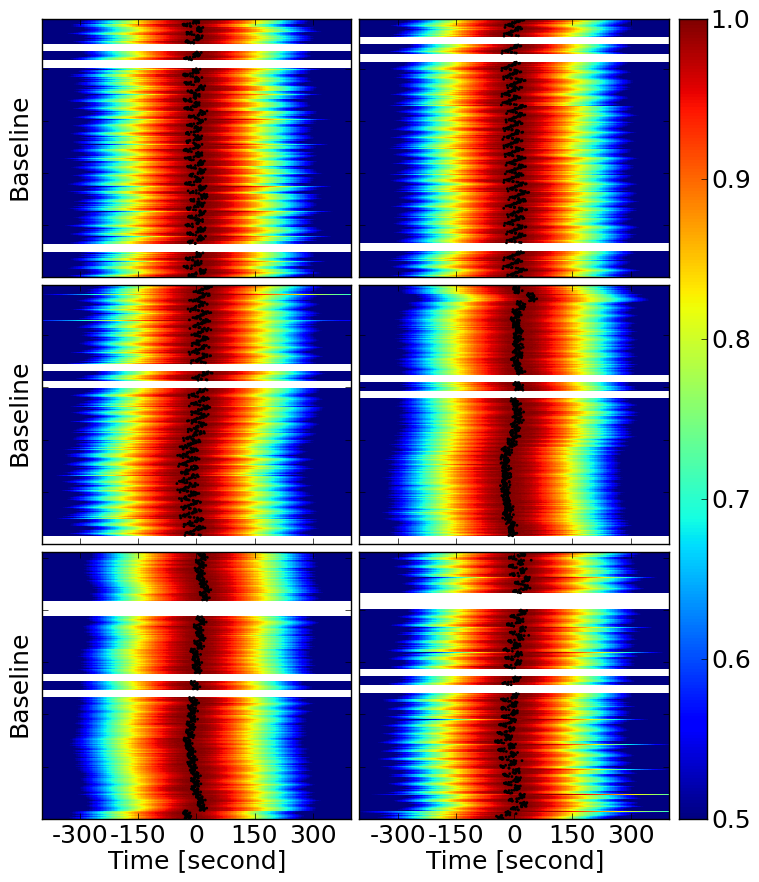}
  \caption{Same as Fig. \ref{fig:transit_2D_cross_V} but for Y polarization.  }
  \label{fig:transit_2D_cross_H}
\end{figure}

For example, the E-W profile for the X-polarization of the $i$-th feed on cylinder A can be obtained by averaging over its cross-correlations with the B cylinder as
\begin{equation}
   P_i^X(A_B)= \sum_{\nu} \sum_{j}  V^{XX}_{(Ai),(Bj)}(\nu).
\end{equation}
We collect such profiles for cylinder A and denote it as $A_B$. In Fig.~\ref{fig:transit_2D_cross_V} and Fig.~\ref{fig:transit_2D_cross_H}, we plot these amplitudes.

The general shapes of the profiles for the same cylinder obtained with different cross-correlation cylinders (e.g. $A_B$ and $A_C$) are quite similar with each other, giving confidence in this method. Also, the X- and Y-polarizations for the same cylinder are very similar to each other, suggesting that the deviations are due primarily to the misalignment of the installation angle for the feeds. However, we note that this misalignment does introduce a small pointing error in the data, which should be accounted for when combing visibilities for map-making.

\begin{figure}[H]
  \centering
  \includegraphics[width=0.21\textwidth]{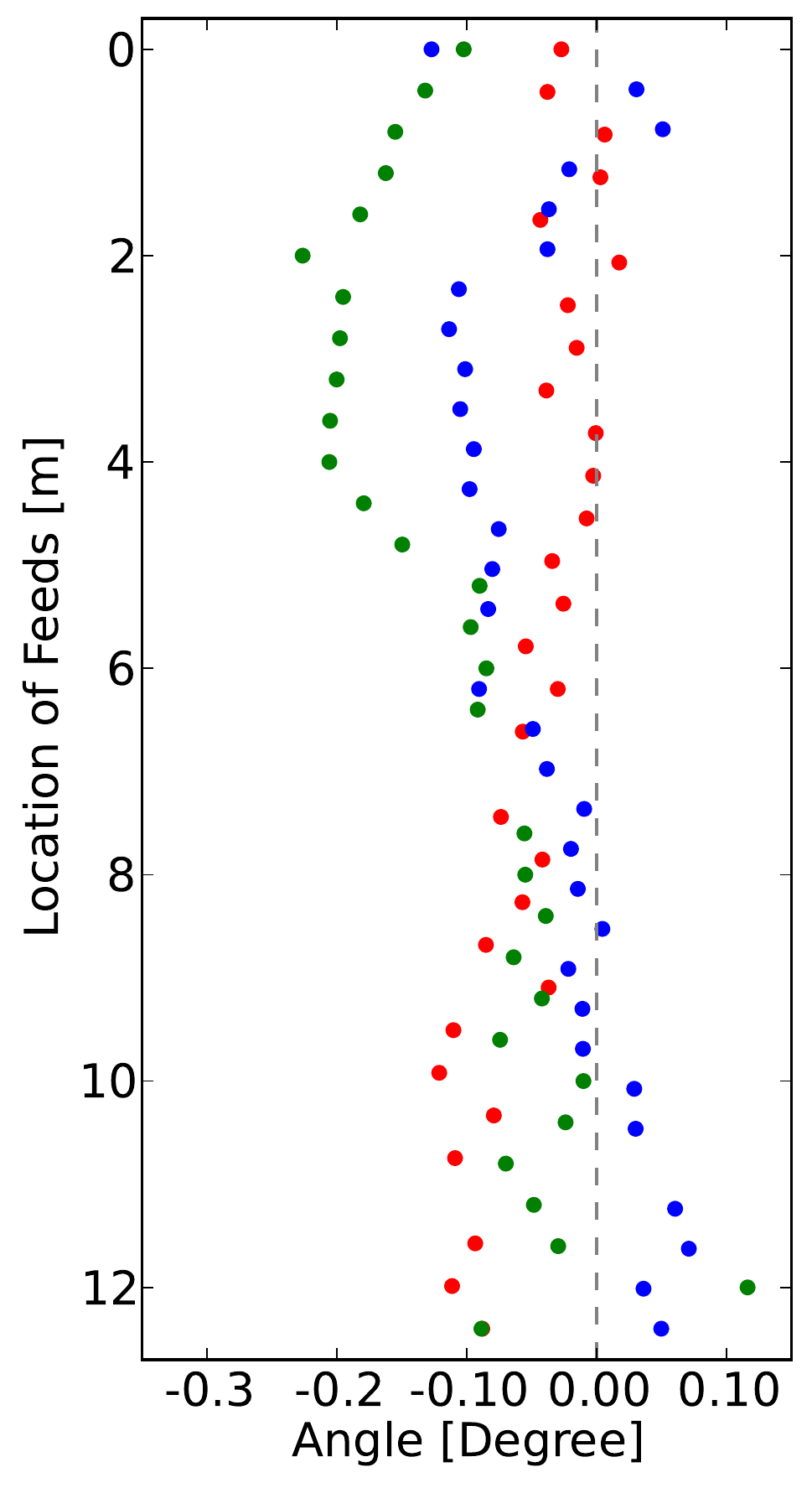}
  \includegraphics[width=0.21\textwidth]{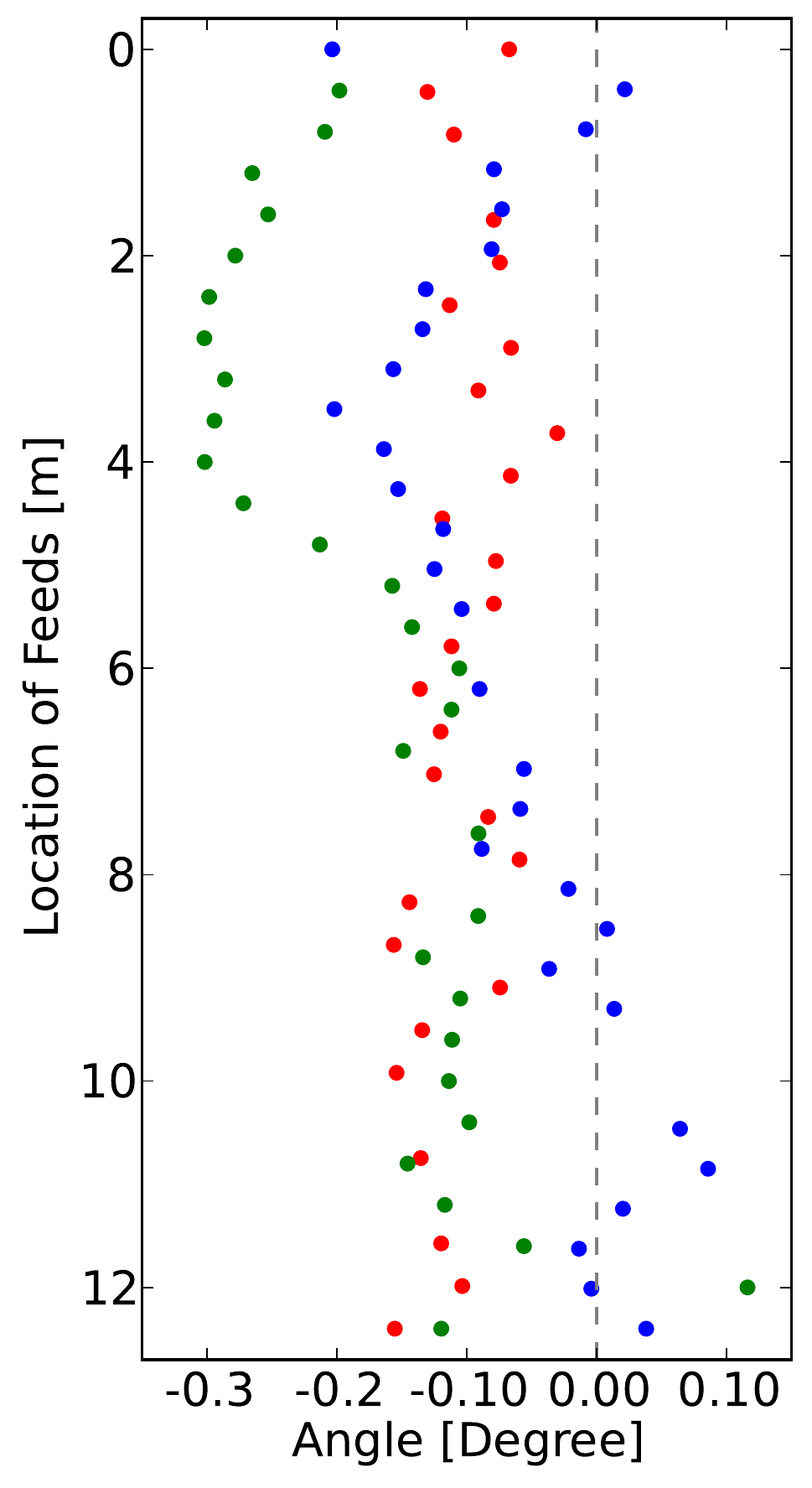}
  \caption{The solved antenna pointings of the X (left) and Y (right) polarizations. Red, green, blue are for Cylinder A, B, C, respectively. The pointings are extracted from the cross-correlations with the baselines in the E-W direction. The baselines in the N-S direction are not used due to the low SNR of Cyg~A in the visibility amplitude. }
  \label{fig:ant_pointing_all}
\end{figure}

The beam center of each feed can be estimated from the center or peak position of the cross-correlations. Assuming the beam center to be $a_l$ for feed A$l$ and $b_m$ for feed B$m$, the cross-correlation beam center will be $\frac{1}{2} (a_l + b_m)$. The centers of the cross-correlation beams are shown as  black dots in Fig. \ref{fig:transit_2D_cross_V} and Fig. \ref{fig:transit_2D_cross_H}. The minimum variance solution for the beam center of the feeds is then given by (see the Appendix for derivation)
\begin{eqnarray}
  a_l &=& \frac{1}{M}\sum_{m=1}^M \theta_{l,m}^\mathrm{AB} + \frac{1}{N}\sum_{n=1}^N \phi_{l,n}^\mathrm{AC} - \frac{1}{MN}\sum_{m=1}^M \sum_{n=1}^N \gamma_{m,n}^\mathrm{BC} \\
  b_m &=& \frac{1}{N}\sum_{n=1}^N \theta_{m,n}^\mathrm{BC} + \frac{1}{L}\sum_{l=1}^L \phi_{m,l}^\mathrm{BA} - \frac{1}{NL}\sum_{n=1}^N \sum_{l=1}^L \gamma_{n,l}^\mathrm{CA} \\
  c_n &=& \frac{1}{L}\sum_{l=1}^L \theta_{n,l}^\mathrm{CA} + \frac{1}{M}\sum_{m=1}^M \phi_{n,m}^\mathrm{CB} - \frac{1}{LM}\sum_{l=1}^L \sum_{m=1}^M \gamma_{l,m}^\mathrm{AB}
  \label{eq:beam_pointing_sol}
\end{eqnarray}
where $a_l, b_m, c_n$ are the solved beam pointing for feed A$l$, B$m$, C$n$ in cylinder A, B, C. $L = 31, M = 32, N = 33$ are the numbers of feeds. $\theta, \phi, \gamma$ are the observed cross-correlation beam centers. The distribution of the solved beam centers is shown in Fig. \ref{fig:ant_pointing_all}.

\begin{figure}[H]
  \centering
  \includegraphics[width=0.45\textwidth]{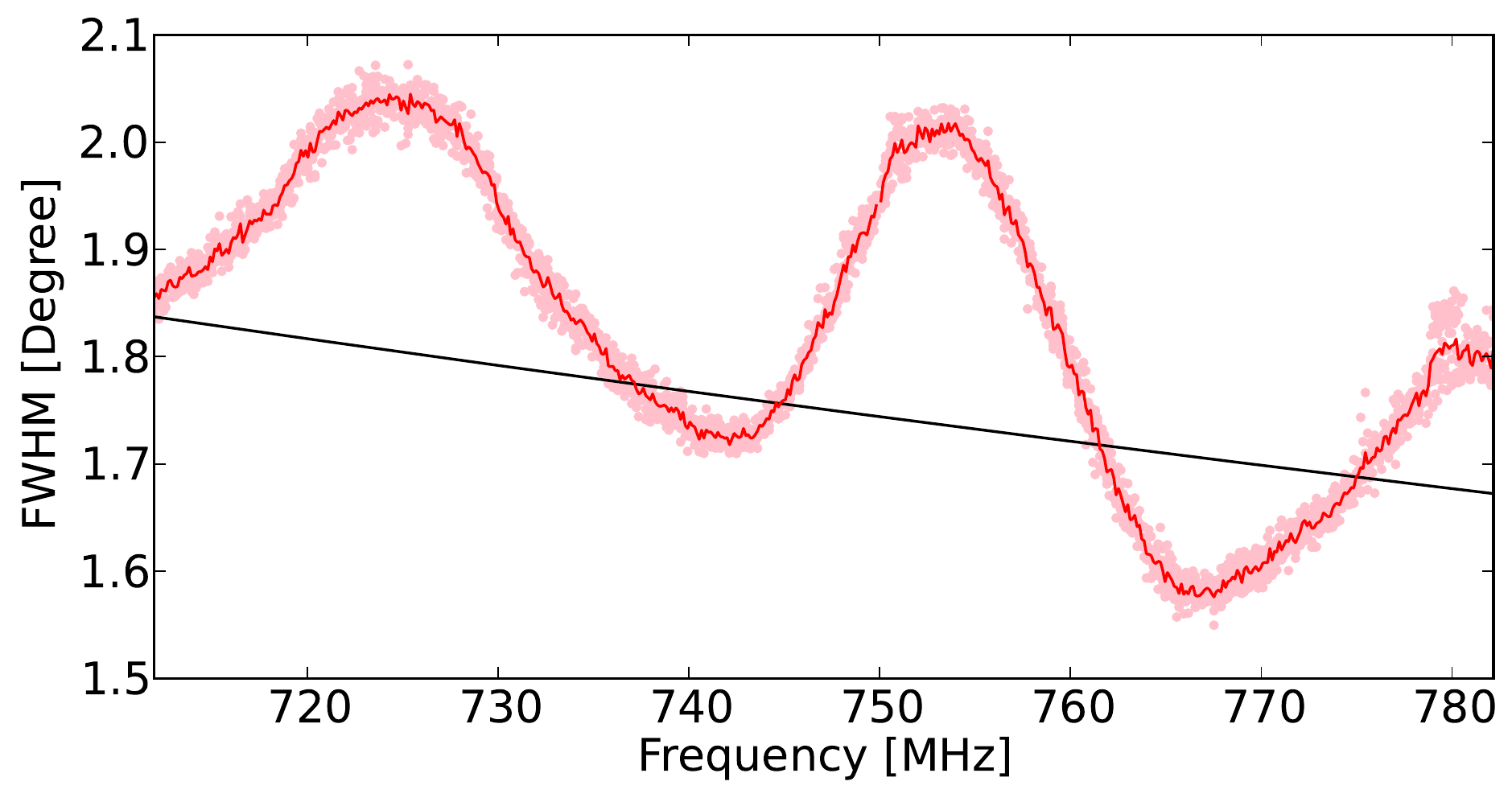}\\
  \includegraphics[width=0.45\textwidth]{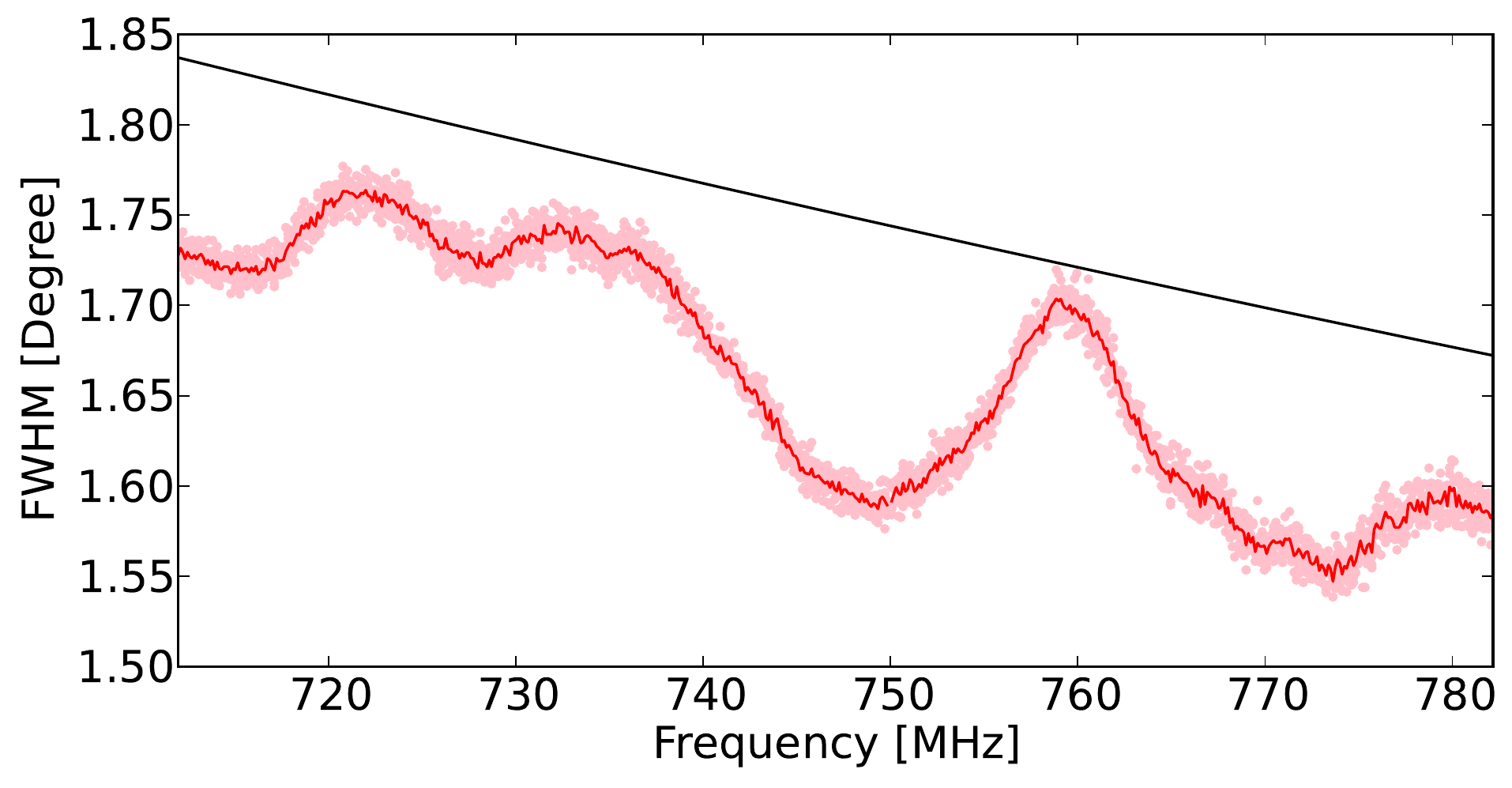}\\
  \caption{The mean FWHM of the primary beam in the E-W direction as a function of frequency using daily transits of Cyg A over 7 days. The measurement is for two fairly typical baselines A14Y-C18Y (top) and A2X-C17X (bottom). The dots are the fitted beam width at each frequency channel for all days, and the red line is the mean. The black line shows the FWHM of a uniformly-illuminated Airy disk with an effective diameter equal to 90\% of the cylinder width ($0.9\times  15$~m).}
  \label{fig:beam_freq}
\end{figure}

Finally, in Fig. \ref{fig:beam_freq}, we show the FWHM of one cut through the beam pattern from a pair of antennas as a function of frequency. The pattern is measured repeatedly in the E-W direction by observations of the transit of Cyg~A on 7 successive days. For each transit and each visibility we fit a Gaussian shape to the magnitude of the visibility as a function of time. The measured pattern is effectively the geometric mean of the patterns of two dishes, which are nominally coaligned. Day-to-day fluctuations of the FWHM are less than 3\%. We also plot the case of a diffraction - limited circular aperture $(1.028\lambda / D_{\rm eff}$ with $D_{\rm eff} = 0.9D)$. Compared with the ideal case, there are apparent sinusoidal variations in the beam size. A similar phenomenon was also found to occur on the CHIME cylinders  \cite{2014SPIE.9145E..4VN}. It may induce frequency-dependent modulations due to mode-mixing effect which complicates the detection of 21~cm signal.

From the beam profile, we can derive the directivity and effective area of the antenna. The directivity is given by
\begin{equation}
  D=\frac{P_{\rm max}(\theta,\phi)}{P_{\rm av}},
\end{equation}
where $P_{\rm max}(\theta, \phi)$ is the maximum power of the antenna, while the average power is
\begin{equation}
  P_{\rm av}=\frac{1}{4\pi}\int P(\theta,\phi) \mathrm{d}\Omega
\end{equation}
so that $D=4\pi/\Omega_A$. The effective area of the antenna is
\begin{equation}
  A_e = \frac{D \lambda^2}{4\pi}.
  \label{eq:area_effective}
\end{equation}

In our case, for a single feed on the cylinder, the simulated directivity is about 25.2 dBi for X polarization and 24.2 dBi for Y-polarization, so that, $A_e^X=4.22 \m^2 $, and $A_e^Y = 3.35 \m^2$ at 750 MHz. These relatively small effective areas are due to the fact that despite the large total area of the cylinder, for observing one particular direction, each feed receives only the wave reflected from a narrow strip of the reflecting surface.

\section{Calibration}
\label{sec:calibration}

Accurate calibration is essential for determining the complex gain (particularly the phase) of each array element of an interferometer array. Below, we first calibrate the frequency response (bandpass), and then calibrate the complex gains of different feed channels using the transit of a strong astronomical source such as Cyg~A \footnote{Strictly speaking, Cyg A is not a point source but a double lobe radio galaxy, but for our purpose, at the low angular resolution of the Tianlai pathfinder, it can be used as one.} (referred as the absolute calibration in this paper). Finally, when a strong astronomical point source is not available, calibration can be made with the help of the artificial CNS (referred to as relative calibration in this paper). The calibration algorithm is described in \cite{Zuo2019}.

\subsection{The Bandpass}
\label{subsec:bandpass}

\begin{figure*}
  \centering
  \includegraphics[width=0.45\textwidth]{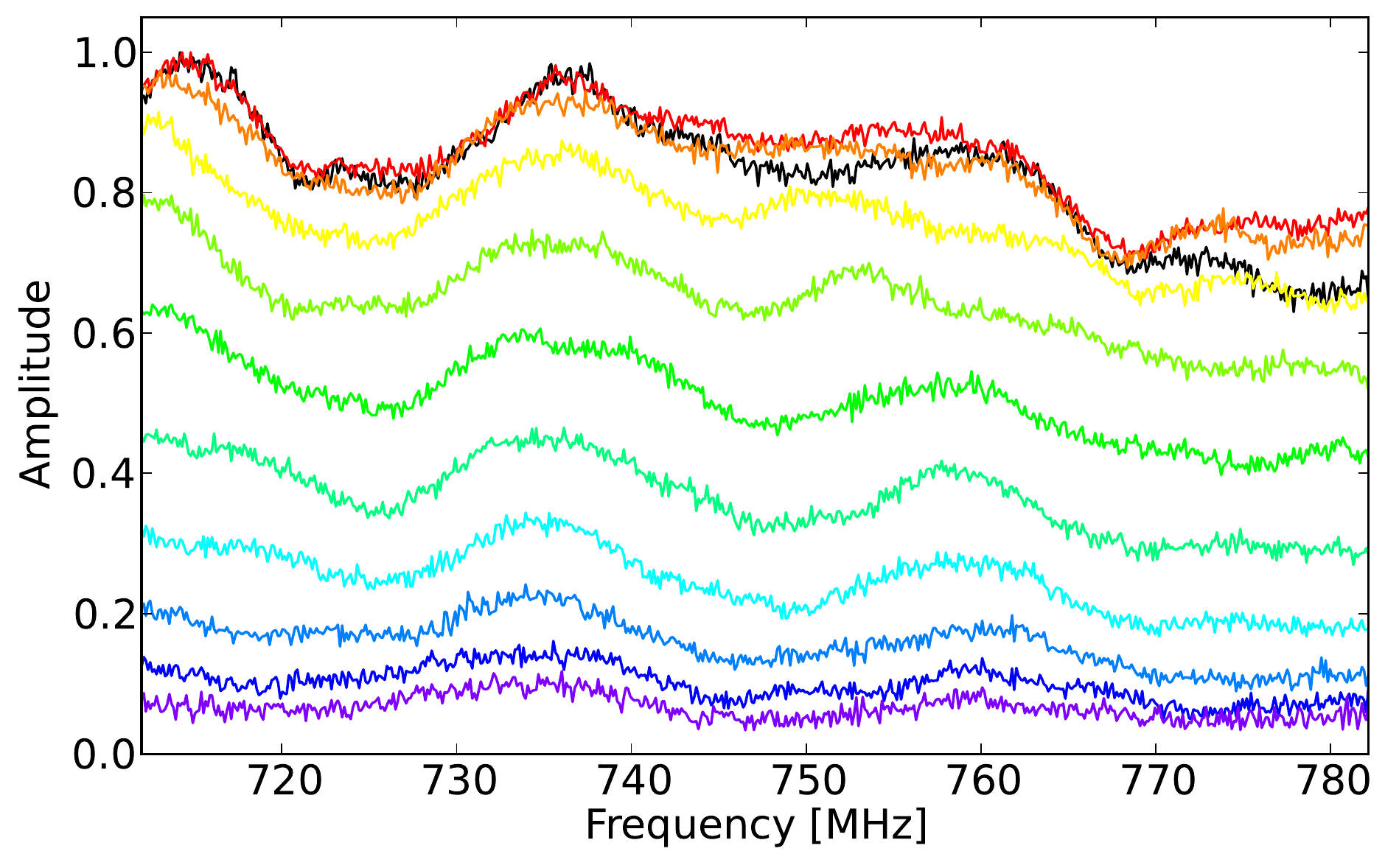}
  \includegraphics[width=0.45\textwidth]{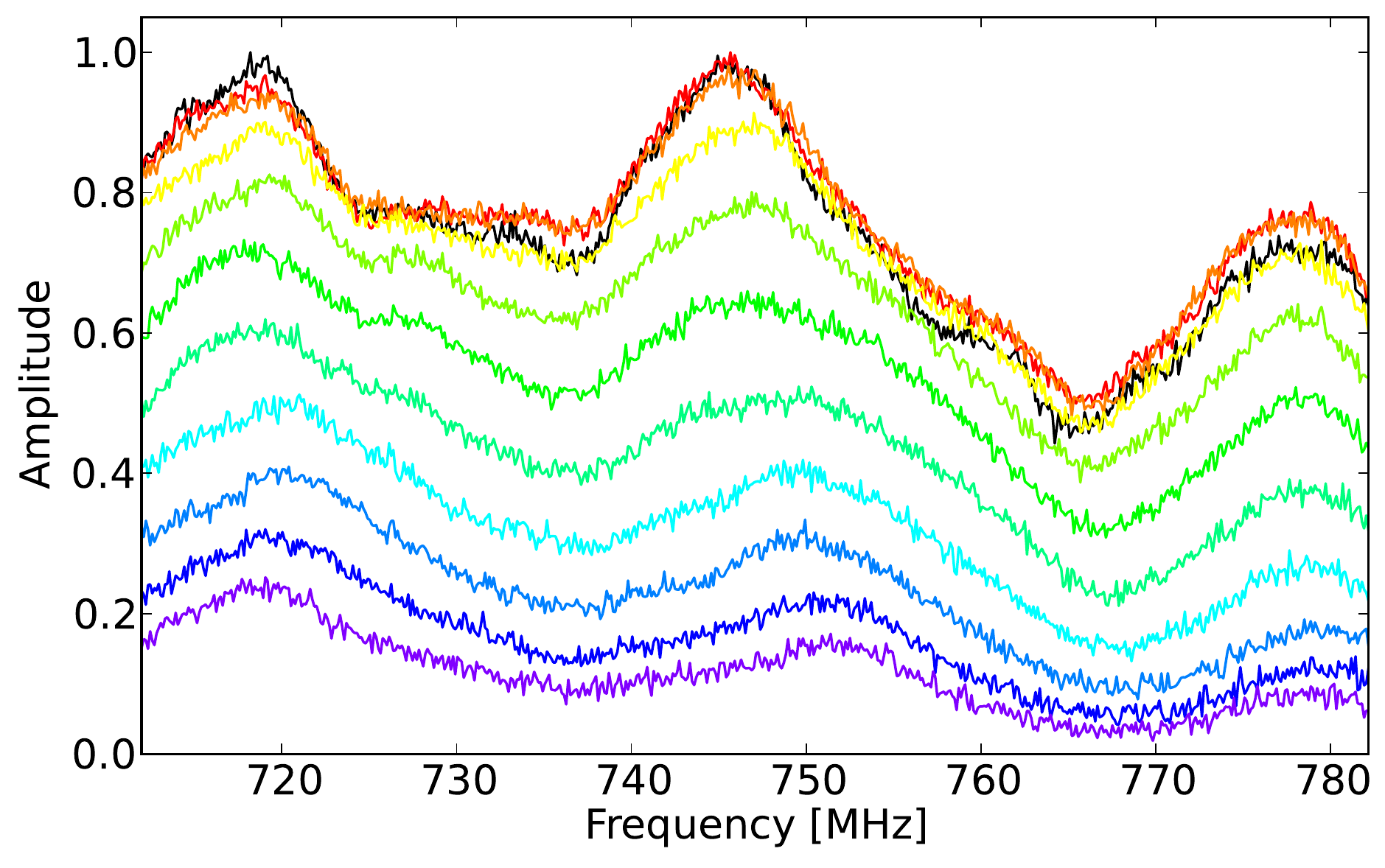}\\
  \includegraphics[width=0.45\textwidth]{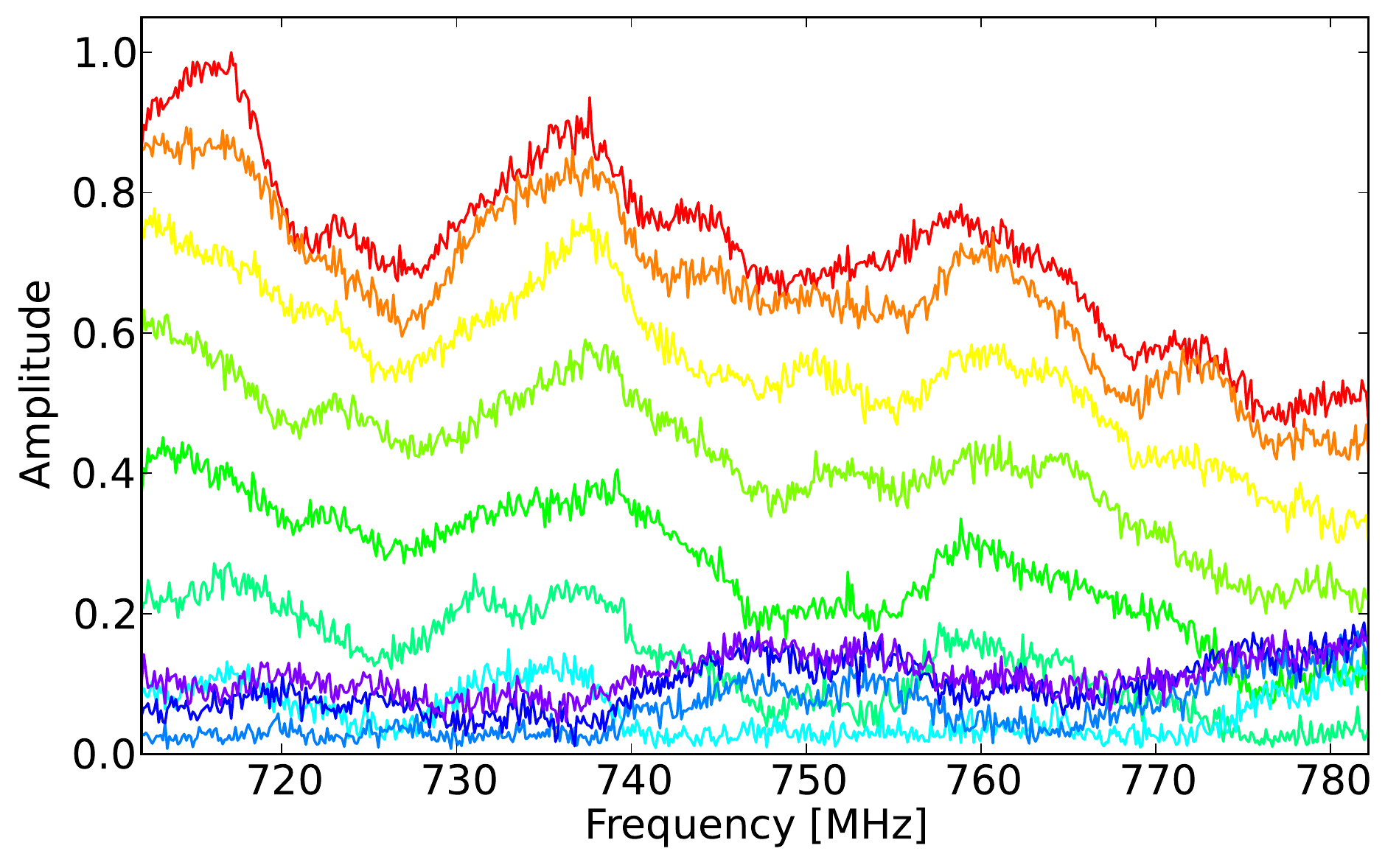}
  \includegraphics[width=0.45\textwidth]{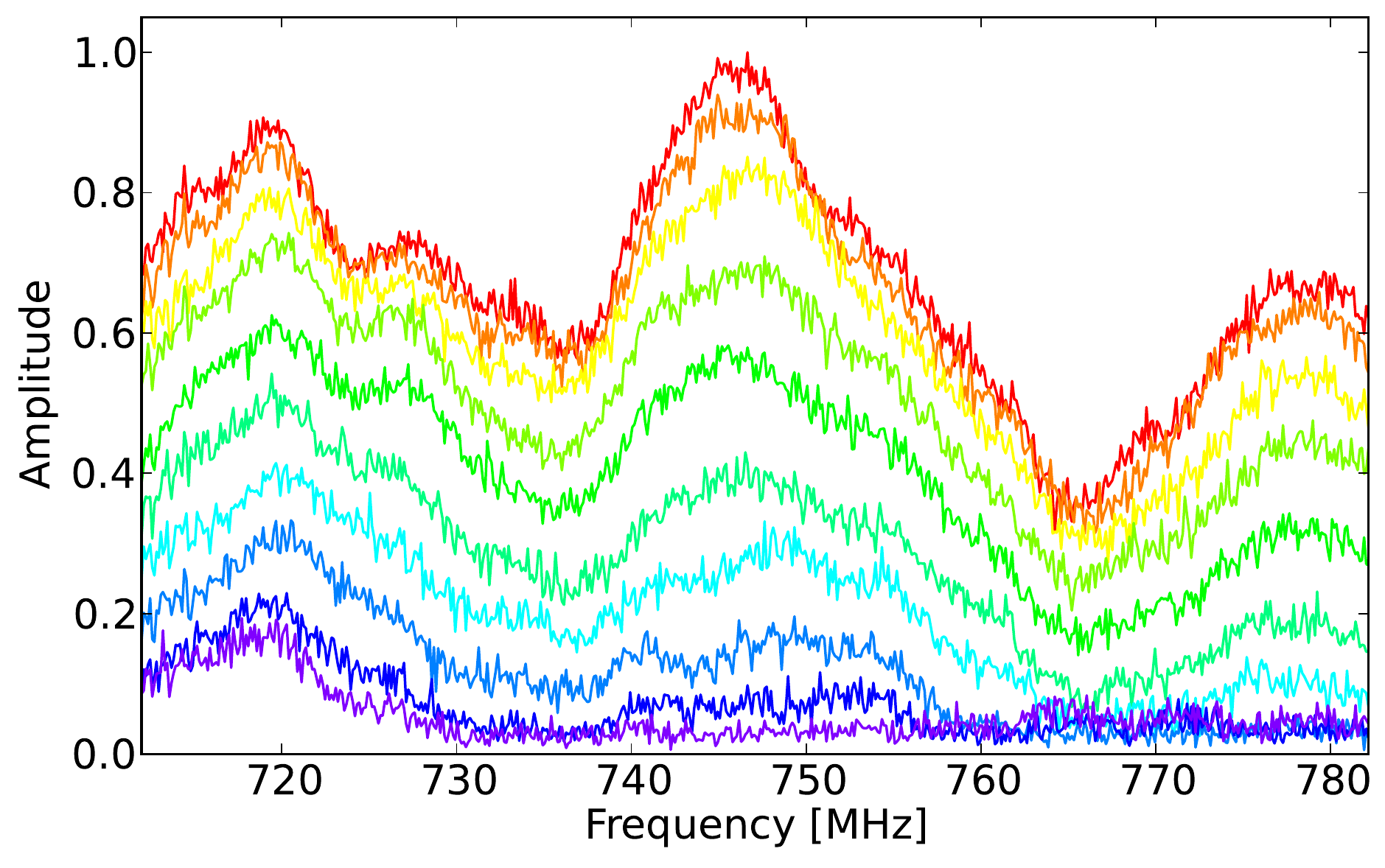}\\
  \includegraphics[width=0.45\textwidth]{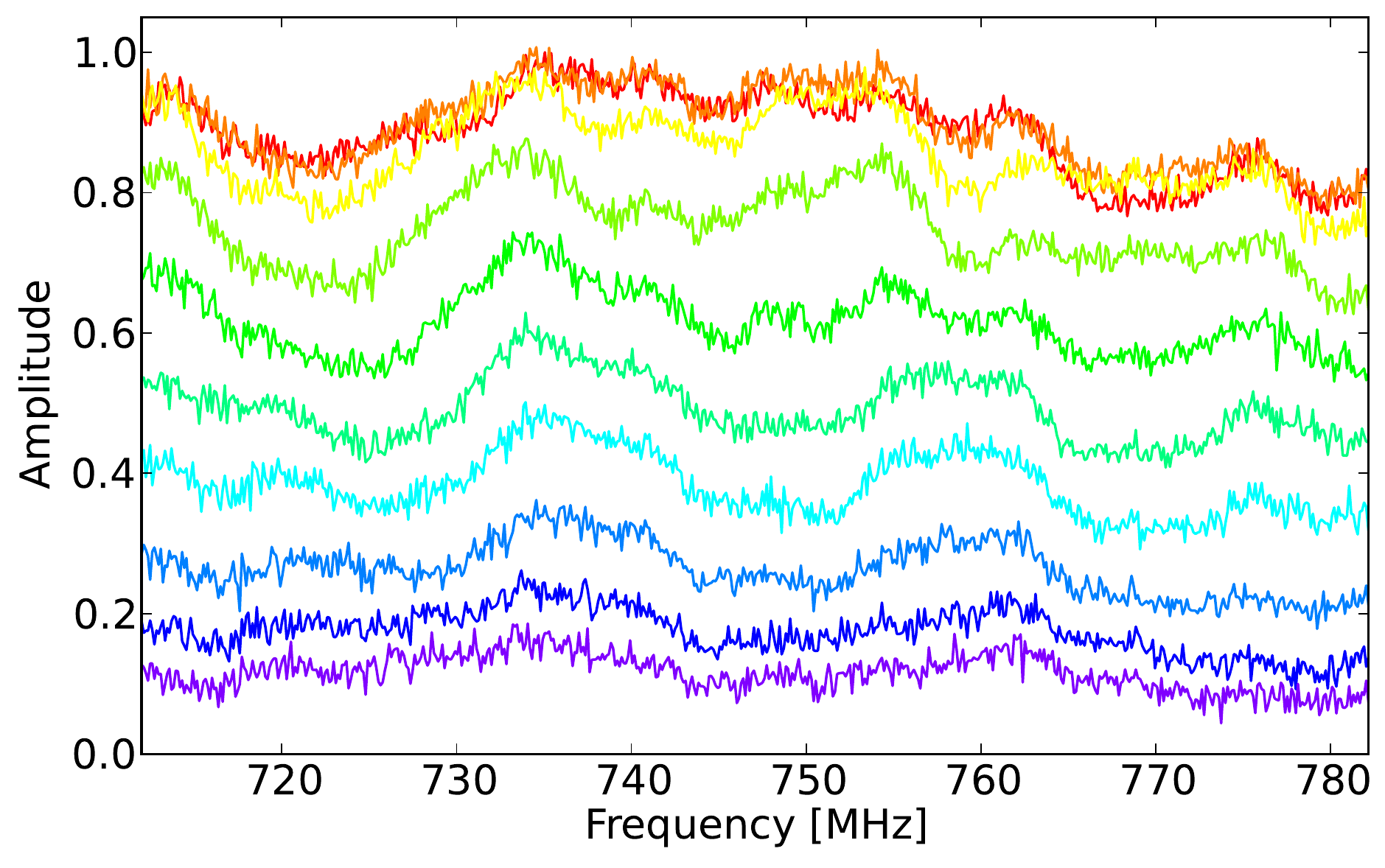}
  \includegraphics[width=0.45\textwidth]{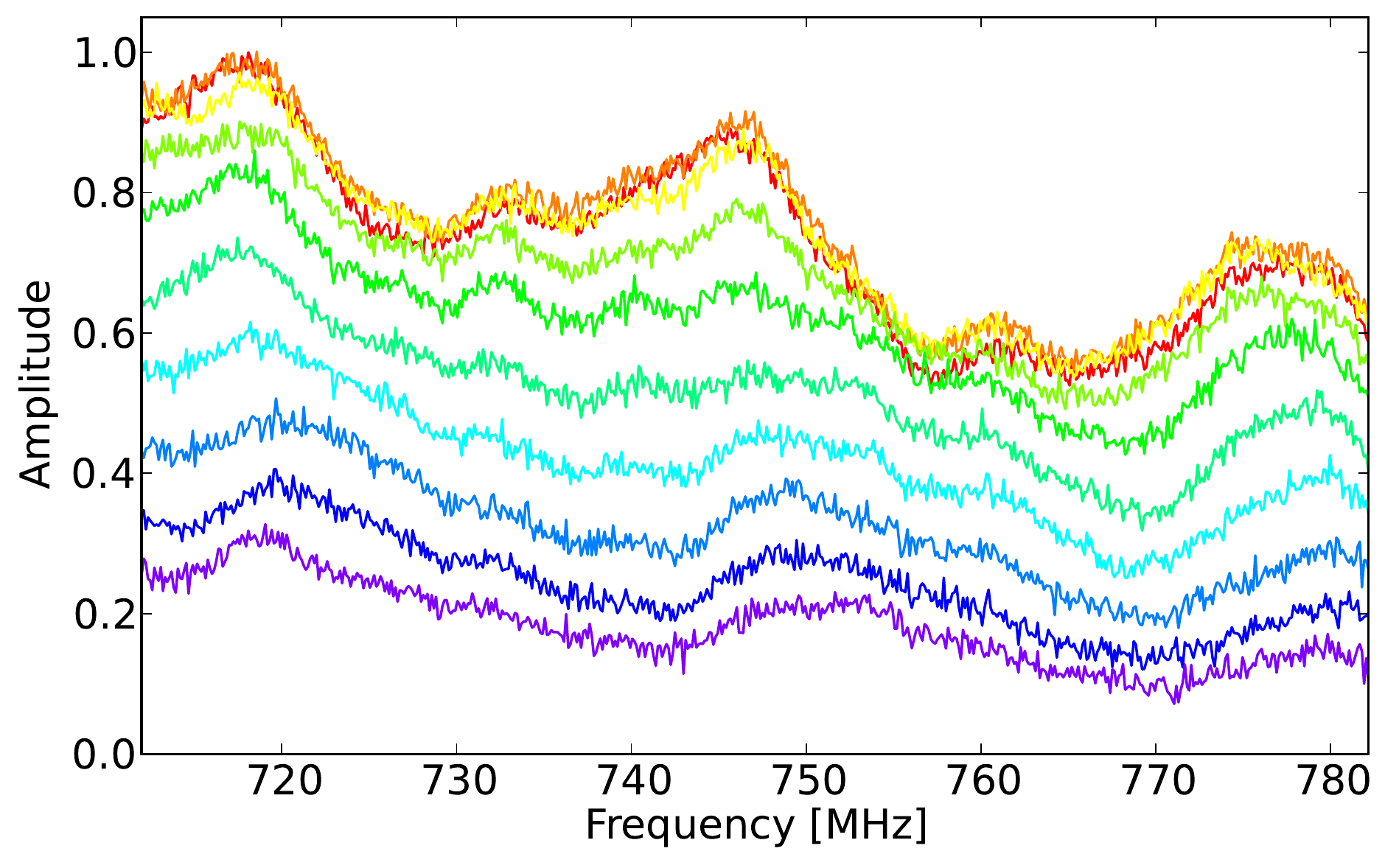}\\
  \caption{Frequency response as obtained from Cyg~A transit, for a typical cross-correlation visibility A4-B9 (top) and two auto-correlations, A4-A4 (middle) and B9-B9 (bottom). Note that for auto-correlation, the background 
has already been subtracted.  The left column shows XX polarization and right column shows YY polarization. The colors from purple to red indicate the different times during the transit process, each separated by 1 minute. The black line in the top panel corresponds to the square root of the product of the two auto-correlation frequency responses at the peak of the transit.}
  \label{fig:band_var_short}
\end{figure*}

\begin{figure*}
  \centering
  \subfigure[A2Y-B27Y from the data set of 2018/03/22.]{ 
  \includegraphics[width=0.45\textwidth]{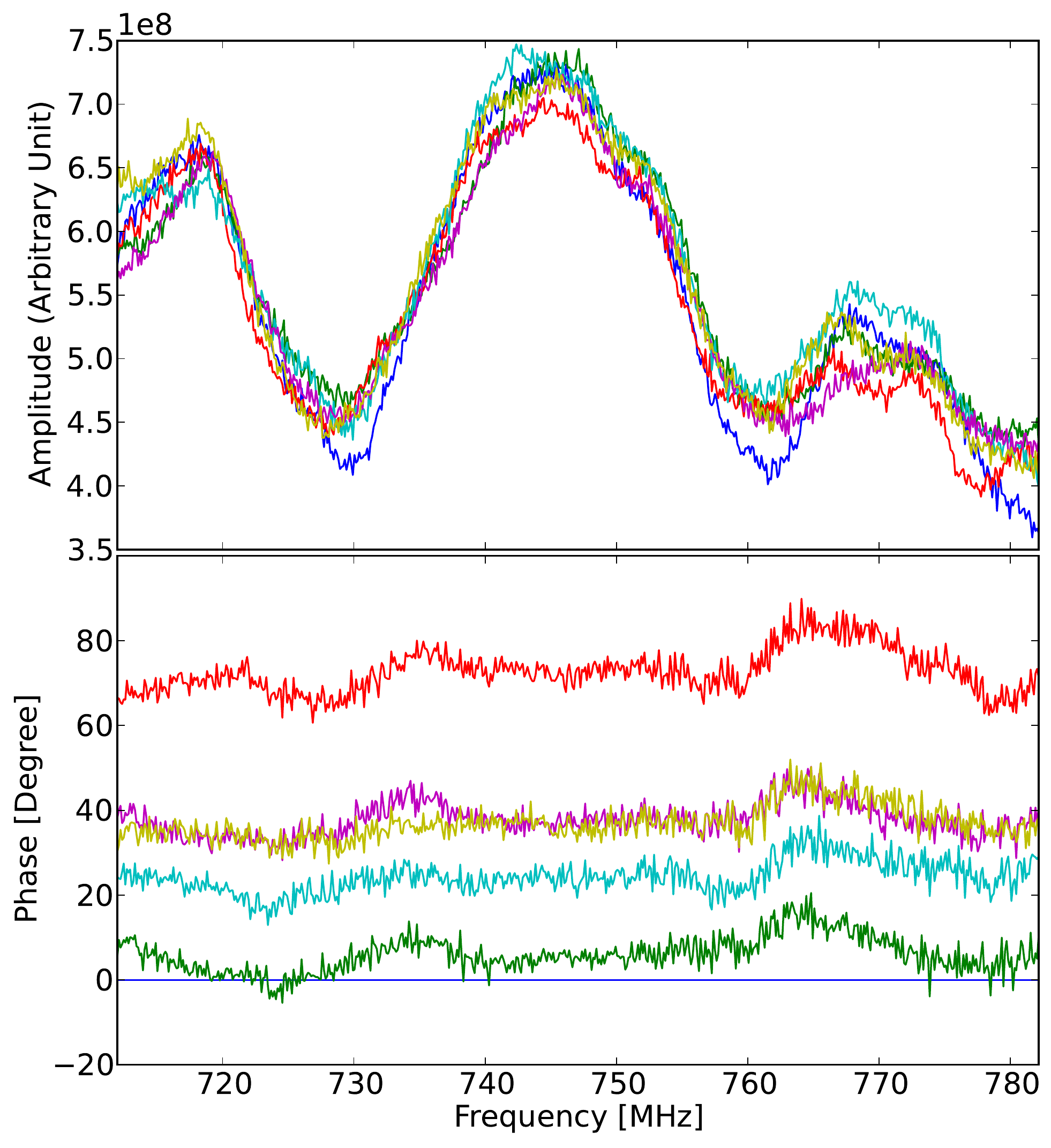}}
  \subfigure[A2Y-B3Y from data set of 2016/09/27]{
  \includegraphics[width=0.45\textwidth]{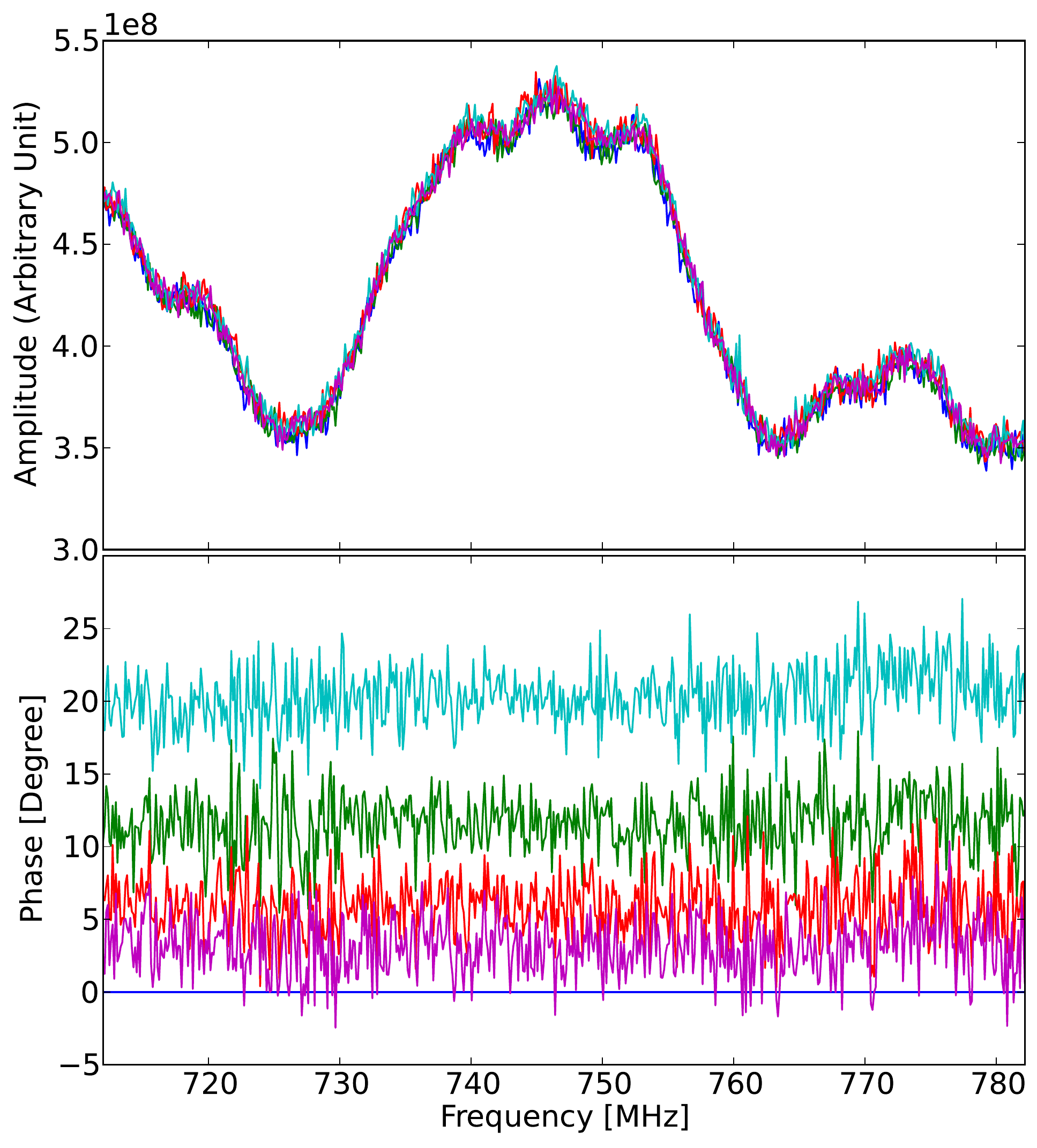}}\\
  \subfigure[A13X-B31X from the data set of 2018/03/22]{
  \includegraphics[width=0.45\textwidth]{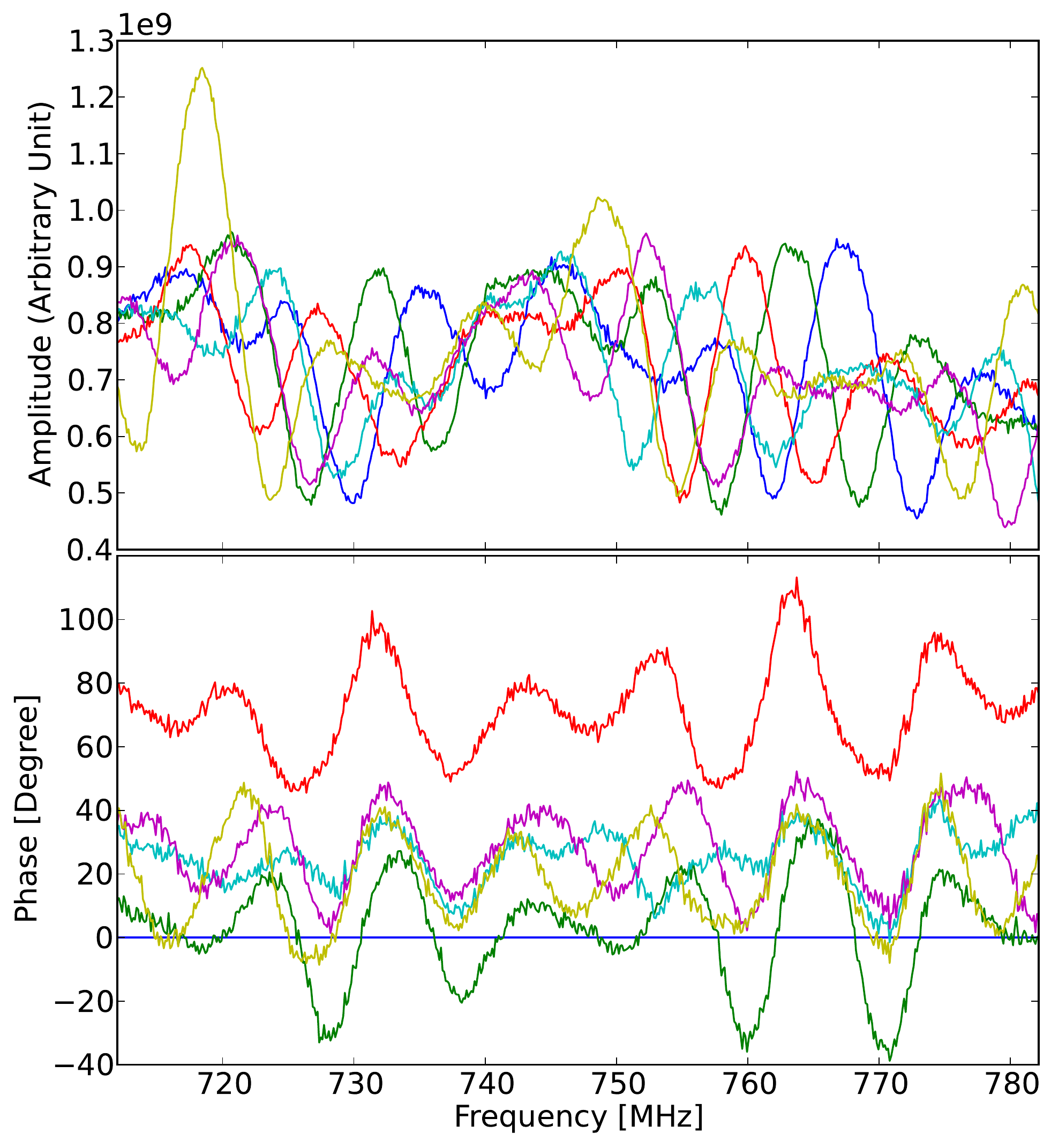}}
  \subfigure[A6X-B13X from data set of 2016/09/27]{
  \includegraphics[width=0.45\textwidth]{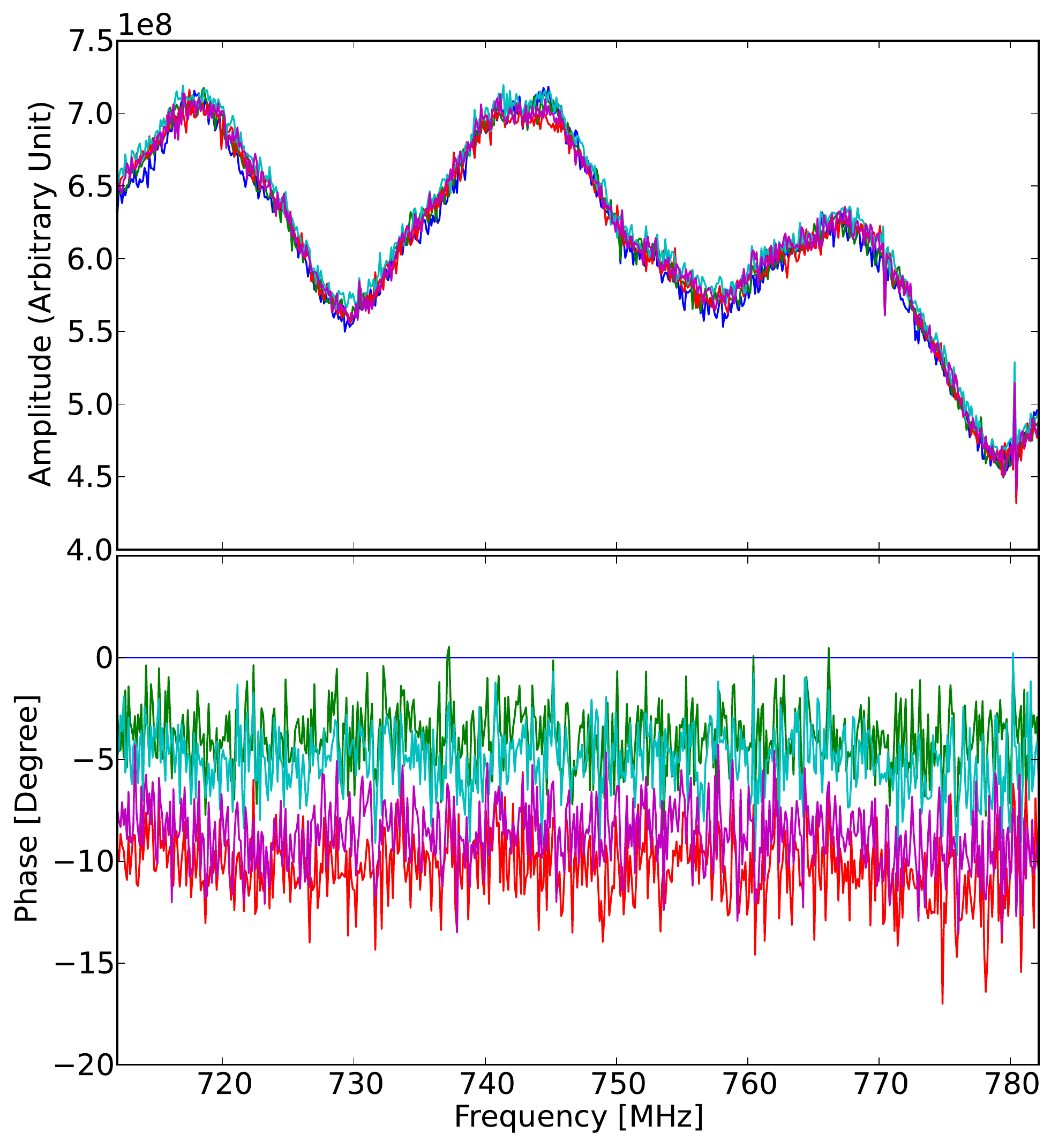}}\\
  \caption{Typical uncalibrated visibility amplitude and residual phase w.r.t. the first day as a function of frequency at the peak of the Cyg~A transits. Each colored curve represents the response for the transit on a different day. }
  \label{fig:band_var}
\end{figure*}

We calibrate the bandpass response by analyzing the transit of strong sources. The bright source Cyg~A has a flux of 2961~Jy at 750~MHz \cite{2017ApJS..230....7P}, and in the frequency range of interest,
\begin{equation}
  \log{S} = \sum_{n=0}^5 a_n [\log(\nu)]^n
\end{equation}
where  $a_0 = 3.3498, a_1 = -1.0022, a_2 = -0.2246, a_3 = 0.0227, a_4 = 0.0425$. In our observation band (700--800 MHz) the flux difference between the low frequency side and high frequency side is about 5\%.

We first check whether it is a good assumption that Cyg A dominates the antenna temperature. In Fig.~\ref{fig:band_var_short}, we plot the auto-correlations for the B9 and A4 feeds, as well as the cross-correlations between them, during the different stages of Cyg~A transit. There is a non-zero background level for the auto-correlations, while for cross-correlations this is expected to be zero (c.f. Eq.\ref{eq:radiometer}). We have therefore subtracted the background for the auto-correlations. As we come closer to the peak of the transit, the amplitude of the visibility increases, and the shapes of the bandpasses become more similar to each other with better SNRs. The bandpasses (as derived from the auto-correlation) for different feeds are not the same, as illustrated in Fig.~\ref{fig:band_var_short}, where the shape of the bandpass for B9 and A4 are different. However, the shape of the cross-correlation is generally consistent with the square root of the product of the bandpasses obtained from the two auto-correlations at the peak of transit, as shown in the top panels by the black curve. From this trial, we see that Cyg~A indeed dominates the visibility at the peak of its transit, and can be used to calibrate the bandpass.

In Fig.~\ref{fig:band_var} we show the amplitudes and phases of the visibilities as a function of frequency at the peak of the Cyg A transit from the observational data set 2018/03/22 (left column) and 2016/09/27 (right column). The cross-correlations of Y-polarization are plotted in the top two rows, and X-polarizations in the bottom two rows. In each plot, the visibilities are taken from the peak of the transit.  Each colored curve corresponds to one day's transit; there are 6 days for the 2018/03/22 data set and 5 days for the 2016/09/27 data.

The amplitude of the visibility as a function of frequency reflects the bandpass response of the system. As shown in the figure, the amplitudes are fairly stable and consistent for the Y-polarization on each day; the day-to-day variation is very small. However, for the X-polarization of the 2018/03/22 data set (shown in the left column of Fig.~\ref{fig:band_var}), the amplitudes are different on each day. After some study, we conclude that this is caused by contamination from the Sun, as the Cyg~A transits occurred at about 01:30 UTC (about 07:30 local civil time) on these days, and the radiation from the Sun, which is above the horizon in the East direction, may enter the beam from the side lobes, because the feeds are directly illuminated. For the Y-polarization, which is oriented in the E-W direction, the projected area is at a minimum towards the East, so it is much less affected than the X-polarization, whose projected area is maximum in the E-W directions. In the 2016/09/27 data set (results shown on the right column) we see the amplitudes have much less variation as the Cas~A transit occurred at about 13:40 UTC (about 19:40 local civil time) after sunset.

The phase of the visibility induced by a strong point source is given by $-2\pi \frac{\nu}{c} \vec{n}\cdot \vec{b}_{ij} + \varphi_{ij}(\nu)$, where $\vec{n}$ is the direction of the source, and $\vec{b}_{ij}$ is the baseline. After subtracting out the term proportional to the frequency, $\nu$, the residual reflects the instrument phase $\varphi(\nu)$. We have compared the phase with the first day's phase. As expected, the phase drifted away from this value on subsequent days, but during each day the instrument phase is nearly constant over frequency; the fluctuations are due to noise. Remarkably, the phase frequency structures are quite similar to each other on different days. Again, the X-polarization for the 2018/03/22 observation is an exception, where the influence of the Sun can not be neglected.

\subsection{Absolute Calibration}
\label{subsec:absolute_cali}

When a strong point source is transiting through the primary beam of the telescope and dominating the visibility, the visibility for a pair of elements $i$ and $j$ is given by
\begin{equation}
  V_{ij}^{0}  = S_{c} \, G_{i} G_{j}^{*} ,
  \label{eq:Vps}
\end{equation}
where $S_c$ is the flux of the source,
\begin{equation} \label{eq:G_i}
  G_i = g_i A_i(\uvec{n}_0) e^{-i 2 \pi \uvec{n}_{0} \cdot \vec{u}_{i}};
\end{equation}
where $A_i$ is the beam pattern of array element $i$, $\uvec{n}_0$ is the unit vector in the direction of the source and $\vec{u}_i$ is the coordinate of array element $i$ in wavelength units. In matrix form,
\begin{eqnarray} \label{eq:V0ps}
  \mat{V}_{0} &=& S_{c} \, \mat{G} \mat{G}^{\dagger}.
\end{eqnarray}
$\mat{G}$ is an eigenvector of $\mat{V}_0$.
If noise is present but small compared with the calibrator source and statistically equal in all elements, i.e. $\mat{V}=\mat{V}_0 + \mat{N}$, where $\mat{N}$ is the noise co-variance matrix, the vector $\mat{G}$ could be obtained by principal component analysis (PCA):  solving for the eigenvectors of matrix $\mat{V}$, with the eigenvector associated with the largest eigenvalue identified as $\mat{G}$.  This is also the least squares solution of the form $\mat{V}= \mat{g} \mat{g}^\dagger$  (for proof and more details see \cite{Zuo2019}). Thus, for each time and frequency point, a solution of the complex gain can be obtained using this eigenvector decomposition method. This method of calibration is, however, only applicable when a strong point source is transiting, which restricts its usage.

\begin{figure}[H]
  \centering
  \includegraphics[width=0.45\textwidth]{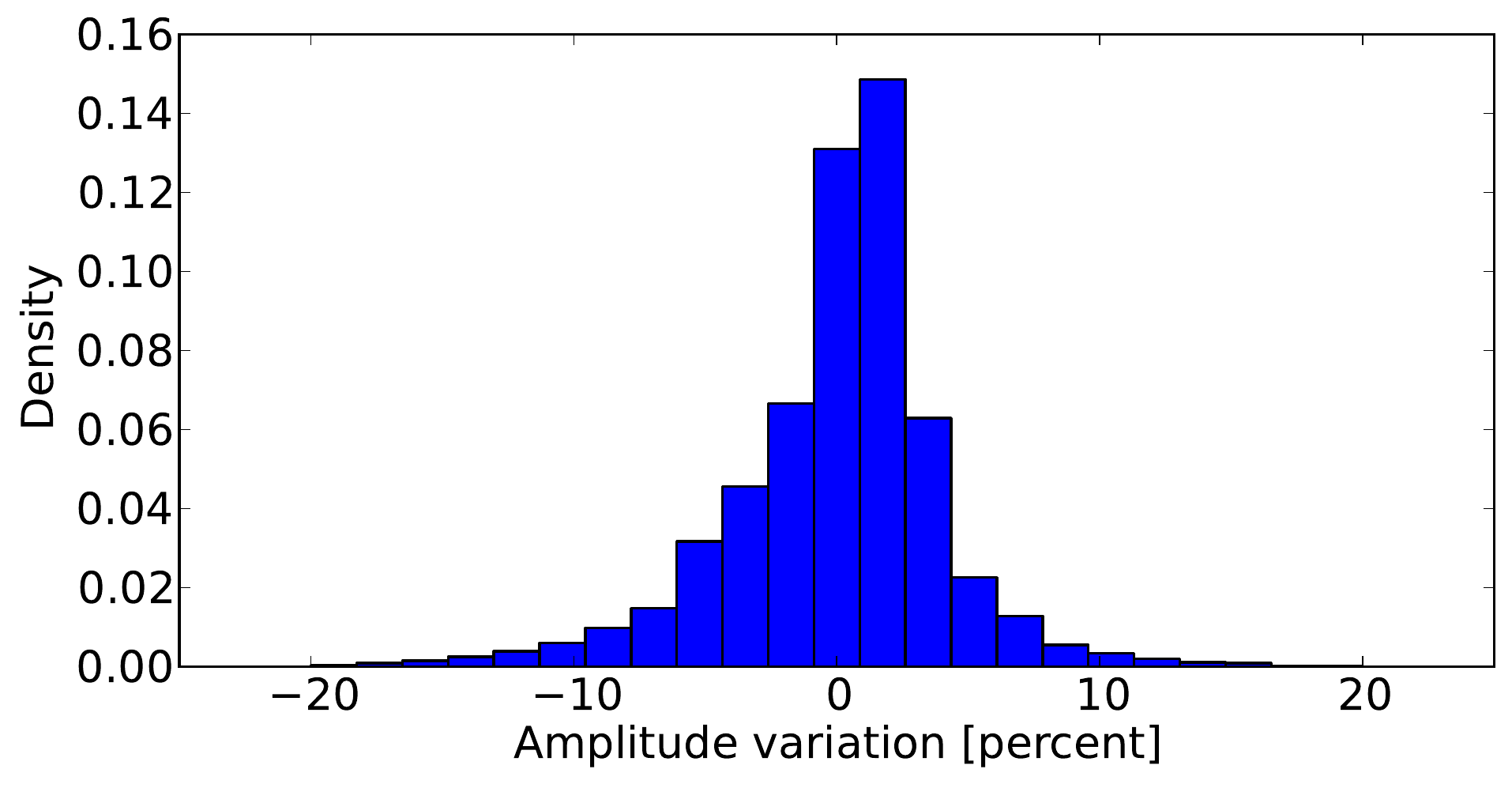}
  \includegraphics[width=0.45\textwidth]{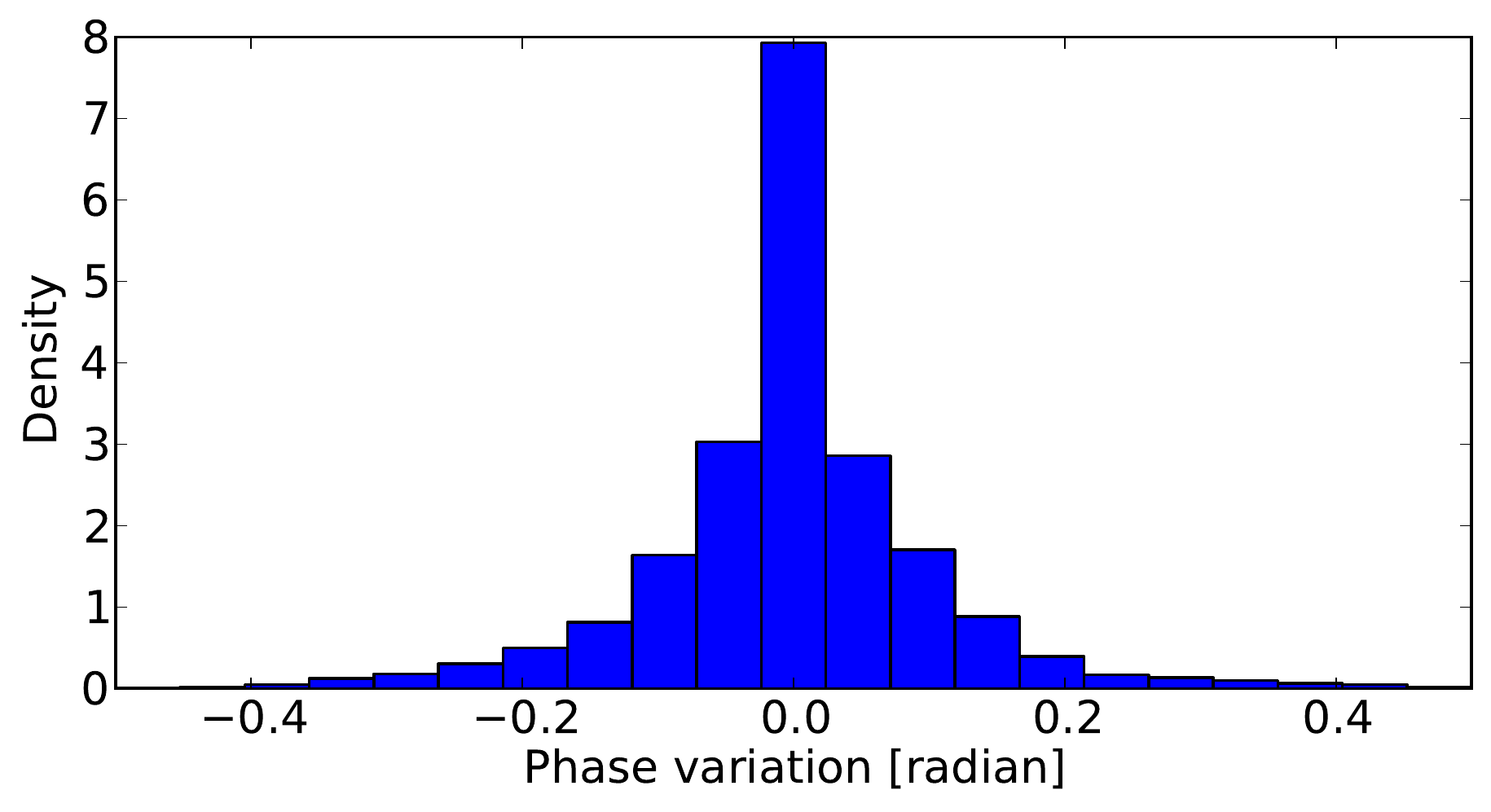}
  \caption{The distribution of receiver gains is shown as a percentage variation in amplitude (top) and  radian variation in phase (bottom), for all polarization channels and frequencies at the peak of Cyg A transits over five days.}
  \label{fig:cyg_gain_hist}
\end{figure}

In Fig.~\ref{fig:cyg_gain_hist}, we plot the variation of the gains during 5 days with respect to the first day, as obtained from the absolute calibration with Cyg~A transits. The top panel shows the variation in amplitude as a percentage, and we find that most of the variations are within  $\pm5\%$. The phase variations are mostly within 0.2 radians. Part of these variations are true variations in the gain of the receivers, while part may be due to the error in the calibration process, which has a limited precision. At present, these two can not be distinguished.

\subsection{Relative Calibration}
\label{subsec:relative_cali}

The CNS, which broadcasts periodically, can be used for relative calibration at any time. The CNS is located near the array and the contribution to the visibility caused by its radiation is
\begin{equation} \label{eq:vnf}
  V_{ij}^{\text{CNS}} \propto  S_n
  e^{-i 2\pi  (\vec{n}_i\cdot \vec{r}_{i} - \vec{n}_j \cdot \vec{r}_{j})},
\end{equation}
where $\vec{r}_i$ and $\vec{r}_j$ are the locations of feeds $i$ and $j$ respectively, and $\uvec{n}_i$ and $\uvec{n}_j$ are the displacement vectors from the noise source to the positions $\vec{r}_{i}$ and $\vec{r}_{j}$. The noise source induced visibility can be obtained by subtracting the noise on and noise off values:
\begin{eqnarray}
  V_{ij}^{\text{on}} - V_{ij}^{\text{off}}
  &\approx& C |G_{ij}| e^{-i 2\pi \nu \Delta \tau_{ij}} e^{-i 2\pi (r_{i} - r_{j})/\lambda}.
\end{eqnarray}
Here, $C$ is a constant and $\Delta \tau_{ij}$ is the equivalent instrument delay difference between the channels $i$ and $j$, which is mostly due to the variation in the cable length for the two channels. We see that, unlike the absolute calibration with the astronomical source, in the relative calibration only the variation of the difference between the phase of a pair of signal channels is directly determined.

We have used two approaches to deal with this problem: calibration based on individual feed channels, and calibration based on feed pairs.

{\bf Individual Feed Channels.} In the approach based on individual feed channels, one tries to obtain the complex gains for all the feed channels by making a best fit solution to all baselines. As there are many more baselines than the number of array elements, it is possible to obtain a best fit solution to the phase of the individual elements. However, note that there is an inherent degeneracy: if we add a time-dependent but identical instrument phase to all elements, there is no observable effect. This problem can be solved by imposing a physically motivated ``gauge condition'': assume that as $t \rightarrow 0$, the changes of phase for all elements should go to zero. So we adopt the condition that the total phase variation for all elements is to be minimized at short time intervals. In this way, a solution can be obtained for each feed input channel.

\begin{figure}[H]
  \centering
  \includegraphics[width=0.45\textwidth]{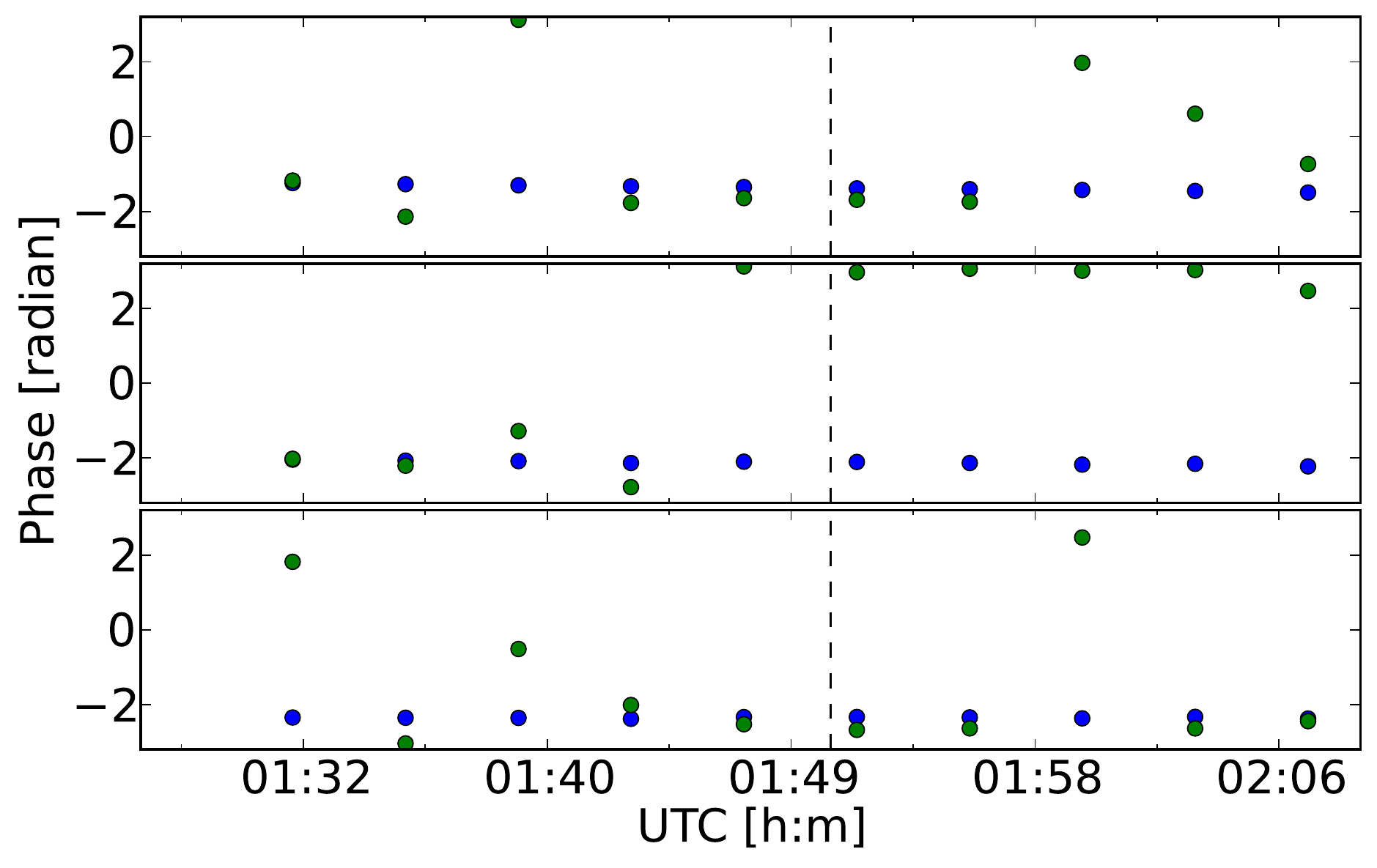}
  \caption{A comparison of the phase of the gain calibrated by the CNS (blue) and Cyg~A (green) for a few typical channels. From top to bottom, the feeds are A20X, B22X, C23X. The vertical dashed line indicates the transit time of Cyg~A.}
  \label{fig:cns_vs_ps}
\end{figure}

In Fig.\ref{fig:cns_vs_ps} we compare the phases obtained by relative calibration and absolute calibration during the transit of Cyg~A. The phases obtained from the relative calibration stay nearly constant. The phases obtained by the absolute calibration change rapidly as Cyg~A comes to the center of the beam. This variation is not a real variation in phase, but simply an error in the absolute calibration: the absolute calibration is only viable when the point source is dominating, and when the source is outside the beam center the result is not reliable. However, once in the center, the absolute calibration also yields nearly constant phases. The absolute and relative calibrations agree with each other for the duration of the peak of transit.

{\bf Feed Pairs.} In the approach based on feed pairs, instead of trying to solve for the complex gain $g_i$ of each channel $i$, we simply calibrate each pair of channels. In principle, the ``gain'' for pair $i,j$ should simply be $g_i g_j^*$, but as discussed above, this can not be obtained directly and has to be solved for in the relative calibration. Instead, one may define a complex gain associated with each pair of input channels,
\begin{equation}
  G_{ij} =  |G_{ij}| e^{-i 2\pi \nu \Delta \tau_{ij}},
\end{equation}
which can be directly inferred from the visibility of the CNS. Here the change of instrumental phase for the visibility of each pair of input channels can be obtained directly by monitoring the CNS signal, without using the complex gain of individual input channels. The computation is also much simpler in this approach.

To check the precision of this approach, we can employ the closure phase relation. If a strong radio point source or the CNS at position $\uvec{n}$ with flux $S_0$ dominates the received signal, the visibility is given by
\begin{equation}
  V_{ij} = g_i^* g_j A_i^*(\uvec{n}) A_j(\uvec{n}) S_0 e^{-i 2\pi \vec{u}_{ij} \cdot \uvec{n}} .
\label{eq:Dominant_Src}
\end{equation}
We see that the quantity $V_{ij} V_{jk} V_{ki}$ should be a real number. Thus, if we write the visibility in the form
$V_{ij} = |V_{ij}| e^{i\phi_{ij}}$, then we shall have
\begin{equation}
  \phi_{ij}+\phi_{jk}+\phi_{ki} = 2 N \pi ,
\end{equation}
where $N$ is an integer. In the relative calibration based on feed pairs, each $V_{ij}$ is individually calibrated, so the closure phase relation may break if the calibration fails. We can use this closure phase relation to check the validity of relative calibration.

\begin{figure}[H]
  \centering
  \includegraphics[width=0.45\textwidth]{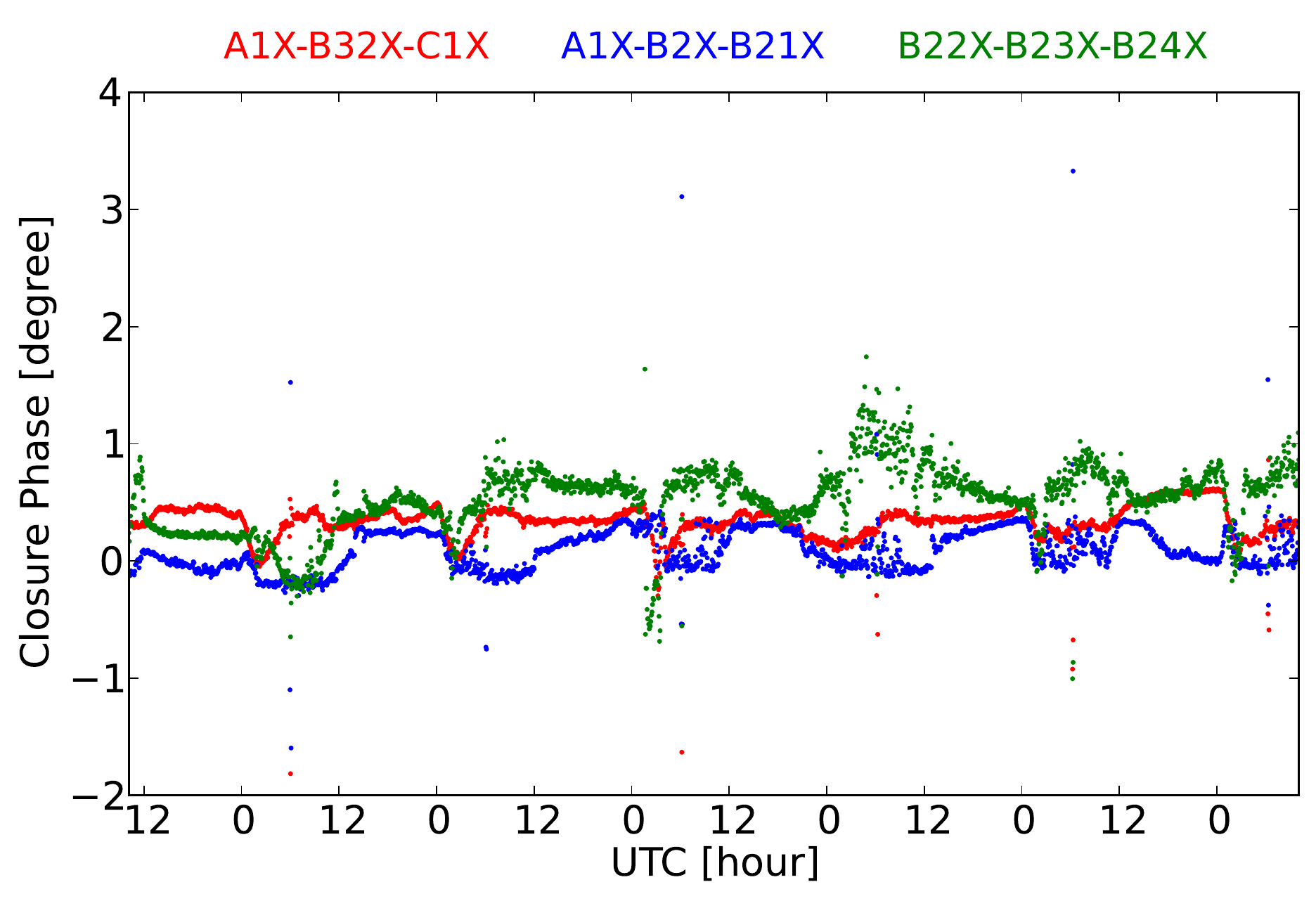}
  \caption{The closure phase for CNS induced visibility of three typical pairs of channels. red: A1X-B32X-C1X; blue: A1X-B2X-B21X; green:B22X-B23X-B24X. }
  \label{fig:closure_phase_ns_cal}
\end{figure}

In Fig. \ref{fig:closure_phase_ns_cal} we plot the closure phase for a few feed channel pairs using the visibilities induced by the CNS over 5 consecutive days. Here we consider three cases:
(1) three feeds are all located on different cylinders (red curve);
(2) two of the three feeds are located on the same cylinder, but are some distance away from each other (blue curve);
(3) the three feeds are adjacent ones on the same cylinder (green curve).
We see the deviations from zero are all relatively small ($<0.02~\mathrm{rad}$) for most of the time, but become larger around the local noon time when the Sun is transiting through the beam, during which the single dominant source approximation for the CNS is slightly broken.

\begin{figure}[H]
  \centering
  \includegraphics[width=0.45\textwidth]{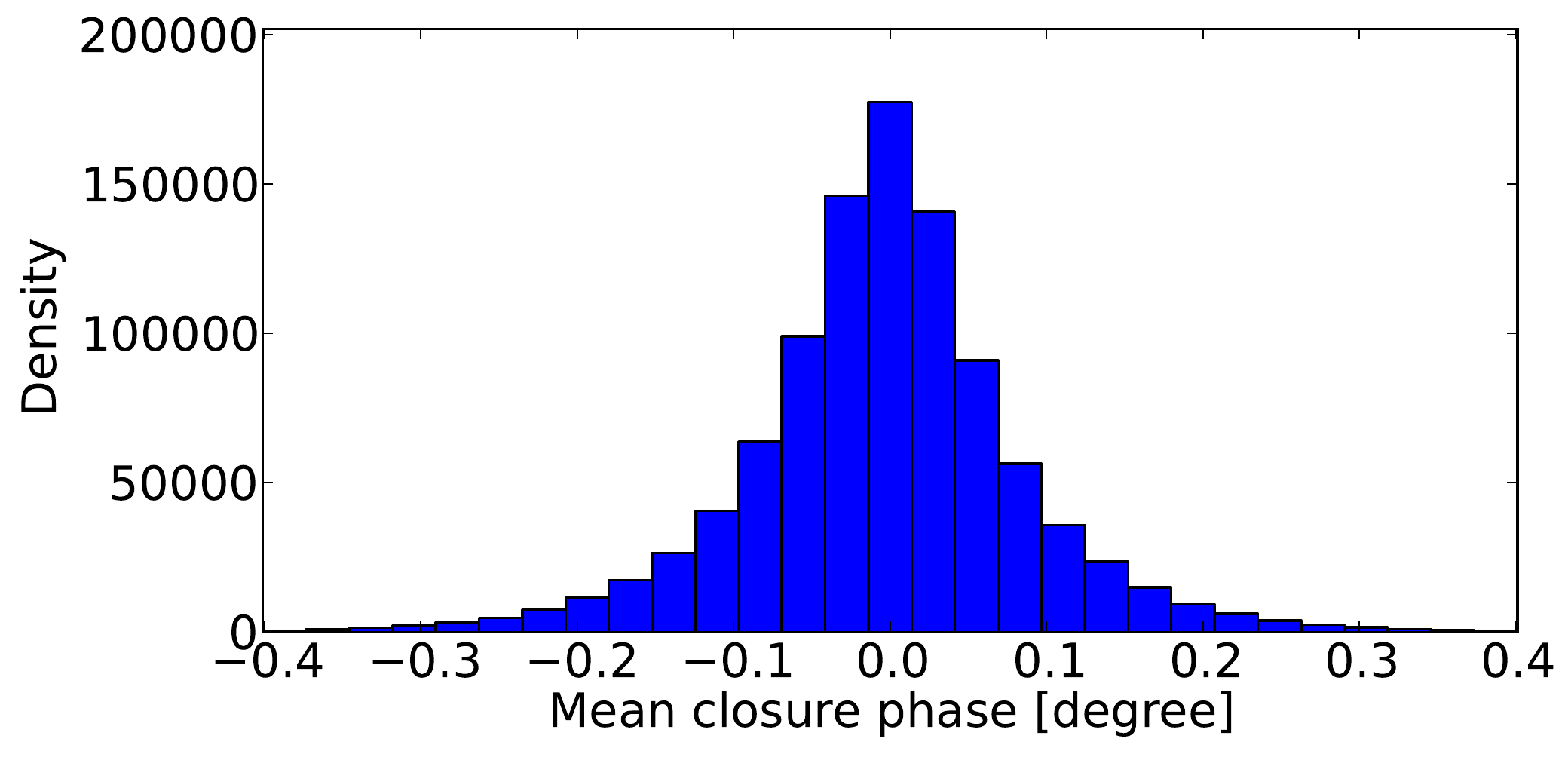}
  \caption{The distribution of closure phase of baseline gains $G_{ij}$ for all combinations of baselines and 100 frequencies of the calibrated data set 2016/09/27.}
  \label{fig:closure_phase_all}
\end{figure}

In Fig.~\ref{fig:closure_phase_all}, we plot the distribution of the closure phase of feed pair gains $G_{ij}$ calibrated with the CNS for all combination of baselines and 100 frequencies. This shows that the closure phase is centered around 0, with small deviations. $\sigma_\phi = 0.0015$ rad, which provides an estimate on the achievable precision of the calibration of feed pair gains.

\subsection{Stability}
\label{subsec:stability}

\begin{figure}[H]
  \centering
  \includegraphics[width=0.45\textwidth]{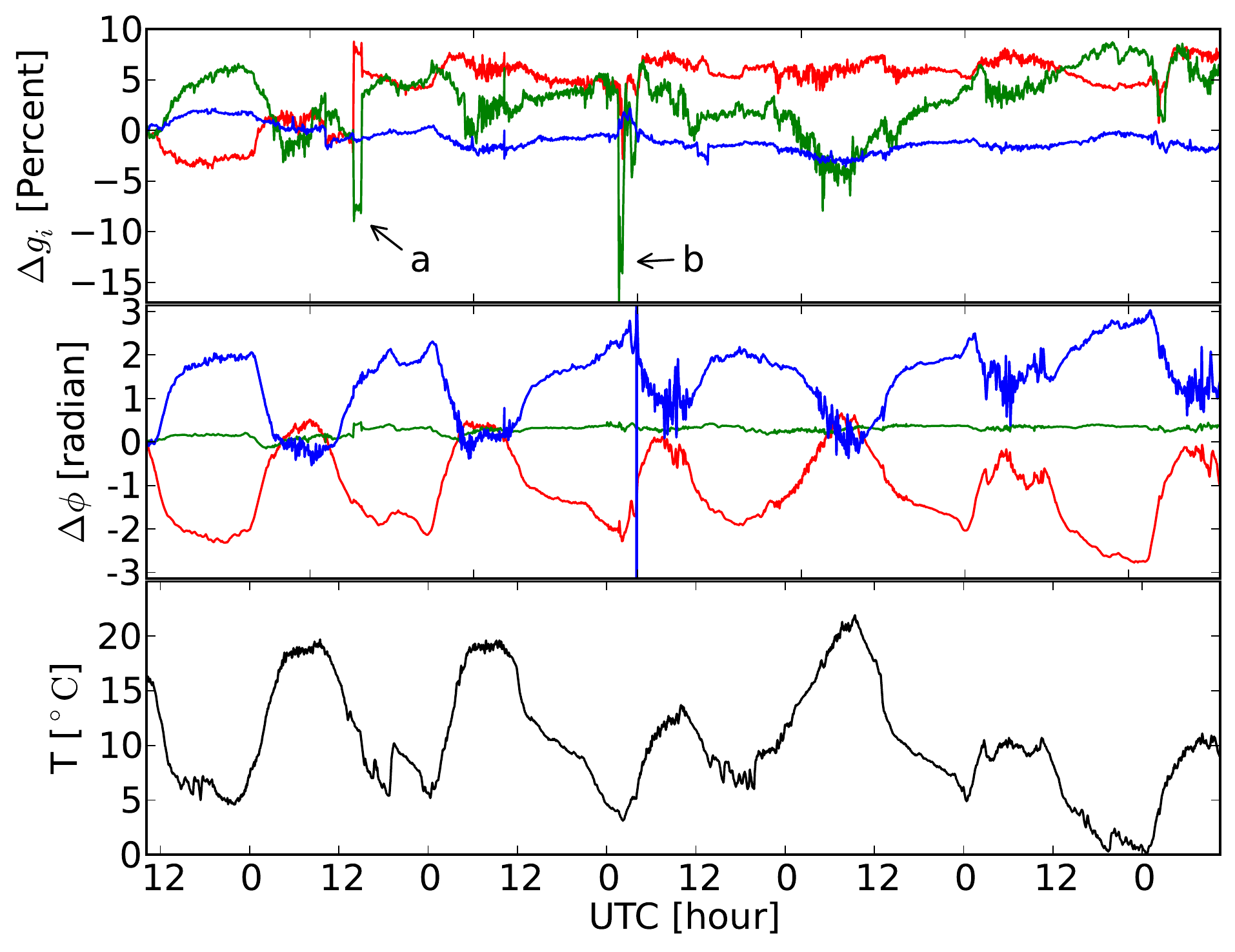}
  \caption{The variation of the gain amplitude (top panel, shown as a percentage) and the phase (middle panel) of 3 typical feed channels A15X (red), A28Y(green), C22Y(blue), and the corresponding temperature (bottom) for 6 consecutive days starting from 2018/03/22.}
  \label{fig:cali_feed_gain_ampphs_180322}
\end{figure}

Using the relative calibration method, the variation of complex gains throughout the day can be tracked. In Fig.~\ref{fig:cali_feed_gain_ampphs_180322}, we show the gain of a few typical individual feed channels during 6 consecutive days, with the top panel showing the change in amplitude as a percentage and the middle panel showing phase. For comparison, the ambient environmental temperature variation during this measurement is plotted in the bottom panel.

The variation of the gain seems correlated with that of temperature, especially for the phase. For different polarization channels, the variations of phase are quite diverse, though appear to be well-correlated with the temperature. For example, the phase of A28Y (green curve) varies very little, while for A15X (red curve) and C22Y (blue curve) the phase variation is much larger and in the opposite sense. These variations are probably induced by the thermal expansion and contraction of the optical fiber length. As the optical cables are about 8 km long, the change in their length could be significant. The optical fibers within the cable are in different positions and have slightly different temperatures, so it is also not too surprising that each of them undergoes different amounts of expansion or contraction, and some even stay constant. To verify that the majority of the instrument phase change comes from the optical fibers, we measured the round trip delay of the optical cable continuously throughout a whole day and indeed found the changes similar to that of the phase of the complex gains.

The changes of the amplitudes of the gains are more irregular, though some correlation with temperature change can still be seen. The gains during the night (lower temperature) are higher than during the day (higher temperature), consistent with our expectation for the gain of amplifiers.

Some rapid changes are understood to be caused by some external events. For example, the sudden changes at the position marked with ``a'' in the figure is probably associated with the change of the dish array pointing at that time. The dish array motor generates some RFI, and the change in pointing of the dishes may also change their reflection of some radio waves. The position marked with ``b'' may be associated with the Sun. However, surprisingly, the variations of the phases at those times are still regular. Still, the causes of some changes remain unknown.

We note here that the gains are obtained by relative calibration with the CNS. The CNS signal is received in the far side lobes of the receiver feeds. The amplitude may be more susceptible to small changes, e.g. the slight swing of the feed by the wind which could shift the side lobe significantly. This suggests that the amplitude of the gain calibrated with the CNS should be used with caution.

\begin{figure}[H]
  \centering
  \includegraphics[scale=0.38]{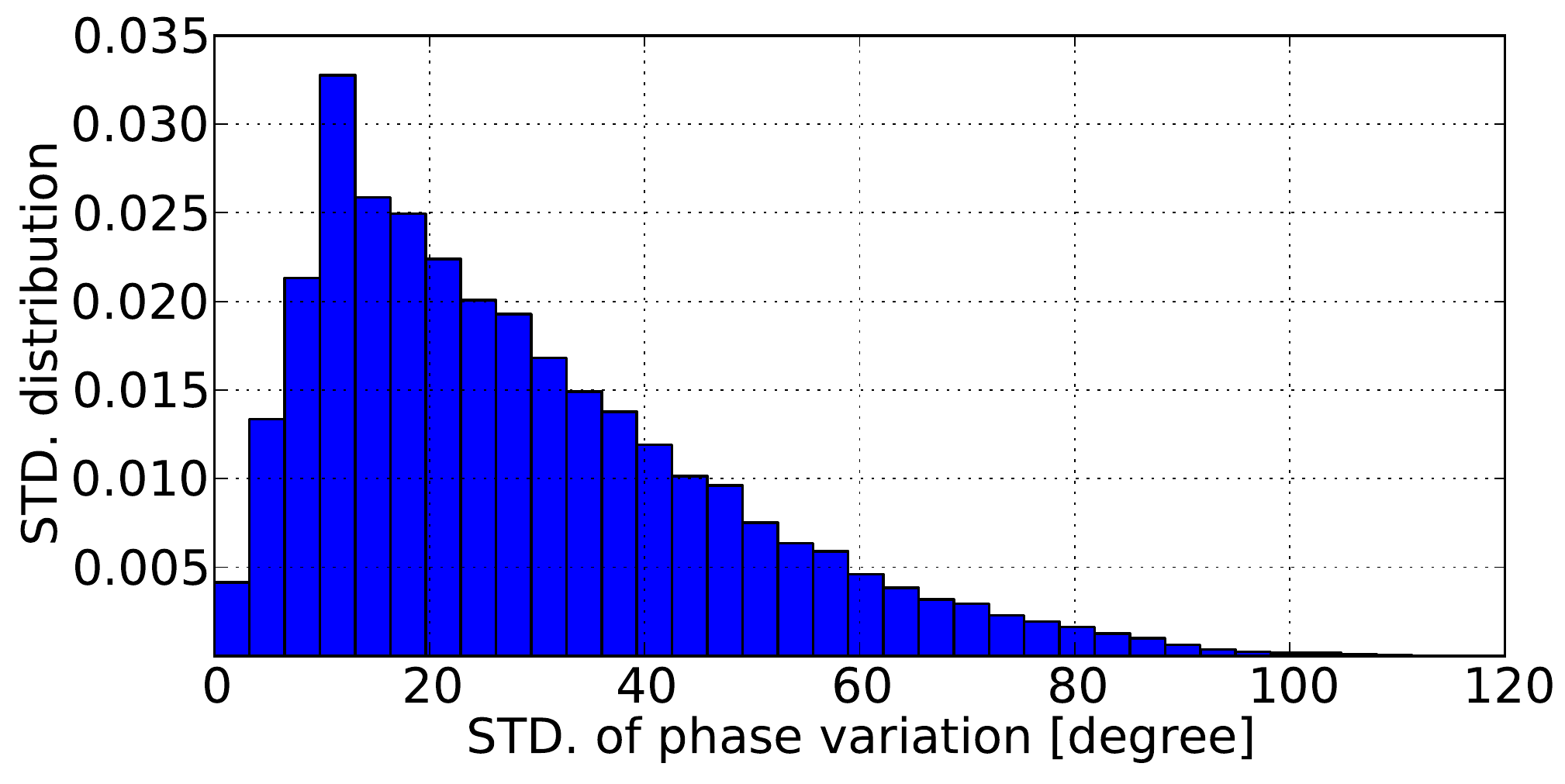}
  \includegraphics[scale=0.38]{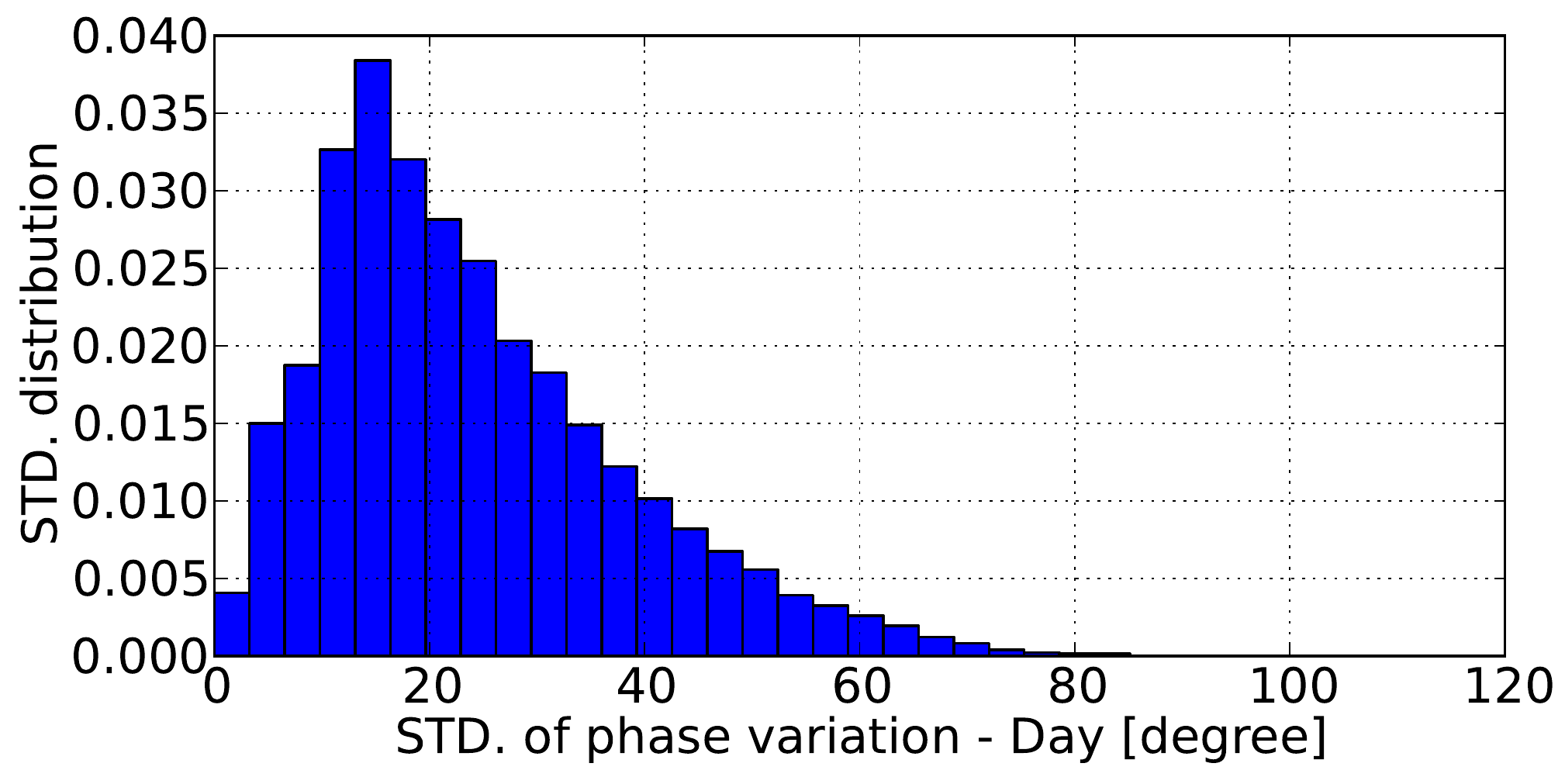}
  \includegraphics[scale=0.38]{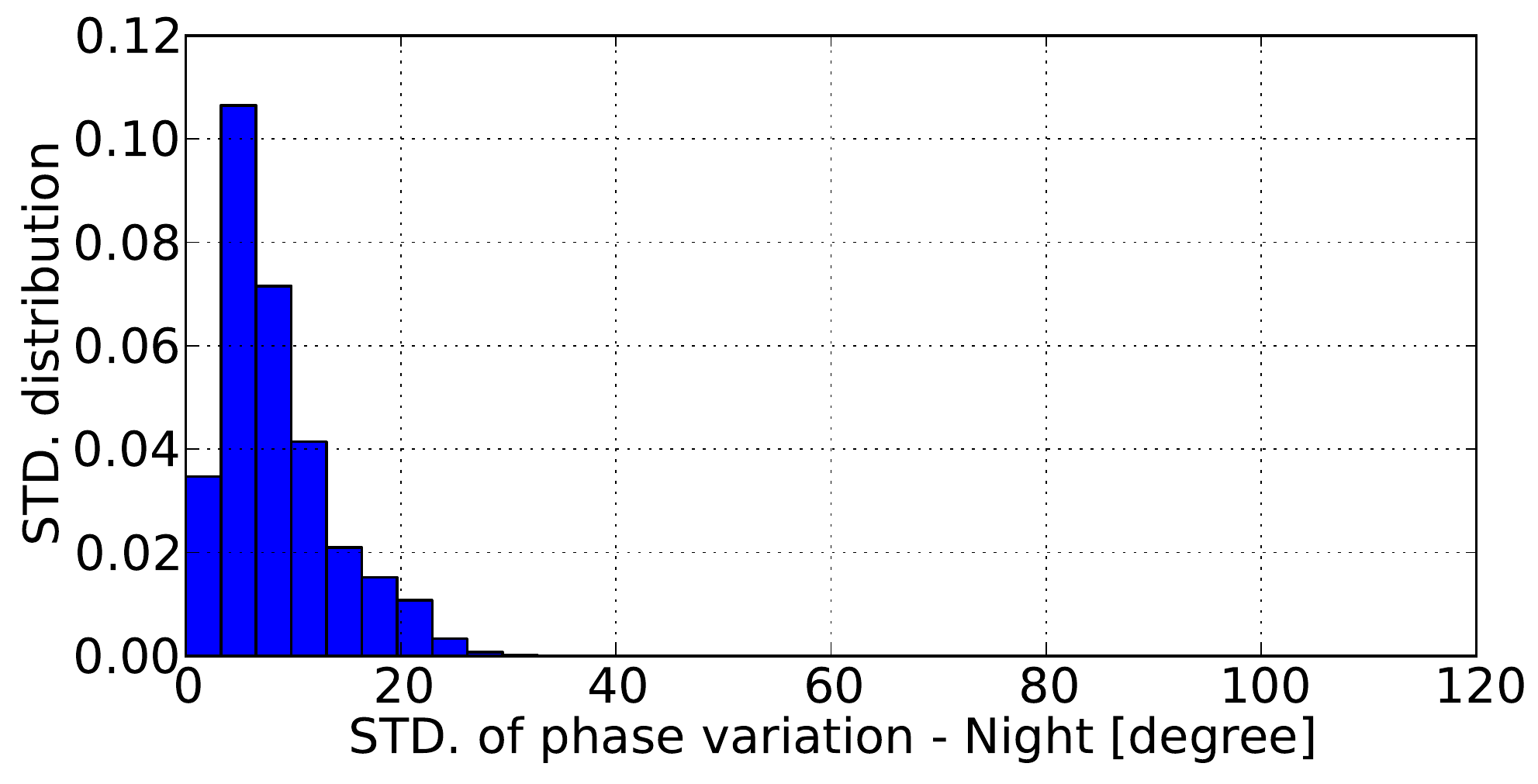}
  \caption{Distribution of the STD of the phase of the complex gain variations for whole day(top), daytime (middle), and nighttime (bottom) over 5 days.}
  \label{fig:gain_phs1d_ns_5day_std_hist}
\end{figure}

Given the relatively calibrated complex pair gains, the variation of each pair over time can be characterized by the standard deviation (STD). The distributions of the STDs of the phase variation in the pair gain for different baselines and frequencies during the course of 5 days are plotted in Figure \ref{fig:gain_phs1d_ns_5day_std_hist}, where we also distinguish the whole day (top), day time (middle panel), and night time (bottom channel) cases. For this figure, the day time is defined as 6 AM to 6 PM local civil time, and  night time 6 PM to 6 AM local civil time.

The peak value of the daytime STD is around 0.3 rad and it extends to 1.0 rad, while the peak value of nighttime STD is around 0.1 rad, and extends to less than 0.5 rad. This shows that the phase variation in nighttime is much smaller than in daytime.

\section{Sensitivity and System Temperature}
\label{sec:sensitivity}

The visibilities obtained after the above calibration procedure are still in ADC units. In order to determine the sensitivity and system temperature, we use a bright point source with known flux density as a calibrator to convert the visibility to the antenna temperature.

If we observe a strong point source such as Cyg~A at the peak of its transit of the beam, in the direction of $\vec{n}$, with flux density $S_0$, the spectral power received in one polarization by the receiver is
\begin{equation}
  P_\nu=\frac{1}{2} \eta A_e f_{\rm ps} S_0.
\end{equation}
Here, $A_e$ is the effective area of the antenna in the direction of maximum response,
$f_{\rm ps} \equiv f(\vec{n})$ is the beam profile in the direction of the point source, normalized to $f_{\rm max}=1$, and $\eta \le 1$ is the efficiency factor. The radiation from the point source induces an antenna temperature $T^{\rm ps}_A$, with $P_{\nu} = \eta k T^{\rm ps}_A$, where $k$ is the Boltzmann constant. The raw visibility (in ADC units) of channels $a,b$ is related to the antenna temperature by $V^{\rm ps}_{ab}= C T^{\rm ps}_A$, where the 
constant $C$ is the calibration coefficient, which can be determined from the observation of bright source transits, assuming the flux density and effective area of the antenna is known. 

The system temperature can then ben obtained for each feed using either the mean auto-correlation or its variance 
outside bright source transits, and neglecting the sky contribution. It is also possible to obtain the geometric average 
of system temperature for two feed channels using the variance of corresponding cross-correlation.

The mean auto-correlation of receiver $a$ is given by
\begin{equation}
  \bar{V}_{aa} = \langle n_a^* n_a \rangle = C\, T^{aa}_{\rm sys},
\end{equation}
where we have assumed that the receiver noise dominates over the signal induced by the astronomical source, which is true in our case. The system temperature is then given by
\begin{equation}
  T_{\rm sys}^{aa} = \frac{ \bar{V}_{aa} T_A^{\rm ps}}{\Delta V_{aa}^{\rm ps}}
  \label{eq:autoT}
\end{equation}
where $\Delta V_{aa}^{\rm ps}= V_{aa}^{ps} -\bar{V}_{aa}$ is the visibility change induced by the source, which can be derived from the difference of the auto-correlation at the peak of transit $V_{aa}^{ps}$ and the  blank sky average $\bar{V}_{aa}$ before or after the transit.

Alternatively, we can also use fluctuations in the auto-correlation visibility to estimate the system temperature using the radiometer equation.
The variance $\sigma_{aa}^2 \equiv \langle |V_{aa} -\bar{V}_{aa}|^2 \rangle$ is given by
\begin{equation}
  \sigma_{aa} = C\, \frac{T^{aa}_{\rm sys}}{\sqrt {\delta \nu \, \delta t}},
\end{equation}
where $\delta t$ is the integration time and $\delta \nu$ is the bandwidth. One can then infer $T_{\rm sys}$ as
\begin{equation}
  T_{\rm sys}^{aa}= \frac{\sigma_{aa}}{\Delta V_{aa}^{\rm ps}} T_A^{\rm ps} \sqrt{\delta\nu \delta t}.
\label{eq:flucT}
\end{equation}

For the cross-correlation, the variance of the visibility is defined as $\sigma_{ab}^2 \equiv \langle |V_{ab} -\bar{V}_{ab}|^2 \rangle$, which is given by
\begin{equation}
  \sigma_{ab} = C\, \frac{T^{ab}_{\rm sys}}{\sqrt {\delta \nu \, \delta t}}.
\end{equation}
for a simple correlator (c.f. \S 6.2 of Ref.\cite{2017isra.book.....T}). One can then infer $T_{\rm sys}$ as
\begin{equation}
  T_{\rm sys}^{ab}= \frac{\sigma_{ab}}{V_{ab}^{\rm ps}} T_A^{\rm ps} \sqrt{\delta\nu \delta t}.
\label{eq:crossT}
\end{equation}
The system temperature associated with the $a,b$ channels is a geometric average of the system temperature for the $a, b$ channels,
\begin{equation}
  T^{ab}_{\rm sys} = \sqrt{T^{a}_{\rm sys} T^b_{\rm sys}}.
\end{equation}
From the system temperature of the pairs, $T^{ab}_{\rm sys}$, one can solve for the system temperature associated with individual feeds. A method of solution using only  different cylinder cross-correlations is given in the Appendix, and it is the same formula as the least square solution for the determination of the center of the beam from cross correlations in Sec.\ref{sec:beam}.

\begin{figure*}
  \centering
  \includegraphics[width=0.9\textwidth]{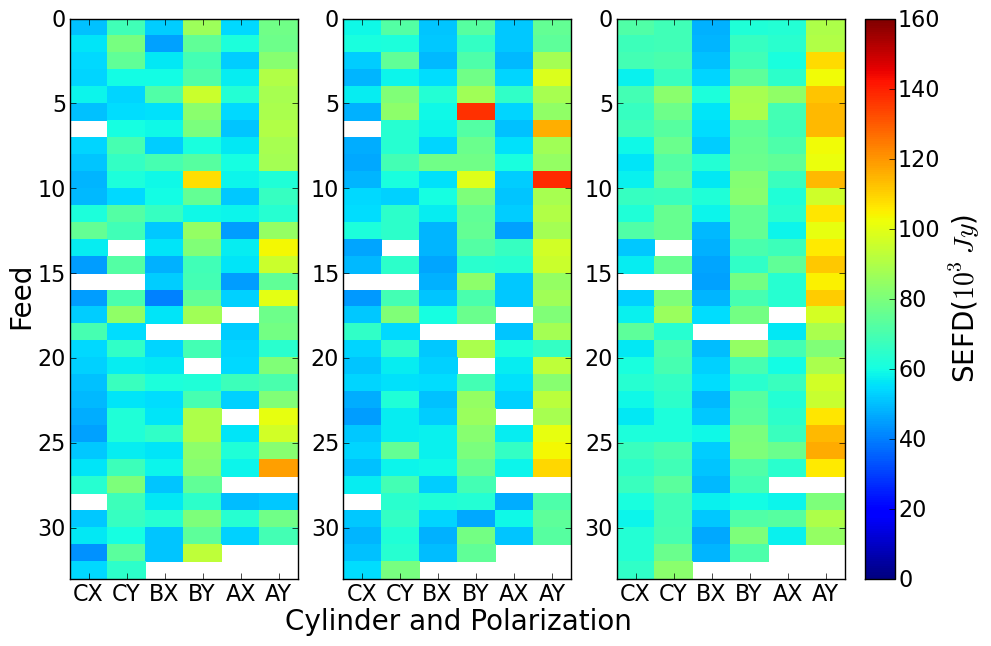}
  \caption{The $f_{\rm ps}^{-1}$ SEFD derived from auto-correlation background (left), auto-correlation fluctuation (center), and cross-correlation fluctuation (right) during the Cyg~A transit from 2016/09/27 to 2016/10/02.}
  \label{fig:SEFD}
\end{figure*}
\begin{figure*}
  \centering
  \includegraphics[width=0.9\textwidth]{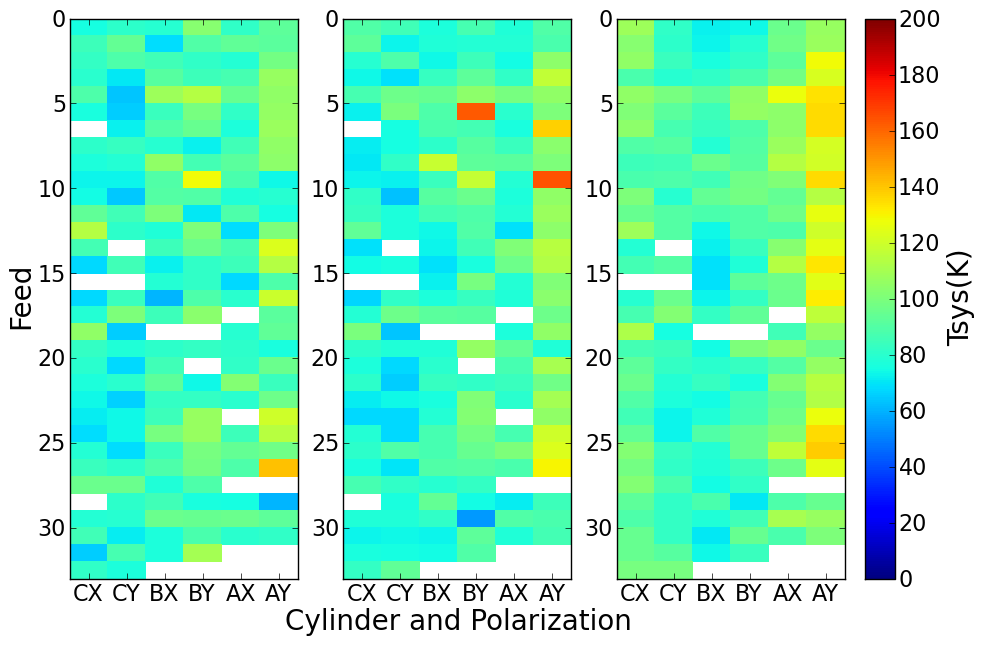}
  \caption{The system temperature derived from the 2016/09/27 to 2016/10/02 data set. Left: derived from auto-correlation background; Center: derived from auto-correlation fluctuation; Right: derived from cross-correlation fluctuation. }
  \label{fig:system_temperature}
\end{figure*}

To derive the system temperature using either the auto-correlation background (Eq.~\ref{eq:autoT}), or the auto-correlation  (Eq.~\ref{eq:flucT}) or cross-correlation fluctuation (Eq.~\ref{eq:crossT}) requires knowing the antenna temperature, which depends on the  effective area of the antenna. The oft-used system equivalent flux density (SEFD, in units of Jansky) is defined as
\begin{equation}
  \mathrm{SEFD} \equiv \frac{2k T_{\rm sys}}{A_e} = \left\{
  \begin{matrix}
    &  f_{\rm ps} S_0  \frac{\bar{V}_{aa}}{\Delta V_{aa}^{\rm ps}} \\
    &\\
    & f_{\rm ps} S_0
    \frac{\sigma_{aa} \sqrt{\delta\nu \delta t}}{V_{aa}^{ps}}.\\
    &\\
    & f_{\rm ps} S_0
    \frac{\sigma_{ab} \sqrt{\delta\nu \delta t}}{V_{ab}^{ps}}.
  \end{matrix}\right.
\label{eq:SEFD}
\end{equation}

Regardless of the value of the antenna effective area and beam response, we can determine $(f_{\rm ps})^{-1}$ SEFD, which is given entirely by the observational data.

Using the CST simulation described in Sec. \ref{sec:beam} we find $f_{\rm ps}^X= 0.9795$ and $f_{\rm ps}^Y=0.9908$, which are fairly close to 1. The effective area for the two polarizations are different: $A_e^X=4.22 \m^2 $, and $A_e^Y = 3.35 \m^2$.

\begin{table}[H]
  \centering
  \caption{The mean SEFD of different  polarizations.(Unit: $10^3 \Jy$) }
  \centering
  \small
  \begin{tabular}{c|c|c|c}
    \hline
    polarization & method 1 & method 2 & method 3 \\
    \hline
    X & $55.1\pm 6.4$ &  $53.9\pm 5.83$ & $60.1 \pm 7.92$\\
    \hline
    Y & $75.1 \pm 13.3$ & $77.5 \pm 16.0$ & $81.4 \pm15.3$\\
    \hline
  \end{tabular}
  \label{SEFD_table}
\end{table}

Fig.~\ref{fig:SEFD} shows the SEFD for each channel. These are derived using the data collected during the transit  of Cyg~A from 2016/09/28 to 2016/10/02. The three panels show the SEFD derived using the three methods given in Eq.(\ref{eq:SEFD}): the left panel shows the one derived using the ratio of background value and Cyg~A
transit value; the center panel shows the one obtained from the fluctuation of the auto-correlation, while the right panel shows the one obtained from the fluctuation of the cross-correlation. In each panel, we arrange the feeds according to their cylinder and polarization (horizontal axis) and feed number (vertical axis). The malfunctioning feed channels are masked (shown as white blocks in the figure);  the A (B) cylinder has only 31 (32) feeds, hence the 2 (4) white blocks at the bottom of their respective columns.

Table.~\ref{SEFD_table} lists the average and standard deviation for the X- and Y- polarization of all feeds (excluding malfunctioning ones). The mean SEFD values obtained with the three methods agree with each other within the error range. The Y-polarization value is greater than the X-polarization. Taking into account the larger effective area for the X-polarization, we obtain similar system temperatures for the two polarizations as 
explained below. 

\begin{table*}
  \centering
  \caption{The mean $T_{\rm sys}$ of different cylinders and polarizations.(Unit: K) }
  \begin{tabular}{c | p{0.7cm}<{\centering} p{0.7cm}<{\centering}
  p{0.7cm}<{\centering} p{0.7cm}<{\centering}
  p{0.7cm}<{\centering} p{0.7cm}<{\centering}|
  p{0.7cm}<{\centering} p{0.7cm}<{\centering}
  p{0.7cm}<{\centering}  p{0.7cm}<{\centering} p{0.7cm}<{\centering}  p{0.7cm}<{\centering} }
    \hline
    \multirow{2}{*}{Method}&\multicolumn{6}{c|}{Cylinder and Polarization} & \multicolumn{6}{c}{All}\\
    \cline{2-13}
    & AX & AY & BX & BY & CX & CY & A & B & C & X & Y & All\\
    \hline
    1 & 82.8 & 105.1 & 82.1 & 90.8 & 79.5 & 77.9 & 94.3 & 86.4 & 78.7 & 81.4 & 91.0 & 86.3\\
    2 & 83.4 & 106.3 & 82.9 & 92.3 & 78.8 & 77.2 & 95.2 & 87.5 & 78.0 & 81.6 & 91.7 & 86.7\\
    3 & 99.7 & 119.0 & 79.4 & 87.4 & 94.2 & 84.0 & 109.5 & 83.4 & 89.2 & 90.9 & 96.7 & 93.8\\
    All & 88.6 & 110.1 & 81.5 & 90.2  & 84.2 & 79.7 & 99.7 & 85.8 & 82.0 & 84.8 & 93.3 & 88.9\\
    \hline
  \end{tabular}
  \label{Tsys_table}
\end{table*}

Using the effective area from simulation, the system temperature can be estimated. The system temperatures for all feed channels as determined from the three methods are shown in Fig.~\ref{fig:system_temperature}, and the averages for different cylinders and polarizations are listed in Table.~\ref{Tsys_table}. The results obtained by the three methods are generally similar with each other. This is especially true for method 1 and method 2, both derived from auto-correlations. The average for all feed and polarization is 86.3 K and 86.7 K respectively, and on individual feeds they also agree with each other in most cases, though there are also some exceptions. For method 3 the average is 93.8 K, slightly larger than method 1 and 2. In method 3 the SEFD and $T_{\rm sys}$ were not directly obtained for each feed channel, but were instead obtained for pairs of channels,  then solved from these. Perhaps for this reason, it appears to be more smoothly distributed  along each cylinder-polarization. As a whole, the relative difference of $T_{\rm sys}$ between different polarizations is less than that of the SEFD. It seems that the system temperature is more uniform, and part of the SEFD difference comes from the different effective area.

There are obvious differences in the system temperatures for the feeds on the three cylinders. Many feeds on Cylinder A, especially the Y (E-W) polarization ones, have high system temperatures, with an average of 105.1, 106.3, 119.0 K for methods 1, 2, 3 respectively. The average system temperature for both polarization is 94.3, 95.2, 109.5 K for the three methods, and the average of the three methods is 110.1 K. By comparison, the feeds on Cylinders B and C have lower temperature, with an average of only 86.4, 87.5, 83.4 K for the three methods respectively for B, and 78.7, 78.0, 89.2 K respectively for C. If we take the average of the three methods, the system temperature for the A, B, C cylinders are 99.7, 85.8 and 82.0 K respectively.

There are also some variations within each cylinder, with some particularly high and lower ones. It is not known why the system temperatures differ so much. One speculation is that  Cylinder A on the East is closer to the dish array, which may have some impact on the system temperature by its reflection, though this not confirmed in any way.

Generally speaking, the system temperatures we found for the cylinder array feeds are comparable with what we see on the dish array (F. Wu et al.,{\it  in preparation}), which shares similar electronics.  The variation of system temperature for different feeds should be taken into account when making synthesized images from the visibility data to make optimal maps.

\begin{figure}[H]
  \centering
  \includegraphics[width=0.4\textwidth]{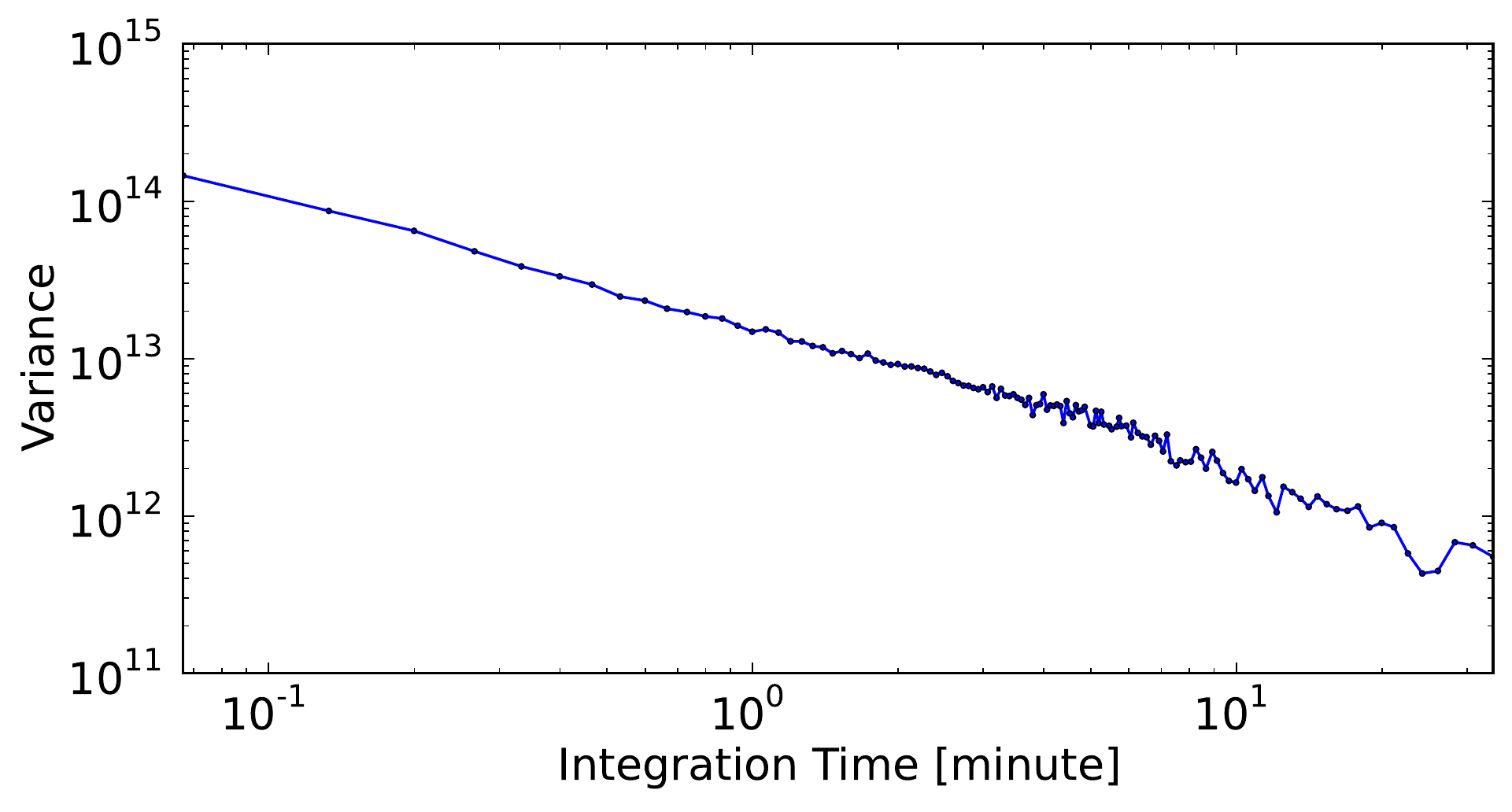}
  \includegraphics[width=0.4\textwidth]{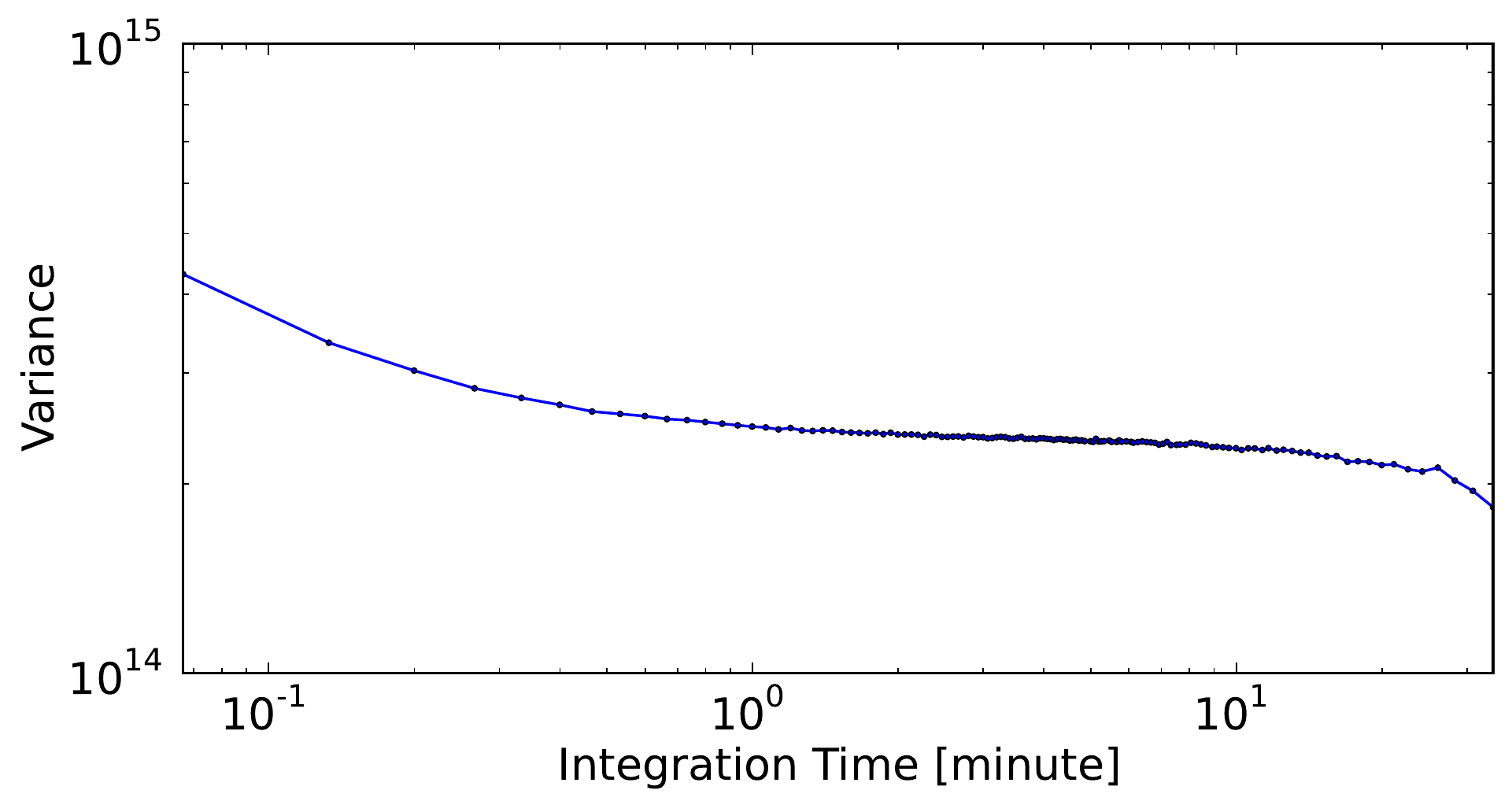}
  \caption{The variance for different integration time scales while observing a relatively blank sky using data on 2018/03/22. Top: variance of baseline A1Y-B2Y versus integration time (averaged visibility). Bottom: for baseline A1Y-A2Y.}
  \label{fig:var_vs_inttime}
\end{figure}

Finally, we consider how the variance of the visibilities scales as a function of integration time. In Fig.~\ref{fig:var_vs_inttime} we plot the variance for different integration time scales while observing a relatively blank sky using data on 2018/03/22. For a typical baseline, for example A1Y-B2Y(top panel), the variance drops steadily as the integration time increases. However, for the feed pairs which are very close to each other, such as A1Y-A2Y (bottom panel), the variance hits a floor after just about a minute of integration, showing the strong coupling and correlated noise in the short baselines. Thus, at least for the longer baselines, the sensitivity can be improved by accumulating longer integration time.

\section{Conclusion}
\label{sec:conclusion}

Neutral hydrogen, ubiquitous throughout cosmic history since recombination, is a potentially powerful tool for cosmological observations. To achieve high precision measurements of the redshifted 21~cm emission, a number of dedicated 21~cm array concepts have been built or are under construction.  Even more ambitious arrays, involving tens of thousands of antennas \cite{Ansari2018,Slosar2019}, have been proposed. Dedicated 21~cm arrays have some common design features determined largely by the nature of the observation and current technology. They all use large numbers of relatively small, inexpensive antennas. The antennas are either fixed, or can only move in elevation, and the observations use either a drift scan mode, or track the target with electronically steered beams.  With computing power following Moore's law scaling and price per computing operation and of data storage  continuously dropping, this approach enables arrays of very large scales to be built with a very moderate cost. Based on simple forecasts, the projected capabilities of these arrays are very impressive (see e.g. Ref. \cite{2010ApJ...721..164S,2012A&A...540A.129A,Xu2015}). However, extracting scientific results from these arrays poses a number of challenges. For example, these arrays produce a huge amount of data.  In addition, with small antennas and uncooled receivers the SNR of the raw data is relatively low; only a few bright sources are available for simple point source calibration, and as the antennas are not movable, calibration with strong point sources can only be performed occasionally. Furthermore, while for forecasting it is customary to assume that the antenna responses are identical, in reality each unit is somewhat different. It is crucial to develop technologies to handle the data from such arrays, and to test the key technologies in order to gain some concrete experience and expose possible problems. The Tianlai pathfinder arrays were built expressly for this purpose.

In this paper, we first described the system functions of the Tianlai Cylinder Pathfinder (Sec. \ref{sec:overview}). The hardware (Sec. \ref{subsec:instrument}) consists of three cylinder reflectors and a total of 96 feeds, the optical analog signal transmission system,  the down-converters which convert the RF signal to IF, and the digital FX correlator which  produces the visibilities.  We also briefly introduced the observational data sets (Sec. \ref{subsec:observations}) and basic procedures for data processing (Sec. \ref{subsec:data_process}). The noise figure of the LNAs and the linearity of the system is discussed in Sec. \ref{sec:system_checks}.

A first analysis of data from 2016 and 2018 observations was then presented in Sec. \ref{sec:visibility_preview}). System overall stability has been show cased through plots of typical raw visibilities. In particular, simple repeating patterns, day after day, corresponding to solar system or astrophysical sources are clearly visible,  
or become visible after a noise background is removed. We also find that the correlated noise is 
stronger for feeds close to each other on the same cylinder, making it harder to use the visibilities 
from nearby pairs. While we could only show the data for a few channels in the paper, the pattern found above is quite common in most channels.

In Sec. \ref{sec:beam} we studied the beam profile of our telescope. We used electromagnetic simulations to obtain the beam profile of the cylinder in the 700--800~MHz frequency range. We then measured the beam profile in the E-W direction by analyzing auto-correlations and cross-correlations using the strong source Cyg~A, which passes very close to the zenith during its transit of the meridian. The general shape of the beam is consistent with the simulation, but a few feeds are slightly misaligned, with an error at the level of $\sim 0.15^\circ$, consistent with what one would expect for the precision of installation using mechanical tools. This is an example of the small non-uniformity which occurs naturally in the construction of large radio arrays.  Not taking into account such differences can induce errors in the final data.

Calibration is a crucial step for the processing of interferometer array data, and we presented our method in Sec. \ref{sec:calibration}. First, we performed bandpass calibrations for individual visibilities, and found relatively stable response. The bandpasses obtained for the cross-correlation visibilities are consistent with the product of the bandpasses of individual feeds, as obtained from the auto-correlations. More important is the calibration of complex gains. For an array with a large field of view and low sensitivity, such as the cylinders, a special challenge is that on the one hand, there are few sources which are bright enough to be ``seen'' directly in individual visibilities, and on the other hand, it is very difficult to construct a sky model that has enough precision over the large field of view to include all sources that contribute to the visibility.  In our experiment we tried two methods of calibration. In what we termed  absolute calibration, we use the transit of the strong astronomical source Cyg~A to solve for the complex gain of each input channel, using the eigen-decomposition method we developed earlier \cite{Zuo2019}. At other times, we use a calibration signal broadcast from an artificial source to perform what we call relative calibration. The results of the two methods are compared for the period when they overlap, and distributions of the complex gains are given. The distribution and variations of the gain are plotted, and the phase of the gain is found to be strongly correlated with the environmental temperature. We also checked the precision of this method using the closure phase relation.
The relative calibration process using the noise source will be used to track short time scale drift of the instrument response, mainly the complex gain, while the absolute calibration process will provide more precise and absolute response, but only up to a few times per day.  

Finally, in Sec. \ref{sec:sensitivity}, we estimated the SEFD and system temperature of the array feeds using the ratio of visibility and its fluctuation during the transit of Cyg~A. We found that feeds on cylinder A have generally higher system temperature, averaging 104.2 K, while feeds on cylinder B have the lowest system temperature, averaging 80.7 K. The average system temperature for all feeds is 90.9 K for X polarization and 96.7 K for Y polarization. This is in general agreement with the system temperature obtained for the Tianlai Dish Array, which has similar electronics, and the lower bound of $\sim$ 50 K contribution from the LNAs.

We also examined how the variance of the visibility scales with time. For longer baselines the variance decreases steadily as integration time increases, showing that the measurement error could be reduced by accumulating more data. However, for shorter baselines the variance hits a floor after a short time, presumably caused by the coupling between nearby feeds. At least for the baselines across different cylinders, higher sensitivity can be achieved with longer integration time within the range we tested.

\begin{figure}[H]
  \centering
  \includegraphics[width=0.41\textwidth]{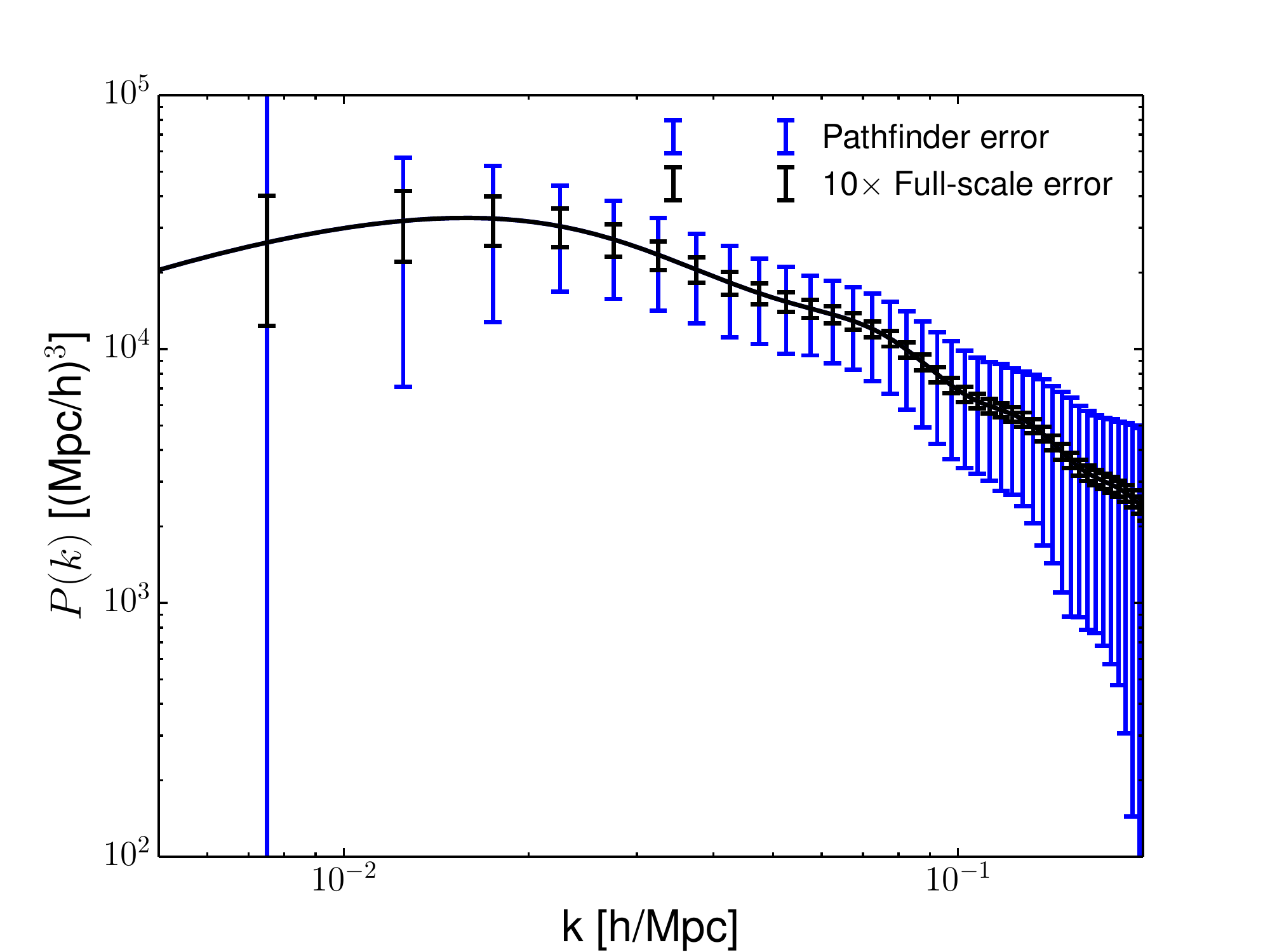}
  \caption{The projected measurement error on the power spectrum for the current Tianlai pathfinder (blue) and for the full-scale cylinder array (black, error bar multiplied by 10). For the full-scale experiment we assume 8 cylinders, each with 256 receiver feeds.}
  \label{fig:ps_noise}
\end{figure}

Using the formalism presented in \cite{Xu2015}, and with the realistic system temperature of about 90 K, we can make forecasts on the  measurement errors on the matter power spectrum. Assuming a survey area of 10000 ${\rm deg}^2$ and an integration time of one year, the expected measurement errors are plotted in Fig.~\ref{fig:ps_noise}. We find that the current pathfinder has the nominal sensitivity to detect the power spectrum, though the complexity in the foreground may make this quite challenging. The full scale Tianlai experiment could in principle measure the power spectrum very precisely (the error bars in Fig.~\ref{fig:ps_noise} for the full scale experiment are enlarged 10 times to make it visible), though it would depend on successful extraction of the 21~cm signal from the foreground, which in turn requires a high level of instrument stability and high precision response model and calibration process.

In this work,  we have only studied a small fraction of the data set collected so far, and presented only some basic performance characteristics. There remain many problems to be solved, for example, the cross-coupling between feeds, the determination of the beam profile in the N-S direction, calibration methods of better precision, map-making, foreground subtraction. More thorough analyses of these problems will be investigated in a series of future work.

{\bf Acknowledgements.}
The Tianlai arrays are operated with the support of the Astronomical Technology Center of National Astronomical Observatories of China (NAOC). The Tianlai cylinder array is built with the support of Ministry of Science and Technology (MOST) grant 2012AA121701, and its survey is supported by MOST grants 2016YFE0100300 and 2018YFE0120800, the National natural Science Foundation (NSFC) grants 11633004, 11473044, 11653003, and the Chinese Academy of Sciences (CAS) QYZDJ-SSW-SLH017. Some on-site experiments are performed with support of the Joint Research Fund in Astronomy (U1631118) under cooperative agreement between NSFC and CAS. The data analysis work is partially supported by the National Key R\&D Program 2017YFA0402603, and the CAS Interdisciplinary Innovation Team (JCTD-2019-05). Part of the computations are performed on the Tianhe-2 supercomputer (with the support of NSFC grant U1501501) and the Tianhe-1 supercomputer. Work at UW-Madison and Fermilab is partially supported by the US National Science Foundation (NSF) Award AST-1616554. Fermilab is operated by Fermi Research Alliance, LLC, under Contract No. DE-AC02-07CH11359 with the US Department of Energy. Authors affiliated with French institutions acknowledge partial support from Centre national de la recherche scientifique (CNRS) via IN2P3 \& INSU, Observatoire de Paris and from Irfu/CEA. XLC acknowledges the support of the Australia-ChinA ConsortiuM for Astrophysical Research (ACAMAR) visiting fellowship, and the University of West Australia, University of Curtin, University of Melbourne, and the University of New South Wales for their hospitality during his visit. 



\begin{appendix}

\section{System Temperature of Channels from Cross Correlation Pairs}
We have assumed the variance of the visibility is given by
\begin{equation}
  \sigma_{ab}^2=\langle |V_{ab}-\bar{V}_{ab}|^2 \rangle = \langle n_a^2 n_b^2 \rangle = C^2 \frac{(T_{\rm sys}^{ab})^2}{\delta t \delta \nu}.
\end{equation}
However, the system temperature here is actually the geometric mean of a pair,
\begin{equation}
  (T_{\rm sys}^{ab})^2 = T_a T_b.
\end{equation}
How do we obtain each system temperature, $T_a$, for the individual feeds?

If we take $\ell_{ab}= \log(T_{s~ ab})$, $\ell_a = \log T_a$, $\ell_b=\log T_b$, we have
\begin{equation}
  \ell_{ab} = \frac{\ell_a +\ell_b}{2}.
\end{equation}
Now $\ell_{ab}$ for all pairs is known, and we need to solve for $\ell_{a}$. We use the average over many pairs to reduce fluctuations, and, to avoid the effect of cross-coupling between feeds, we only use correlations for pairs on different cylinders. (Note that, this method can also be used to solve for the beam pointing in Eq. \ref{eq:beam_pointing_sol} for individual feed from the baseline beam pointing. One can simply regard $\ell_{a}$ as the pointing of individual feed and $\ell_{ab}$ as the transit peak time of the pair.) The solution of this mathematical problem is as follows.

Without loss of generality, we consider the feed $a$ on cylinder $A$,
\begin{eqnarray}
  \label{eq:def1}
  \hat{\ell}_{Aa}(B)&=&\frac{1}{N_B} \sum_{b} \ell_{(Aa) (Bb)}\\
  &=&\frac{1}{2}(\ell_{Aa} + \frac{1}{N_B} \sum_{b} \ell_b).
  \end{eqnarray}
Introducing the notation
\begin{equation}
  L_B = \frac{1}{N_B} \sum_{b} \ell_b,
\end{equation}
we have
\begin{equation}
  \hat{\ell}_{Aa}(B)=\frac{1}{2} (\ell_{Aa} + L_B), \qquad \hat{\ell}_{Aa}(C)=\frac{1}{2} (\ell_{Aa} + L_C).
\end{equation}
Similar equations can be obtained by cycling through $A, B, C$. Thus, the formula for getting the $\ell_{Aa}$ using all different cylinder cross correlations is
\begin{equation}
  \ell_{Aa} = \hat{\ell}_{Aa}(B) + \hat{\ell}_{Aa}(C) - \frac{1}{2}(L_B + L_C).
  \label{eq:sol1}
\end{equation}
If we know $L_B$ and $L_C$ we can obtain the desired results.

To obtain $L_A, L_B, L_C$, we note that
$$\sum_{a} \hat{\ell}_{Aa} (B) =\frac{1}{2}(\sum_{a}\ell_{Aa} + N_A L_B), $$
using Eq.~(\ref{eq:def1}), and we have
\begin{equation}
  \sum_{a} \hat{\ell}_{Aa} (B)  =\frac{N_A}{2}(L_A+L_B).
\end{equation}
Thus, we have
\begin{eqnarray}
  L_A + L_B &=& \frac{2}{N_A N_B} \sum_{a,b} \hat{\ell}_{(Aa)(Bb)} \\
  L_B + L_C &=& \frac{2}{N_B N_C} \sum_{b,c} \hat{\ell}_{(Bb)(Cc)} \\
  L_C + L_A &=& \frac{2}{N_C N_A} \sum_{c,a} \hat{\ell}_{(Cc)(Aa)}.
\end{eqnarray}
Note the R.H.S. is entirely known from the observations.

Substituting these expressions into Eq.(\ref{eq:sol1}), we finally find the solution to be
\begin{eqnarray}
  \ell_{Aa} &=& \frac{1}{N_B} \sum_{b} \ell_{(Aa)(Bb)} + \frac{1}{N_C} \sum_{c} \ell_{(Cc)(Aa)} -\frac{1}{N_B N_C} \sum_{b,c} \ell_{(Bb)(Cc)}\nonumber\\
  \ell_{Bb} &=& \frac{1}{N_C} \sum_{c} \ell_{(Bb)(Cc)} + \frac{1}{N_A} \sum_{a} \ell_{(Aa)(Bb)} -\frac{1}{N_C N_A} \sum_{c,a} \ell_{(Cc)(Aa)}\nonumber\\
  \ell_{Cc} &=& \frac{1}{N_A} \sum_{a} \ell_{(Cc)(Aa)} + \frac{1}{N_B} \sum_{b} \ell_{(Bb)(Cc)} -\frac{1}{N_A N_B} \sum_{a,b} \ell_{(Aa)(Bb)}\nonumber.\\
  \label{eq:solabc}
\end{eqnarray}
Thus, applying Eq.~(\ref{eq:solabc}) we can get the system temperature for each feed from the cross-correlation of feed pairs.\\\\

\renewcommand{\thesection}{Appendix}
\end{appendix}

\end{multicols}

\end{document}